\documentclass{vldb}
\usepackage[export]{adjustbox}
\usepackage{pbox}
\usepackage{caption,graphicx,newfloat}
\usepackage{afterpage}
\usepackage{pifont}
\usepackage{algorithm,algpseudocode}
\usepackage{amsmath}
\usepackage{soul}
\usepackage{listings}
\usepackage{empheq}
\usepackage{etoolbox}
\usepackage{fancyvrb}

\usepackage{amsthm}
\usepackage{caption}
\usepackage{subcaption}
\usepackage{times}
\usepackage{amsfonts}
\usepackage{colortbl}
\usepackage{paralist}
\usepackage{enumitem}
\usepackage[dvipsnames]{xcolor}
\usepackage{epsfig}
\usepackage{amsmath}
\usepackage{forest}
\usepackage{graphicx}
\usepackage{multicol}
\usepackage{multirow}
\usepackage{pgf}
\usepackage{pdflscape}
\usetikzlibrary{automata,positioning}
\usepackage[normalem]{ulem}
\usepackage{relsize}
\usepackage{textcomp}
\usepackage{tikz}
\usepackage[T1]{fontenc}
\usepackage{wrapfig}
 \usepackage{xspace}
\usepackage{balance}
\usepackage{thmtools,thm-restate}
\usepackage[normalem]{ulem}
\usepackage{url}
\usepackage{amsmath}
\usepackage{amssymb}
\usepackage{amsthm}
\usepackage[normalem]{ulem}
\usepackage{tikz}
\usetikzlibrary{matrix,positioning}

\definecolor{redi}{RGB}{255,38,0}
\definecolor{redii}{RGB}{200,50,30}
\definecolor{yellowi}{RGB}{255,251,0}
\definecolor{bluei}{RGB}{0,150,255}
\definecolor{blueii}{RGB}{135,247,210}
\definecolor{blueiv}{RGB}{115,244,253}
\definecolor{bluev}{RGB}{1,58,215}
\definecolor{orangei}{RGB}{240,143,50}
\definecolor{yellowii}{RGB}{222,247,100}
\definecolor{greeni}{RGB}{166,247,166}

\definecolor{blueiii}{rgb}{0.94, 0.92, 0.84}

\tikzset{ 
	table/.style={
		matrix of nodes,
		row sep=-\pgflinewidth,
		column sep=-\pgflinewidth,
		nodes={rectangle,draw=black,text width=3.0ex,align=center,	font=\large},
		nodes in empty cells
	},
	texto/.style={font=\large\sffamily},
	title/.style={font=\large\sffamily}
}

\newcommand\CellText[2]{%
	\node[texto,left=of mat#1,anchor=east]
	at (mat#1.west)
	{#2};
}

\newcommand\SlText[2]{%
	\node[texto,left=of mat#1,anchor=west,rotate=75]
	at ([xshift=3ex]mat#1.north)
	{#2};
}

\newcommand\RowTitle[2]{%
	\node[title,left=6.3cm of mat#1,anchor=west]
	at (mat#1.north west)
	{#2};
}

\renewcommand{\paragraph}[1]{\vspace{2pt} \noindent {\bf #1}}

\newcommand{\rev}[1]{{\leavevmode{#1}}}

\newcommand{\cut}[1]{}

\newcommand{\smallcaption}[1]{{\small{#1}}}

\renewcommand{\paragraph}[1]{\vspace{2pt} \noindent {\bf #1}}
\renewcommand{\paragraph}[1]{\vspace{2pt} \noindent {\bf #1}}
\newcommand{\eat}[1]{}
\newcommand{\later}[1]{}

\newcommand{\fixlater}[1]{}
\newcommand{\tar}[1]{}

\renewcommand{\paragraph}[1]{\vspace{2pt} \noindent {\bf #1}}

\newcommand{\system}{\textsc{Compare}\xspace}

\newcommand{\db}{\textsc{SQL Server}\xspace}
\newcommand{\dbb}{\textsc{SQL Server}\xspace}
\newcommand{\mdw}{\textsc{Middleware}\xspace}

\newcommand{\constraint}{{constraint}}
\newcommand{\constraints}{{constraints}}
\newcommand{\indexi}{{grouping}}
\newcommand{\measure}{{measure}}
\newcommand{\group}{{trend}}
\newcommand{\groups}{{trends}}
\newcommand{\scorer}{{scorer}}
\newcommand{\groupset}{{trendset}}
\newcommand{\groupsets}{{trendsets}}
\newcommand{\seto}{{\sf Set1\xspace}}
\newcommand{\sett}{{\sf Set2\xspace}}

\newcommand{\optr}{\textsc{Compare}\xspace}

\newcommand{\cw}{{\fontfamily{qpl}\selectfont 
		week%
}}

\newcommand{\cdc}{{\fontfamily{qpl}\selectfont 
		sales%
}}

\newcommand{\cdr}{{\fontfamily{qpl}\selectfont 
		R%
}}

\newcommand{\cf}{$\mathcal{F}$\xspace}

\newcommand{\cg}{$\mathcal{G}$\xspace}

\newcommand{\cdcc}{{$m_1$}}
\newcommand{\cdca}{{$m_2$}}

\newcommand{\cdk}{$K$}

\newcommand{\cdt}{{\fontfamily{qpl}\selectfont 
		T%
}}

\newcommand{\cp}{{\fontfamily{qpl}\selectfont 
		product%
}}

\newcommand{\cdiff}{DIFF($m_1, m_2, 2$)\xspace}

\newcommand{\cs}{\textcolor{brown}{\bf \texttt{<->}}}

\newcommand{\diff}{\sf DIFF\xspace}

\newcommand{\pqp}{$\mathcal{PQ}_\mathcal{P}$\xspace}
\newcommand{\pqs}{$\mathcal{PQ}_\mathcal{S}$\xspace}

\newcommand{\ptstate}{{\em{TState} \xspace}}
\newcommand{\segagg}{{\em SegAgg \xspace}}

\newcommand{\amax}{MAX\xspace}
\newcommand{\amin}{MIN\xspace}
\newcommand{\asum}{SUM\xspace}
\newcommand{\aavg}{AVG\xspace}
\newcommand{\acount}{COUNT\xspace}

\makeatletter
\def\thmheadbrackets#1#2#3{%
  \thmname{#1}\thmnumber{\@ifnotempty{#1}{ }\@upn{#2}}%
  \thmnote{ {\the\thm@notefont[#3]}}}
\makeatother

\newtheoremstyle{brakets}% Name
  {}% space above
  {}% space below
  {\itshape}% body font
  {}% indent
  {\bfseries}% head font
  {.}% punctuation after head
  { }% space after head (has to be space or dimension!)
  {\thmheadbrackets{#1}{#2}{#3}}% head spec

\newtheoremstyle{defbrakets}% Name
  {}% space above
  {}% space below
  {\normalfont}% body font
  {}% indent
  {\bfseries}% head font
  {.}% punctuation after head
  { }% space after head (has to be space or dimension!)
  {\thmheadbrackets{#1}{#2}{#3}}% head spec

\theoremstyle{brakets}

\newtheorem{definition}{Definition}

\newtheorem{theorem}[definition]{Theorem}
\newtheorem{problem}{Problem}[section]
\newenvironment{denselist}{
    \begin{list}{\small{$\bullet$}}%
    {\setlength{\itemsep}{0ex} \setlength{\topsep}{0ex}
    \setlength{\parsep}{0pt} \setlength{\itemindent}{0pt}
    \setlength{\leftmargin}{0.8em}
    \setlength{\partopsep}{0pt}}}%
    {\end{list}}

\newcounter{enum}
\newenvironment{packed_enum}{
\begin{list}{\textbf{(\arabic{enum})}}{
  \setlength{\itemsep}{0pt}
  \setlength{\parskip}{0pt}
  \setlength{\labelwidth}{-4 pt}
  \setlength{\leftmargin}{0 pt}
  \setlength{\itemindent}{0pt}
  \usecounter{enum}}
}{\end{list}}

\newcommand{\techreport}[1]{{#1}}

\newcommand{\papertext}[1]{}
\newcommand{\stitle}[1]{\vspace{0.25em}\noindent\textbf{#1}}

\newif\iftechreportcode
\techreportcodetrue

\newif\ifpapertext
\algdef{SE}[DOWHILE]{Do}{doWhile}{\algorithmicdo}[1]{\algorithmicwhile\ #1}%

\newcommand{\squishlist}{
   \begin{list}{$\bullet$}
    { \setlength{\itemsep}{0pt}
      \setlength{\parsep}{2pt}
      \setlength{\topsep}{0pt}
      \setlength{\partopsep}{0pt}
      \leftmargin=25pt
\rightmargin=0pt
\labelsep=5pt
\labelwidth=10pt
\itemindent=0pt
\listparindent=0pt
\itemsep=\parsep
    }
}

\DeclareFloatingEnvironment[
  fileext=lob,
  listname={List of Boxes},
  name=Table
]{BOX}

\newcommand{\squishend}{\end{list}}

\makeatletter
\def\@copyrightspace{\relax}
\makeatother

\begin{document}

%\title{Accelerating Comparative Queries in Relational Databases for Data Analytics}

\title{COMPARE: Accelerating Comparative Queries in Relational Databases for Data Analytics}

\title{COMPARE: Accelerating Groupwise Comparison in Relational Databases for Data Analytics}

\title{COMPARE: Accelerating Groupwise Comparison in Relational Databases for Data Analytics\\
(Extended Version)}

\author{
\alignauthor{
Tarique Siddiqui
\hspace{1cm}
Surajit Chaudhuri}
\hspace{1cm}
Vivek Narasayya\\
\vspace{.2cm}
       Microsoft Research\\
       \vspace{.2cm}
       \{tasidd, surajitc, viveknar\}@microsoft.com
}

%\title{Accelerating Groupwise Comparison in Relational Databases for Comparative Analytics}
%\papertext{\input{cover}}
%\pagenumbering{gobble}
\maketitle

\abstract
Data analysis often involves \emph{comparing} subsets of data across many dimensions for finding unusual trends and patterns. While the comparison between subsets of data can be expressed using SQL, they tend to be complex to write, and suffer from poor performance over large and high-dimensional datasets. In this paper, we propose a new logical operator \optr  for relational databases that concisely captures the enumeration  and  comparison between subsets of data and greatly simplifies the expressing of a large class of comparative queries. 
We extend the database engine with optimization techniques that exploit the semantics of \optr  to significantly improve the performance of such queries. We have implemented these extensions inside Microsoft SQL Server, a commercial DBMS engine. Our extensive evaluation on synthetic and real-world datasets shows that \optr results in a significant speedup over existing approaches, including physical plans generated by today's database systems, user-defined functions (UDFs), as well as  middleware solutions that compare subsets outside the databases.

\vspace{-8pt}
\section{Introduction}
\label{sec:intro}
Comparing subsets of data is an important part of data exploration~\cite{agrawal1993efficient, lin1995fast,  zenvisagevldb, ding2019quickinsights, seedb},  routinely performed by data scientists to find unusual patterns and gain actionable insights. For instance, market analysts often compare products over different attribute combinations (e.g., revenue over week, profit over week, profit over country, quantity sold over week, etc.) to find the ones with similar or dissimilar sales. However, as the size and complexity of the dataset increases, this manual enumeration and comparison of subsets becomes challenging.
To address this, a number of visualization tools~\cite{seedb,  wongsuphasawat2017voyager,ding2019quickinsights, zenvisagevldb} have been proposed that automatically compare subsets of data to find the ones that are relevant. Figure 1a depicts an example from Seedb~\cite{seedb} where the user specifies the subsets of population (e.g.,  based on marital status,  race)  and the tool automatically find a socio-economic indicator (e.g.,  education,  income,  capital gains) on which the subsets differ the most. \techreport{ Similarly, Figure 1b depicts an example from Zenvisage~\cite{zenvisagevldb} for finding states with similar house pricing trends.} Unfortunately, \rev{most of these tools perform comparison of subsets in a middleware and as depicted in Figure 2, with the increase in size and number of attributes in the dataset, these tools 
incur large data movement as well as serialization and deserialization overheads, resulting in poor latency and scalability.}
%Additionally, these tools require data scientists to learn and switch to their domain-specific querying interfaces, thereby limiting their broad adoption.}

\techreport{
\begin{figure}[t!]
	\hspace{-0.2cm}
	\vspace{-10pt}
	\centering
	\begin{subfigure}{0.80\columnwidth}
		\centerline {
			\hbox{\resizebox{\columnwidth}{!}{\includegraphics{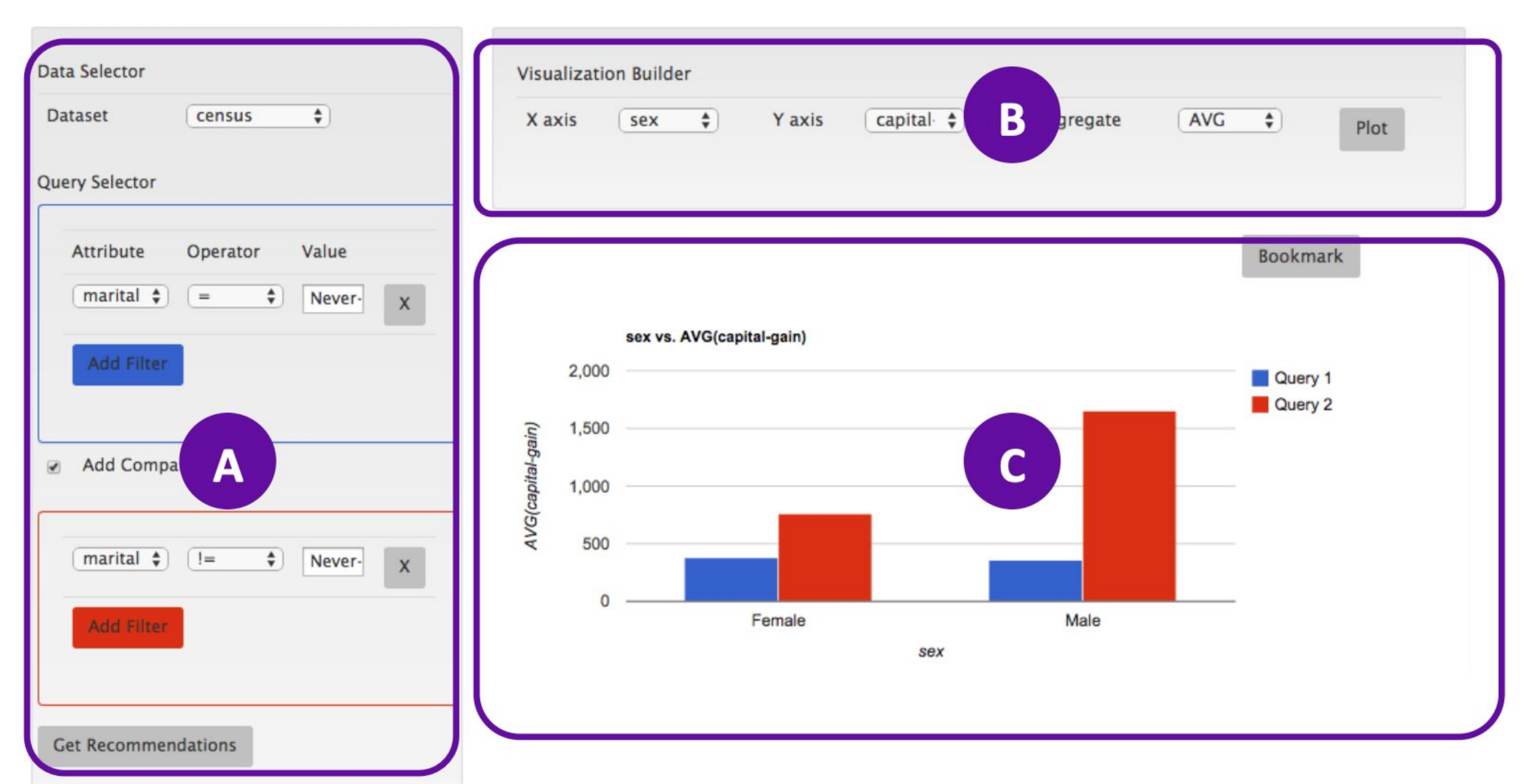}}}}
		\caption{Seedb~\cite{seedb}}
		\caption*{}
		\label{fig:seedbdash}
	\end{subfigure}
	\hspace{0.2cm}
	\begin{subfigure}{0.80\columnwidth}
		\centerline {
			\hbox{\resizebox{\columnwidth}{!}{\includegraphics{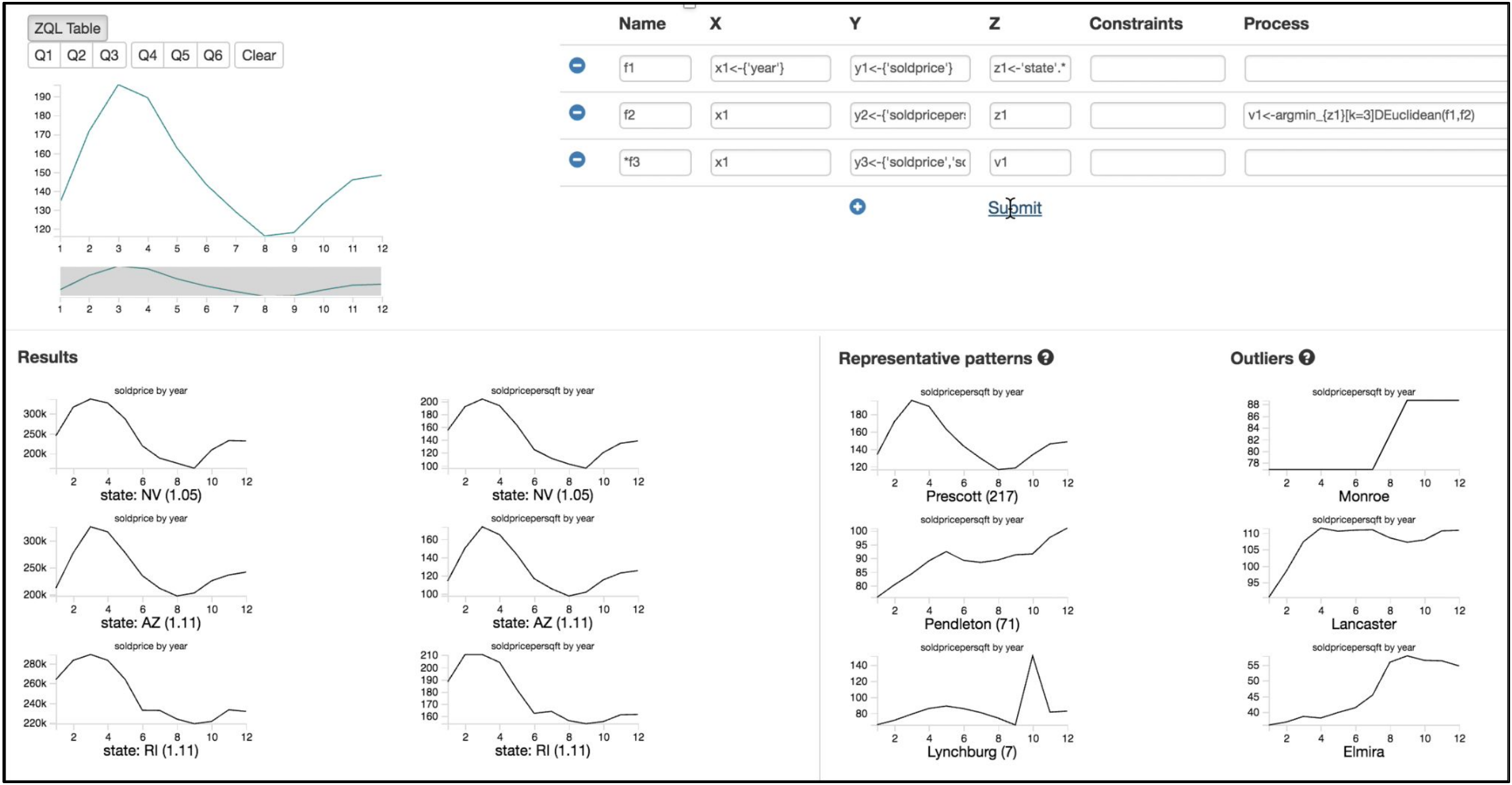}}}}	
		\caption{Zenvisage~\cite{zenvisagevldb}}
		%	\caption{\smallcaption{Searching states with comparable housing price {\groups} in Zenvisage~\cite{zenvisagevldb} }}
		\label{fig:zenvisagedash}
	\end{subfigure} 
	\hspace{0.2cm}
	\caption{\small Examples of comparative queries from visual analytic tools: a) Finding socio-economic indicators that differentiate married and unmarried couples in Seedb~\protect\cite{seedb}.The user specifies the subsets (A) after which the tool outputs a pair of attributes (B) along with corresponding visualizations (C) that differentiates the subsets the most. b) A comparative query in Zenvisage~~\protect\cite{zenvisagevldb} for finding states with comparable housing price trends. \tar{make diagrams bigger?}}
	\label{fig:motiv}	
	\vspace{-10pt}
\end{figure}
}

The question we pose in this work is: \emph{can we efficiently perform comparison between subsets of data within the relational databases to improve  performance and scalability of comparative queries?}
\rev{Supporting such queries within relational databases also makes them broadly accessible via general-purpose data analysis tools such as PowerBI~\cite{powerbi},  Tableau~\cite{tableau},  and Jupyter notebooks~\cite{jupyter}.
All of these tools let users directly write SQL queries and execute them within the DBMS to reduce the amount of data that is shipped to the client.} 
%In contrast, the aforementioned  visualizations tools require data scientists to learn and switch to their domain-specific querying interfaces, thereby limiting their broad adoption.

\begin{figure}
	\centerline {
		\hbox{\resizebox{0.9\columnwidth}{!}{\includegraphics{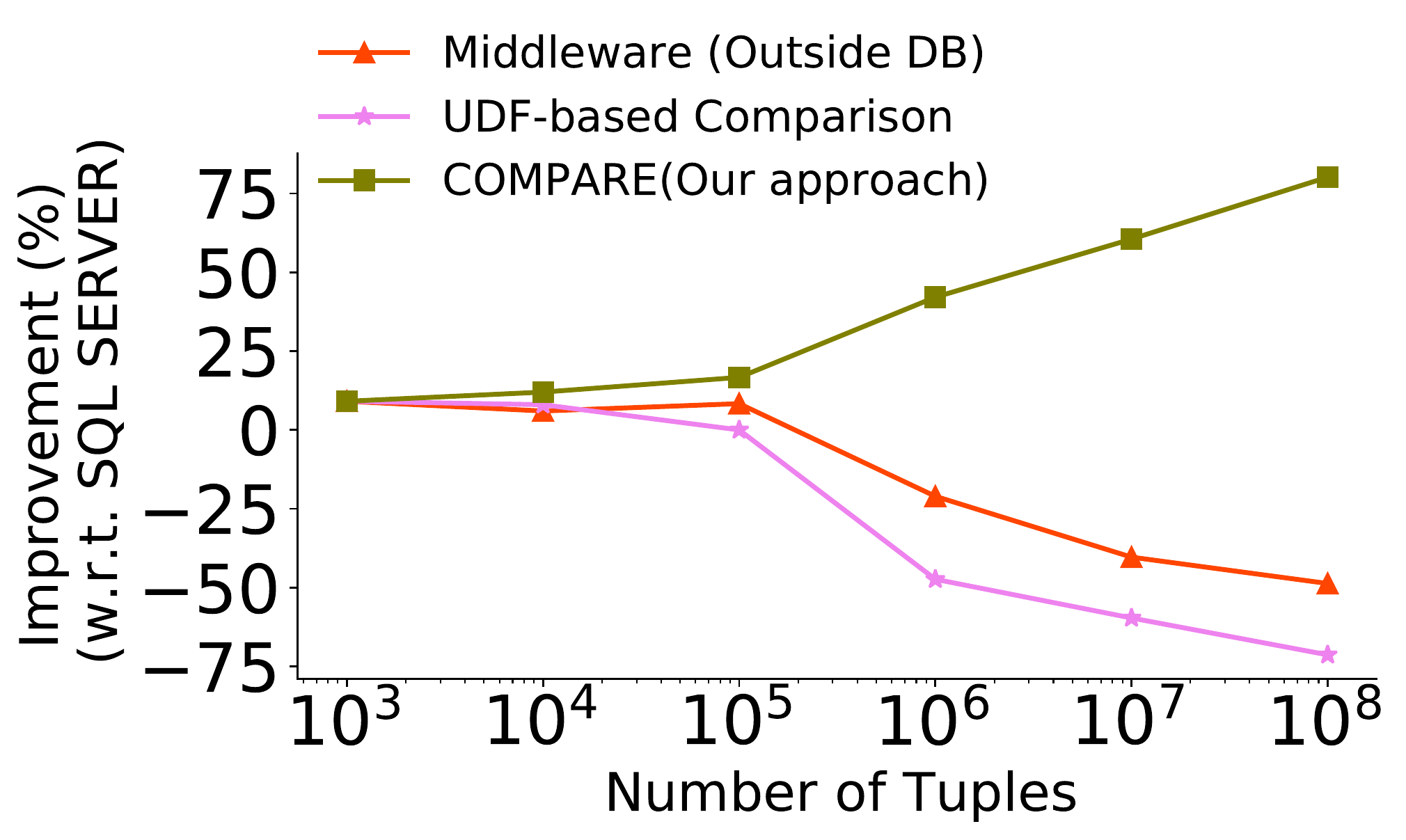}}}}
	\vspace{-10pt}
	\caption{\small Relative performance of different execution approaches for a comparative query w.r.t unmodified \db execution time (higher the better). The query finds a pair of origin airports that have the most similar departure delays over week trends in the flight dataset~\protect\cite{airlinedata}}
	\label{fig:motivperf}
	\techreport{\vspace{-10pt}}
\end{figure}

One option for in-database execution is to extend DBMS with \emph{custom user-defined functions} (UDFs) for comparing subsets of data. However, UDFs incur invocation overhead and are executed as a batch of statements where each statement is run sequentially one after other with limited resources (e.g.,  parallelism,  memory). As such, the performance of UDFs does not scale with the increase in the number of tuples (see Figure 2). Furthermore,  UDFs have limited interoperability with other operators, and are less amenable to logical optimizations, e.g., PK-FK join optimizations over multiple tables. 

While comparative queries can be expressed using regular SQL, such queries require complex combination of multiple subqueries. The complexity makes it hard for relational databases to find efficient physical plans, resulting in  poor performance.  While prior work have proposed extensions  ~\cite{chatziantoniou1996querying, chatziantoniou2007using, gray1997data, cunningham2004pivot, tang2015similarity} such as grouping variables, GROUPING SETs, CUBE; as we discussed in the later sections, expressing and optimizing \emph{grouping} and \emph{comparison} simultaneously remains a challenge. To describe the complexity using regular SQL, we use the following example.

%As creating and examining all possible visualizations for cities and attributes is tedious and time-consuming, 
\vspace{5pt}
\stitle{Example.} {\em Consider a market analyst exploring sales trends across different cities. The analyst generates a sample of visualizations depicting different trends, e.g., average revenue over week, average profit over week, average revenue over country, etc., for a few cities. She notices that trends for cities in Europe look different from those in Asia. To verify whether this observation generalizes, she looks for a counterexample  by searching for pairs of attributes over which two cities in Asia and Europe have most similar trends. Often,  an $L_p$ norm-based distance measure (e.g.,  Euclidean distance,  Manhattan distance) that measures deviation between trends and distributions is used for such comparisons~\cite{zenvisagevldb,seedb, ding2019quickinsights}.}
\vspace{5pt}
%Figure 3 depicts a SQL query that compares trends of products over multiple dimensions (e.g., average revenue over week, average profit over time, average revenue over country, etc.).

Figure 3 depicts a SQL query template for the above  example. The query involves multiple subqueries, one for each attribute pair. Within each subquery, subsets of data (one for each city) are aggregated and compared via a sequence of self-join and aggregation functions that compute the  similarity (i.e.,  sum of squared differences). Finally, a join and filter is performed to output the tuples of subsets with minimum scores. Clearly, the query is quite verbose and complex,  with redundant expressions across subqueries. While comparative queries often explore and compare a large number of attribute pairs~\cite{seedb,  lee2019you},  we observe that even with only  a few attribute pairs,  the SQL specification can become extremely long.

%Due to complex nature of comparative queries, query optimizers~\footnote{ 	Besides Microsoft SQL Server,  we also observed highlighted issues in MySQL and PostgreSQL} often fail to capture the semantics of comparative queries,  resulting in poor performance. 
Furthermore, the number of groups to compare can often be large---determined by the number of possible constraints (e.g., citi- es), pairs of attributes,  and aggregation functions---which grow significantly with the increase in dataset size or number of attributes. This results in many subqueries with each subquery taking substantially long time to execute.
In particular, while there are large opportunities for sharing computations (e.g.,  aggregations) across subqueries,  the relational engines execute subqueries for each attrib- ute-pair separately resulting in substantial overhead in both runtime as well as storage. Furthermore, as depicted in subquery 1 in Figure 3,  while each pair of groups (e.g.,  set of tuples corresponding to each city) can be compared independently,  the relational engines perform an expensive self-join over a large relation consisting of all groups. The cost of doing this increases super-linearly as the number and size of subsets increases (discussed in more detail in Section 4.1). Finally, in many cases,  we only need the  aggregated result for each comparison; however  the join results in large intermediate data---one tuple for each pair of matching tuples between the two sets,  resulting in substantial overheads.

\begin{figure}
	\centerline {
		\resizebox{.95\columnwidth}{!}{\includegraphics{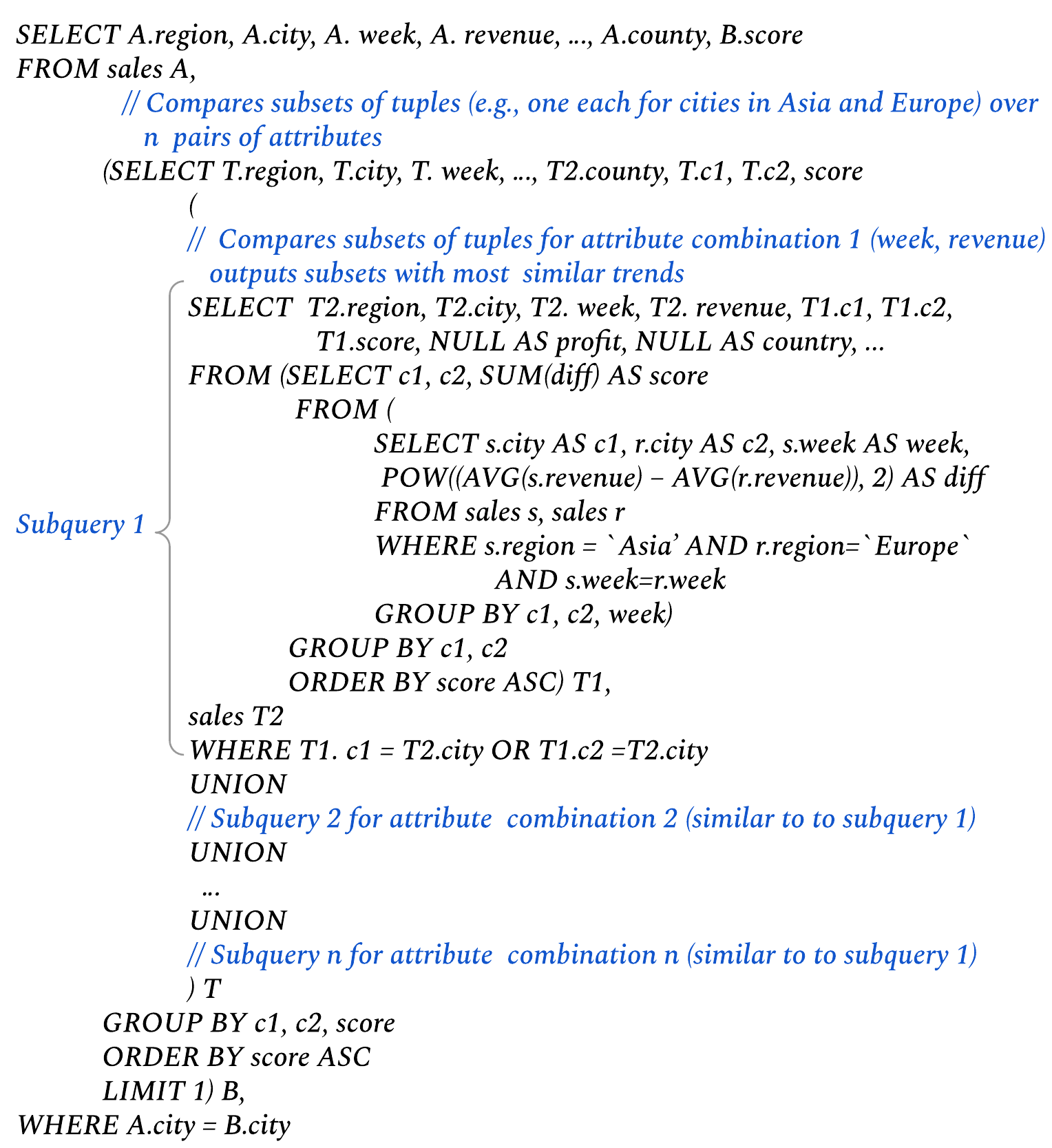}}}
	\vspace{-5pt}
	\caption{\smallcaption{A SQL query for comparing subsets of data over different attribute combinations,  depicting the complexity of specification using existing SQL expressions.}}
	\label{fig:sqlqueryex4}
	\vspace{-15pt}
\end{figure}

\subsection{Overview of Our Approach} 

In this paper,  we take an important step towards making specification of the comparative queries easier and ensuring their efficient processing. To do so,   we introduce
a logical operator and extensions to the SQL language, as well as  optimizations in relational databases, described below.

\stitle{Groupwise comparison as a first class construct (Section 2 and 3)}. We introduce a new logical operation,  \optr ($\Phi$),  as a first class  relational construct,  and formalize its semantics that help capture a large class of frequently used comparative queries. We propose extensions to SQL syntax
that allows intuitive and more concise specification of comparative queries. For instance,  the comparison between two sets of cities $C_1$ and $C_2$ over $n$ pairs of attributes: ($x_1$,  $y_1$),  ($x_2$,  $y_2$),  ...,  ($x_n$,  $y_n$) using a comparison function {\cf} can be succinctly expressed as \optr [$C_1$\cs$C_2$][ ($x_1$,  $y_1$),  ($x_2$,  $y_2$),  ...,  ($x_n$,  $y_n$)] USING \cf. As illustrated earlier,  expressing the same query using existing SQL clauses requires a UNION over $n$ subqueries,  one for each ($x_i$, $y_i$) where each subquery itself tends to be quite complex. Overall,  \rev{while \optr does not give additional expressive power to the relational algebra,  it
reduces the complexity of specifying comparative queries and facilitates 
optimizations via query optimizer and the execution engine.}

\begin{figure*}
	\vspace{-20pt}
	\centerline {
	\papertext{	\hbox{\resizebox{0.85\textwidth}{!}{\includegraphics{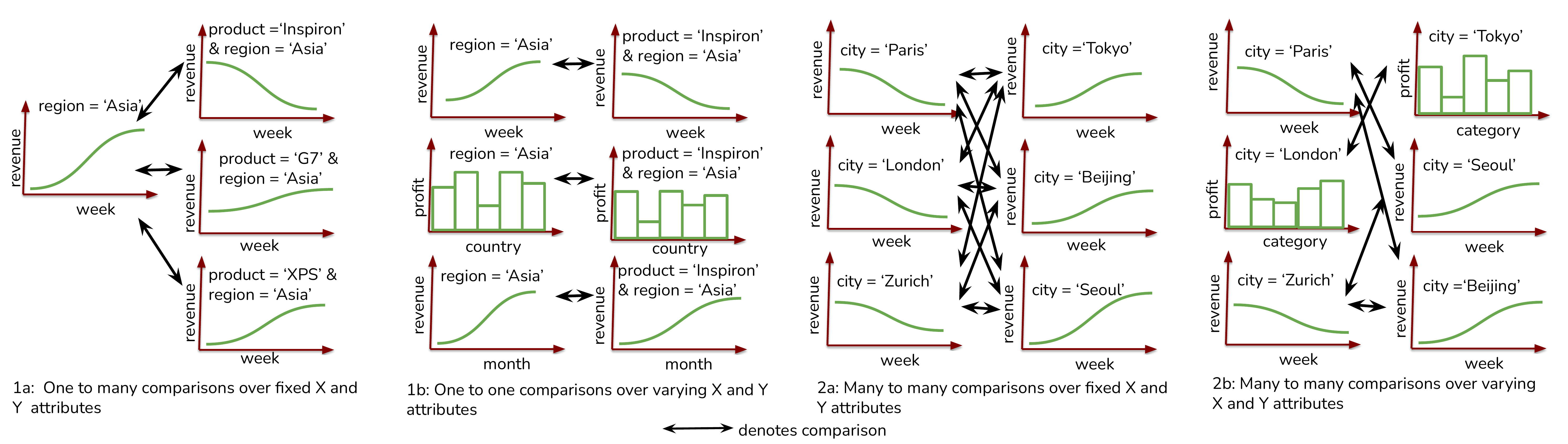}}}}}
		\techreport{	\hbox{\resizebox{\textwidth}{!}{\includegraphics{figs/examples1.pdf}}}}
	\vspace{-10pt}
	\caption{\smallcaption{Illustrating comparative queries described in Section 2.1}}
	\label{fig:examplesmotiv}
	\vspace{-15pt}
\end{figure*}

%While it is also possible to extend the optimizer with algebraic transformation rules to translate comparative subqueries expressed using standard SQL operators to $\Phi$.

%While \optr does not add to the expressivity of relational algebra,   it helps introduce the notion of subset comparison in the optimizer,  making it easier to represent and reason about optimizations,  as well as more effectively inter-operate with other relational operators.
\fixlater{motivate the need for complex query?}

\stitle{Efficient processing via optimizations (Section 4 and 5)}. We exploit the semantics of \optr to share aggregate computations across multiple attribute combinations,  as well as partition and compare subsets in a manner that significantly reduces the processing time.
While these optimizations work for any comparison function,  we also introduce specific optimizations (by introducing a new physical operator) that exploit properties of frequently used comparison functions (e.g.,  $L_p$ norms). These optimizations help prune many subset comparisons without affecting the correctness.

%We introduce a new physical operator,  $\Phi_P$,  for the logical operator \optr,  that ensures  efficient execution of comparative queries. $\Phi_P$ exploits the semantics of \optr operation to partition and compare tuples in a groupwise manner that significantly reduces the number of tuple comparisons as well as the size of intermediate data.  In addition,  $\Phi_P$ incorporates several  physical optimizations that prune a large number of tuple comparisons without affecting the accuracy.
\eat{
	For instance,  for the example 1,  $\Phi_P$ minimizes tuple comparisons and intermediate data by partitioning the input relation and performing comparison independently and in parallel for each product. In addition,  $\Phi_P$ applies several physical optimizations that help in identifying the top $K$ products without computing the exact score. At a high-level,  $\Phi_P$ can prune low scoring products without performing tuple-level comparisons by only comparing a few coarse-grained aggregates of partitions,  that are faster to compute. Among the partitions that are not pruned,  $\Phi_P$ further prioritizes tuple comparisons in an order that leads to faster convergence of the top $K$ partitions.
}
%For example,  $\Phi_P$ summarizes sales of each product using coarse-grained aggregates that help prune out low scoring product without performing their tuple-level comparisons. $\Phi_P$ further prioritizes products that are most likely to be in top K and performs tuple comparison in an order that minimize wastage of tuple comparisons. 

\stitle{Inter-operator optimizations (Section 6)}. We introduce new  transformation rules that  transform the logical tree containing the \optr operator along with other relational operators into equivalent logical trees that are more efficient. For instance, the attributes referred in \optr may be spread across multiple tables, involving PK-FK joins between fact and dimension tables. To optimize such cases, we show how we can push \optr below join that reduces the number of tuples to join. Similarly, we describe how aggregates can be pushed below \optr, how multiple \optr operators can reordered and how we can detect and translate an equivalent sub-plan expressed using existing relational operators to \optr.

%We also introduce a cost model for the physical operator $\Phi_p$ that enables the optimizer to cost plans containing $\Phi_p$. 

%We integrate \optr within the optimizer of \db that enables automatic translation from logical ($\Phi$)  to physical ($\Phi_p$) operator using cost-based optimization.  We further extend the optimizer with algebraic transformation rules that allow logical optimizations when \optr occurs as a subquery in a larger SQL query. 

% of queries where \optr occurs as a sub-component in a larger query. via reordering of \optr to minimize the overall query processing time of queries where \optr occurs as a sub-component in a larger query. 
%\stitle{C4. Extensions (Section~\ref{sec:extensions}).}  We  discuss several important  extensions that build on the core ideas of \optr. For example,  we show how to use multiple groups as references for comparison,  how we can support query-by-sketch systems~\cite{visual2,  timesearcher, zenvisagevldb, leT2019you},  and how we can support additional comparison metrics and user-defined functions (UDF).

\stitle{Implementation inside commercial database engine (Section 7).}  
We have prototyped our techniques in Microsoft \db engine,  including the physical optimizations. %We perform a thorough performance study of the extended \db with that of the unmodified \db on both real world and synthetic datasets. 
Our experiments show that even over moderately-sized datasets (e.g., $10$--$20$ GB) \optr results in up to 4$\times$ improvement in performance relative to alternative approaches including physical plans generated by \db, UDFs, and middlewares (e.g., Zenvisage, Seedb). With the increase in the number of tuples and attributes, the performance difference grows quickly, with \optr giving more than a order of magnitude better performance.
%ranging from $3$$\times$ to $60$$\times$.

%In the rest of the paper, we describe each of the above contributions in detail. We first start by characterizing comparative queries.

\vspace{-5pt}
\section{Characterizing Comparative\\ Queries}
In this section, we first characterize comparative queries with the help of additional examples drawn from visualization tools~\cite{ding2019quickinsights, zenvisagevldb, seedb} and data mining~\cite{simtime, bohm2004k, agrawal1993efficient, lin1995fast}. Then, we give a formal definition that concisely captures the semantics of comparative queries.

\subsection{Examples}
We return to the example scenario discussed in introduction: a market analyst is exploring sales trends of products with the help of visualizations to find unusual patterns. The analyst first looks at a small sample of visualizations, e.g., average revenue over week trends for a few regions (e.g.,  Asia,  Europe) and for a subset of cities and products within each region. She observes some unusual patterns and wants to quickly find additional visualizations that either support or disprove those patterns (without examining all possible visualizations). Note that we use the term "trend" to refer to a set of tuples in a more general sense where both categorical (e.g., country) or ordinal attributes (e.g., week) can be used for ordering or alignment during comparison.
We consider several examples below in increasing order of complexity. Figure~\ref{fig:examplesmotiv} illustrates each of these examples, depicting the differences in how the comparison is performed.

%Often,  an $L_p$ norm-based distance measure (e.g.,  Euclidean distance,  Manhattan distance) that measures deviation between trends and distributions is used for such comparisons.

\stitle{Example 1a.} The analyst notes that the average revenue over week trends for Asia as well as for a subset of products in that region look similar. As a counterexample,  she wants to {\em find a product whose revenue over week trend in Asia is very dissimilar (typically measured using $L_p$ norms) to that of the Asia's overall trend.}  There are visualization systems~\cite{visual2,   timesearcher, zenvisagevldb, lee2019you} that support similar queries.

\stitle{Example 1b.} In the above example, the analyst finds that the trend for product `Inspiron' is different from the overall trend for the region `Asia'. She finds it surprising and wants to see the attributes for which trends or distributions of Inspiron and Asia deviate the most. More precisely,  she wants to \emph{compare `Inspiron' and `Asia' over multiple pairs of attributes (e.g.,  average profit over country, average quantitysold over week,  ...,  average profit over week) and select the one where they deviate the most}. Such comparisons can be found in features such as Explain Data\cite{explaindata} in Tableau and tools such as Seedb~\cite{seedb},  Zenvisage~\cite{zenvisagevldb},  Voyager~\cite{wongsuphasawat2017voyager}.

\stitle{Example 2a.} Consider another scenario: the analyst visualizes the revenue trends of a few cities in Asia and in Europe,  and finds that while most cities in Asia have increasing revenue trends, those in Europe have decreasing trends. Again,  as a counterexample to this,  {\em she wants to find a pair of cities in these regions where this pattern does not hold,  i.e.,  they have the most similar  trends}. Such tasks involving search for similar pair of items are ubiquitous in data mining~\cite{rajaraman2010finding} and time series~\cite{simtime,  bohm2004k, agrawal1993efficient, lin1995fast}. 

\stitle{Example 2b.} In the above example, the analyst finds that the output pair of visualizations look different,  supporting her intuition that perhaps no two cities in Europe and Asia have similar revenue over week trends. To verify whether this observation generalizes when compared over other attributes, she \emph{searches for pairs of attributes (similar to ones mentioned in Example 1b) for which two cities in Asia and Europe have most similar trends or distributions}. 
Such queries are common in tools such as Zenvisage~\cite{zenvisagevldb} that support finding outlier visualizations over a large set of attributes.

In summary,  the comparative queries in above examples \emph{fast-forward} the analyst to a \emph{few} visualizations that depict a \emph{pattern} she wants to verify---thereby allowing her to skip the tedious and time-consuming process of manual comparison of all possible visualizations.
As illustrated in Figure~\ref{fig:examplesmotiv},  each query involves comparisons between two sets of visualizations (henceforth referred as {\seto} and {\sett}) to find the ones which are similar or dissimilar. Each visualization depicting a trend is represented via two attributes (X attribute, e.g.,  week and a Y attribute, e.g.,  average revenue) and a set of tuples (specified via a constraint, e.g., product = `Inspiron'). We now present a succinct representation to capture these semantics.

\subsection{Formalization}

We formalize our notion of comparative queries and propose a concise representation for specifying such queries.
\rev{
\subsubsection{Trend}
A {\group} is a set of tuples that are compared together as one unit. Formally,
\begin{definition}[Trend]
Given a relation R,  a {\group} $t$ is a set of tuples derived from R via the triplet: {\constraint} $c$,  {\indexi} $g$,  {\measure} $m$ and represented as ($c$)($g$,  $m$).
\end{definition}

\begin{definition}[Constraint]
 Given a relation R,  a constraint is a conjunctive filter of the form: $({p_1} = \alpha_1,  p_2 = \alpha_2,  ...,  p_n = \alpha_n)$ that selects a subset of tuples from R. Here,  $p_1,  p_2,  ...,  p_n$ are attributes in $R$ and $\alpha_i$ is a  value of $p_i$ in $R$. One can use `ALL' to select all values of $p_i$,  similar to \cite{gray1997data}.
\end{definition}

\begin{definition}[(Grouping, Measure)]
Given a set of tuples selected via a {\constraint},  all tuples with the same value of {\indexi} are aggregated using {\measure}. A tuple in one {\group} is only compared with the tuple in another {\group} with the same value of {\indexi}.
\end{definition}

In example 1a,  (R.region  =  `Asia')(R.week,  AVG(R.revenue)) is a {\group} in \seto, where (region  =  `Asia') is a {\constraint} for the {\group} and all tuples with the same value of  {\indexi}:`week' are aggregated using the {\measure}: `AVG(revenue)'. We currently do not support range filters for {\constraint}.
}

\subsubsection{Trendset}

A comparative query involves two sets of {\groups}. We formalize this via {\groupset}.

\begin{definition}[Trendset]
A {\groupset} is a set of {\groups}. A {\group} in one {\groupset} is compared with a {\group} in another {\groupset}. 
%a {\groupset} $TS$: \{($c_1$)($g_1$,  $m_1$),  ($c_2$)($g_2$,  $m_2$),  ...,  ($c_n$)($g_n$,  $m_n$) \} consist of {\groups} ($c_1$)($g_1$,  $m_1$),  ($c_2$)($g_2$,  $m_2$),  $...$,  ($c_n$)($g_n$,  $m_n$).
\end{definition}

\noindent
In example 1a,  the first {\groupset} consists of a single {\group}: \{(R.region  =  `Asia')(R.week,  AVG(R.revenue))\},  while the second {\groupset} consists of as many {\groups} as there are are unique products in $R$: \{(R.region  =  `Asia',  R.product  =  `Inspiron') (R.week,  AVG(R.reven-ue)),  (R.region  =  `Asia',  R.product  = `XPS')(R.week,  AVG(R.reven- ue)),   $...$,   (R.region  =  `Asia',  R.product = `G7') (R.week,   
AVG(R.rev- enue))\}.

As is the case in the above example, often a {\groupset} contains one {\group} for each unique value of an attribute (say $p$) as a {\constraint},  all sharing the same ({\indexi},  {\measure}). Such a {\groupset} can be succinctly represented using only the attribute name as \constraint,  i.e.,  [$p$][($g_1$,  $m_1$)]. If $\alpha_1$,  $\alpha_2$,  ..$\alpha_n$ represent all unique values of $p$,  then, 

[$p$][($g_1$,  $m_1$)] 	$\Rightarrow$  \{($p = \alpha_1$)($g_1$,  $m_1$),  ($p = \alpha_2$)($g_1$,  $m_1$),  ...,  ($p = \alpha_n$)($g_1$,  $m_1$)\} ($\Rightarrow$ denotes equivalence)

Similarly,  
[$p_1$,  $p_2  =  \beta$][($g_1$,  $m_1$)] 	$\Rightarrow$  \{($p_1 = \alpha_1$,   $p_2 = \beta$)($g_1$,  $m_1$),  ($p_1 = \alpha_2$,  $p_2 = \beta$)($g_1$,  $m_1$),  ...,  ($p_1 = \alpha_n$,  $p_2  =  \beta$)($g_1$,  $m_1$)\}

Alternatively,  a {\groupset} consisting of different ({\indexi},  {\measure}) combinations but the same constraint (e.g.,  $p = \alpha_1$) can be succinctly written as:

[($p = \alpha_1$)][($g_1$,  $m_1$),  ...,  ($g_n$,  $m_n$)] 	$\Rightarrow$  \{($p = \alpha_1$)($g_1$,  $m_1$),  ...,  ($p = \alpha_1$)($g_n$,  $m_n$)\}

\subsubsection{Scoring}
We first define our notion of `Comparability' that tells when two {\groups} can be compared.

\begin{definition}[Comparability of two \groups]Two {\groups} $t_1$: ($c_1$)($g_1$,  $m_1$) and $t_2$: ($c_1$)($g_2$,  $m_2$) can be compared if  $g_1$ $ = $ $g_2$ and $m_1$ $ = $ $m_2$,  i.e.,  they have the same {\indexi} and {\measure}.
\end{definition}

For example,   a {\group} (R.product = `Inspiron') (R.week,  AVG( R.revenue)) and a {\group} (R.product = `XPS')( R.month,  AVG(R.pro- fit)) cannot be compared since they differ on {\indexi} and {\measure}.

Next,  we define a function {\scorer} for comparing two {\groups}.

\begin{definition}[Scorer]
Given two {\groups} $t_1$ and $t_2$,  a {\scorer} is any function that returns a single scalar value called `score' measuring how $t_1$ compares with $t_2$.
\end{definition}

While we can accept any function that satisfies the above definition as a {\scorer};  as mentioned earlier, two {\groups} are often compared using  distance measures such as Euclidean distance,  Manhattan distance~\cite{macke2018adaptive, seedb, zenvisagevldb}. Such functions are also called aggregated distance functions~\cite{papadias2005aggregate}.  All aggregated distance functions use a function DIFF(.) as defined below.

%that essentially aggregate the differences between the {\measure} (e.g.,  AVG(sales)) value of tuples in two {\groups}.

\begin{definition}[{\diff}($m_1,  m2,  p$)]\footnote{Note that the function {\diff} is distinct from another operator~\cite{abuzaid2018diff} with similar name.}
Given a tuple with {\measure} value $m_1$ and {\indexi} value $g_i$ in {\group} $t_1$ and another tuple with {\measure} value $m_2$ and the same {\indexi}  value $g_i$,  \text {\diff}($m_1$,  $m_2$,  p)  =  $|m_1 - m_2|^p$ where $p \in \mathbb{Z}^+$. Tuples with non-matching {\indexi} values are ignored.
\end{definition}
Since $m_1$ and $m_2$ are clear from the definition of $t_1$ and $t_2$ and tuples across {\groups} are compared only when they have same {\indexi} and {\measure} expressions,  we succinctly represent {\diff}($m_1$,  $m_2$,  $p$)  =  {\diff}($p$)

\begin{definition}[Aggregated Distance Function]
An aggregated distance function compares  {\groups} $t_1: (c_i)(g_i,  m_i)$ and $t_2: (c_j)$ $(g_i,  m_i)$ in two steps: (i) first {\diff}(p) is computed between every pairs of tuples in $t_1$ and $t_2$ with same values of $g_i$,  and (ii) all values of  {\diff}(p) are aggregated using an aggregate function AGG such as SUM, AVG, MIN,  and MAX to return a score. An aggregated distance function is represented as AGG OVER {\diff}(p).
\end{definition}

For example, $L_p$ norms\footnote{We ignore the $p$th root as it does not affect the ranking of subsets.} such as Euclidean distance can be specified using SUM OVER DIFF(2),  Manhattan distance using SUM OVER DIFF(1),  Mean Absolute Deviation as AVG OVER DIFF(1),  Mean Square Deviation as AVG OVER DIFF(2).

\subsubsection{Comparison between Trendsets}
We  extend Definition 5 to the following observation over {\groupsets}.

\stitle{Observation 1 [Comparability between two \groupsets]} Given two {\groupsets} $T_1$ and $T_2$,  a {\group} $(c_i)(g_i,  m_i)$ in $T_1$ is compared with only those {\groups} $(c_j)(g_j,  m_j)$ in $T_2$ where $g_i  =  g_j$ and  $m_i  =  m_j$.

Thus,  given two {\groupsets},  we can automatically infer which {\groups} between the two {\groupsets} need to be compared. We use $T1$\cs$T2$ to denote the comparison between two {\groupsets} $T_1$ and $T_2$. For example,  the comparison  in example 1a can be represented as:

\noindent
[region = `Inspiron'][(week,  AVG(revenue))] \text{\cs} [region = `Asia',  product][ (week,  AVG (revenue))]

If both $T_1$ and $T_2$ consist of the same set of {\indexi} and  {\measure} expressions say \{($g_1$,  $m_1$),  $...$,  ($g_n$,  $m_n$)\} and differ only in {\constraint},  we can succinctly represent $T_1$ \text{\cs} $T_2$ as follows:

\noindent
[$c_1$][($g_1$,  $m_1$),  $...$,  ($g_n$,  $m_n$)] \text{\cs} [$c_2$][($g_1$,  $m_1$),  $...$,  ($g_n$,  $m_n$)]  $\Rightarrow$ [$c_1$  \text{\cs} $c_2$][($g_1$,  $m_1$),  $...$,  ($g_n$,  $m_n$)] 

Thus,  the comparison between {\groupsets} in example 1a can be succinctly expressed as:

\noindent
[(region = `Asia') \text{\cs} (region = `Asia',  product) ][(week,  AVG(reve- nue))] 

\vspace{3pt}
\noindent
Similarly,  the following expression represents the comparison in example 1b.

\noindent
[(region = `Asia') \text{\cs} (region = `Asia',  product = `Inspiron')][(week, \\
AVG(revenue)),  (country,  AVG(profit)),  ... , (month,  AVG(revenue))]

\vspace{4pt}
We can now define a comparative expression using the notions introduced so far.

\begin{definition}[Comparative expression]
Given two {\groupsets} $T_1$ \text{\cs} $T_2$ over a relation $R$,  and a {\scorer} $\mathcal{F}$,  a comparative expression computes the scores between {\groups} $(c_i) (g_i,  m_i)$  in $T_1$ and  $(c_j)$ $(g_j,  m_j)$ in $T_2$ where $g_i  =  g_j$ and  $m_i  =  m_j$.
\end{definition}

\techreport{\vspace{5pt}}
\section{The COMPARE Operator}
\label{sec: comparitiveView}

In this section,  we introduce a new  operator \optr,  that makes it easier for data analysts and application developers to express comparative queries. We first explain the syntax and semantics of \optr and then show how \optr inter-operates with other relational operators to express top-k comparative queries as discussed in Section 2.1.

\subsection{Syntax and Semantics}
%\optr  is a logical operator that takes as input  a relation \cdr \xspace,  and creates two {\groupset}s $t_1$ and $t_2$,  each consisting of one or more {\groups}. Each {\group} in $t_1$ is compared with another {\group} in $t_2$ with the same {\indexi} and {\measure} using a {\scorer}. Finally,  the scores for each each pair of compared {\groups} are returned.
\tar{replace letters with meaningful names}

\optr,  denoted by $\Phi$,   is a logical operator that takes as input a a comparative expression specifying two {\groupsets} $T_1$ \cs $T_2$  over relation $R$ along with a {\scorer} $\mathcal{F}$ and returns a relation $R'$.

\begin{center}
  \vspace{-8pt}
  $\Phi(R,  T_1 \text{\cs} T_2, \mathcal{F})$ $\rightarrow$ $R'$
  \vspace{-8pt}
\end{center}

$R'$ consists of scores for each pair of compared {\groups} between the two {\groupsets}. For instance, the table below depicts the output schema (with an example tuple) for the \optr expression [$c_1$  {\cs}  $c_2$][($g_1$,  $m_1$), ($g_2$,  $m_2$)]. The values in the tuple indicate that the {\group} ($c1$ = $\alpha_1$)($g_1$,  $m_1$) is compared with the {\group} ($c_2$ = $\alpha_2$)($g_1$,  $m_1$) and the score is $10$.

\eat{
 Formally,  given a relation $R$,  and two inputs: 1) a comparison expression between  two {\groupset}s $t_1$ and $t_2$: T1|T2,  and 2) a {\scorer}: $\mathcal{F}$,  \optr,  denoted by $\Phi$,  returns a relation $R'$.

  \vspace{5pt}
  $\Phi(R,  T1  \cs T2, \mathcal{F})$ $\rightarrow$ $R'$
  \vspace{5pt}

$R'$ consists of scores for each pair of {\groups} that were compared. For instance,  for $t_1$  \cs $t_2$: [$c_1$  \cs $c_2$][($g_1$,  $m_1$), ($g_2$,  $m_2$)],  the below table shows the schema and one of its tuples. The values in the tuple indicate that the {\group} ($c_1$ = $\alpha_1$)($g_1$,  $m_1$) in $t_1$ is compared with the {\group} (P2 = $\alpha_2$)($g_1$,  $m_1$) in $t_2$ and the score is $10$.
}

\begin{center}
{
\papertext{\footnotesize}
\centering
\papertext{\vspace{-8pt}}
\begin{tabular}{ |c|c|c|c|c|c|c| } 
		\hline
		$c_1$ & $c_2$ & $g_1$ &  $m_1$ & $g_2$ & $m_2$  & score  \\ \hline 
		$\alpha_1$ & $\alpha_2$ & True &  True & False & False  & $10$  \\ \hline 
  ... & ... & ... &  ... & ... & ...  & ...  \\ \hline 
\end{tabular}
 } 
\end{center}

\vspace{10pt}
We express the \optr operator in SQL using two extensions: \optr  and USING:

\begin{center}
\papertext{\vspace{-12pt}}
\techreport{\vspace{-6pt}}
\begin{lstlisting}[mathescape = true]
@\bf \color{blue}\textbf{COMPARE}@  T1 @\cx@ T2
@\bf \color{blue}\textbf{USING}@  $\mathcal{F}$
\end{lstlisting}
\papertext{\vspace{-12pt}}
\techreport{\vspace{-6pt}}
\end{center}

\noindent
For instance, for example 1a, the comparison between the  AVG(reve- nue) over week trends for the region `Asia' and each of the products in region 'Asia' can be succinctly expressed as follows:

\begin{center}
\footnotesize
\papertext{\vspace{-10pt}}
\begin{lstlisting}[caption =   {\bf \color{black}COMPAREXPR1A},  mathescape = true,  basicstyle = \small]
SELECT R1,  P,  W,  V,  score
FROM sales R
@\bf \color{blue}COMPARE@ [((R.region = Asia) AS R1) @\cx@ (R1,  R.product AS P)]
            [R.week AS W,  AVG (R.revenue) AS V]
@\bf \color{blue}USING@ SUM OVER DIFF(2) AS score 
\end{lstlisting}	
\papertext{\vspace{-8pt}}
\end{center}

\begin{table}
\footnotesize
	\centering
	\caption{Output of \optr in Example 1a}
	\vspace{-10pt}
	\papertext{\resizebox{0.5\columnwidth}{!}}
	\resizebox{0.60\columnwidth}{!}{%
		\begin{tabular}{ |c|c|c|c|c| } 
		\hline
		R1 & P & W & V  & score  \\ \hline 
		Asia & XPS & True & True  & 30 	\\ \hline
		Asia & Inspiron & True & True & 24 	\\ \hline
		 ... & ... & ... & ... & ...  	\\ \hline
		Asia & G8 & True & True & 45 	\\ \hline
		\end{tabular}%
		}
	\label{tab:compex1}
	\papertext{\vspace{-15pt}}
\end{table}

Here $T_1$ = [((R.region = Asia) AS R1)][R.week AS W,  AVG (R.revenue) AS V] and $T_2$ = [((R.region = Asia) AS R1, R.product AS P)][R.week AS W,  AVG (R.revenue) AS V]. Observe that  $T_1$ and  $T_2$ share the same set of ({\indexi}, {\measure}) and the filter predicate (R.region = Asia) in their {\constraints}, thus it is concisely expressed as [((R.region = Asia) AS R1)\cs(R1,  R.product AS P)][R.week AS W,  AVG (R.revenue) AS V]. 

Table~\ref{tab:compex1} illustrate the output of this query. The first two columns R1 and P identify the values of {\constraint} for compared {\groups} in T1 and T2. The columns W and V are Boolean valued denoting whether R.week and AVG(R.revenue) were used for the compared {\groups}. Thus,  the values of (R1,  P,  W,  V) together  identify the pairs of {\groups} that are compared. Since R.week and AVG(R.revenue) are {\indexi} and {\measure} for all {\groups} in this example,  their values are always True. Finally,  the column score specifies the scores computed using Euclidean distance,  expressed as SUM OVER DIFF(2).

Now, consider below the query for example 1b that compares tuples where (R.region = Asia) with tuples where (R.region = Asia) and (R.product = 'Inspiron') over a set of ({\indexi}, {\measure}):
\eat{
\begin{center}
\footnotesize
\begin{lstlisting}[mathescape = true]
@\bf \color{blue}COMPARE@ [(R.product = Inspiron AS P1) @\cx@ (R.product = XPS AS P1)] 
	[(R.week,  AVG(R.revenue)),  (R.country,  AVG(R.profit)), 
	(R.month,  AVG(R.revenue))]
@\bf \color{blue}USING@ SUM OVER DIFF(2) AS score
\end{lstlisting}	
\end{center}
}

\begin{center}
\footnotesize
\papertext{\vspace{-8pt}}
\begin{lstlisting}[caption =  {\bf \color{black}COMPAREXPR1B},  mathescape = true,  basicstyle = \small]
SELECT R1,  P,  W,  C,  V, ..., M,  score
FROM sales R
@\bf \color{blue}COMPARE@ [((R.region = Asia) AS R1) @\cs@ (R1,  (R.product = 'Inspiron') 
	AS P)][(R.week AS W,  AVG(R.revenue) AS V), (R.country AS 
	C, AVG(R.profit) AS O), ..., (R.month AS M,  V)]
@\bf \color{blue}USING@  SUM OVER DIFF(2) AS score
\end{lstlisting}	
\papertext{\vspace{-10pt}}
\end{center}

\begin{table}
\footnotesize
	\centering
	\caption{Output of \optr in Example 1b}
    \vspace{-10pt}
	\resizebox{0.8\columnwidth}{!}{%
			\begin{tabular}{ |c|c|c|c|c|c|c|c| } 
			\hline
			 R1 & P & W &  C & M & V & O  & score  \\ \hline 
			Asia & Inspiron & True & False & False & True& False  & 40 	\\ \hline
			Asia & Inspiron & False & True & False & True& False  & 20 	\\ \hline

			 ... & ... & ... & ... & ... & ... & ... & ... 	\\ \hline
	    	Asia & Inspiron & False & False  & True & True& False  & 10 	\\ \hline
		\end{tabular}%
	}
	\label{tab:compex5}
		\vspace{-10pt}
\end{table}

 Table~\ref{tab:compex5} depicts the output for this query. The columns R1 and P are always set to "Asia" and "Inspiron" since the  {\constraint} for all {\groups} in T1 and T2 are fixed.  W,  C,  M,  V,  and P consist of Boolean values telling which columns among R.week,  R.country,  R.month,  AVG(R.revenue),  and AVG(R.profit) were used as  ({\indexi}, {\measure})  for the pair of compared {\groups}.

From above examples, it is easy to see that we can write queries with \optr expression for examples 2a and 2b as follows:

\begin{center}
\footnotesize
\papertext{\vspace{-8pt}}
\begin{lstlisting}[caption =  {\bf \color{black}COMPAREXPR2A},  mathescape = true,  basicstyle = \small]
SELECT R1, C1, R2, C2, W,  V,  score
FROM sales R
@\bf \color{blue}COMPARE@  [((R.Region  =  Asia) AS R1,  (R.city) AS C1) @\cx@  ((R.Region  
		=  Europe) AS R2,  (R.city) AS C2)][R.week AS W, 
		AVG(R.revenue) AS V]
@\bf \color{blue}USING@  SUM OVER DIFF(2) AS score 
\end{lstlisting}	
\papertext{\vspace{-10pt}}
\end{center}

\begin{center}
\footnotesize
\papertext{\vspace{-5pt}}
\begin{lstlisting}[caption =  {\bf \color{black}COMPAREXPR2B},  mathescape = true,  basicstyle = \small]
SELECT R1, C1, R2, C2, W,  C,  V, ...,  M,  score
FROM sales R
@\bf \color{blue}COMPARE@  [((R.Region  =  Asia) AS R1,  (R.city) AS C1) @\cx@  ((R.Region  
  =  Europe) AS R2,  (R.city) AS C2)][(R.week AS W,  AVG(R.revenue) AS 
  V), (R.country AS C, AVG(R.profit) AS O), ..., (R.month AS M,  V)]
@\bf \color{blue}USING@  SUM OVER DIFF(2) AS score 
\end{lstlisting}	
\papertext{\vspace{-5pt}}
\end{center}

\rev{Note that \optr is semantically equivalent to a standard  relational expression consisting of multiple sub-queries involving union, group-by, and join operators as illustrated in introduction. As such, \optr does not add to the expressiveness of relational algebra SQL language. The purpose of \optr is to provide a  succinct and more intuitive mechanism to express a large class of frequently used comparative queries as shown above. For example, expressing the query in Listing 2 using existing SQL clauses (see Figure 3) is much more verbose, requiring a complex sub-query for each  ({\indexi}, {\measure}). Prior work have proposed similar succinct abstractions such as GROUPING SETs~\cite{groupingsets} and CUBE~\cite{gray1997data} (both widely adopted by most of the databases) and more recently DIFF~\cite{abuzaid2018diff}, which share our overall goal that with an extended syntax, complex analytic queries are easier to write and optimize.

Furthermore, the input to \optr is a relation, which can either be a base table or an output from another logical operator (e.g., join over multiple tables); similarly the output relation from \optr can be an input to another logical operator or the final output. Thus, \optr can interoperate with other operators. In order to illustrate this, we discuss how \optr interoperates with other operators such as join, filter to select top-k {\groups}.
}

\subsection{Expressing Top-k Comparative Queries}
While \optr outputs the scores for each pair of compared {\groups},  comparative queries often involve selection of top-$k$ {\groups} based on their scores (Section 2.1).
 In this section,  we show how
we can use the above-listed \optr sub-expressions (referred by COMPAREXPR1A, COMPAREXPR1B,  COMPAREXPR2A, and COMPAREXPR1B) with LIMIT and join to select tuples for {\groups} belonging to top-$k$.

\stitle{Example 1a.} The following query selects the tuples of a product in region `Asia' that has the most different AVG(revenue) over week trends compared to that of  region `Asia' overall. {\bf \color{black} COMPAREXPR1A} refers to the sub-expression in Listing 1.

\eat{
\begin{center}
\small
\begin{lstlisting}[mathescape = true,  basicstyle = \small]
SELECT T.product,  T.week,  T.revenue,  S.score 
FROM sales T JOIN
(SELECT P2,  score
FROM sales R
@\color{blue}COMPARE@ [(R.region = Asia) AS R1 | (R1,  R.product AS P2)][R.week AS 
W,  AVG (R.revenue) AS V]
@\color{blue}USING@  SUM OVER DIFF(2) AS score 
ORDER BY score DESC
LIMIT 1) AS S
WHERE T.product = S.P2
\end{lstlisting}	
\end{center}
}

\begin{center}
\papertext{\vspace{-10pt}}
\small
\begin{lstlisting}[mathescape = true,  basicstyle = \small]
SELECT T.product,  T.week,  T.revenue,  S.score 
FROM sales T JOIN
(SELECT * FROM @\bf \color{black}COMPAREXPR1A@ 
ORDER BY score DESC
LIMIT 1) AS S
WHERE T.product = S.P
\end{lstlisting}
\papertext{\vspace{-5pt}}
\end{center}

The ORDER BY and LIMIT clause select the top-1 row in Table~\ref{tab:compex1} with the highest score with P consisting of the most similar product. Next,  a join is performed with the base table to select all tuples of the most similar product along with its score.

\stitle{Example 2a.} The query for example 2a differs from example 1a in that both {\groupsets} consist of multiple {\groups}. Here, one may be interested in selecting tuples of both cities that are similar, thus  we use the WHERE condition (T.city = S.C1 AND T.Region =  S.R1) OR (T.city = S.C2 AND T.Region =  S.R2). (S.R1,  S.R2,  S.C1,  S.C2) in SELECT clause identifies the pair of compared {\groups}.

\eat{
\begin{center}
{
\tiny
\begin{lstlisting}[mathescape = true,  basicstyle = \small]
SELECT T.Region,  T.city,  T.week,  T.revenue,  
S.score,  S.R1,  S.C1,  S.R2,  S.C2,  score
FROM sales T
JOIN
(SELECT R1,  R2,  P1,  P2,  score
FROM sales R
\optr [((R.Region  =  Asia) AS R1,  (R.city) AS C1)
| ((R.Region  =  Europe) AS R2,  (R.city) AS C2)][R.week AS W, 
AVG(R.revenue) AS V]
USING SUM OVER DIFF(2) AS score 
ORDER BY score DESC
LIMIT 1) AS S
WHERE  (T.city = S.C1 AND T.Region =  S.R1) OR (T.city = S.C2 AND
T.Region =  S.R2)
\end{lstlisting}	
}
\end{center}
}

\begin{center}
{
\papertext{\vspace{-10pt}}
\tiny
\begin{lstlisting}[mathescape = true,  basicstyle = \small]
SELECT T.Region,  T.city,  T.week,  T.revenue, S.R1,  S.C1,  S.R2,  S.C2,
 S.score
FROM sales T JOIN
(SELECT * FROM  @\bf \color{black}COMPAREXPR2A@ 
ORDER BY score 
LIMIT 1) AS S
WHERE  (T.city = S.C1 AND T.Region =  S.R1) OR (T.city = S.C2 AND
T.Region =  S.R2)
\end{lstlisting}	
}
\papertext{\vspace{-5pt}}
\end{center}

\stitle{Examples 1b and 2b.} These examples extend the first two examples to multiple attributes. We show the query for example 2b; it's a complex version of (example 1b) where {\groups} in each {\groupsets} are created by varying all three: {\constraint},  {\indexi},  {\measure} (example 1b has a fixed {\constraint} for each {\groupset}). 

\eat{
\begin{center}
\small
\begin{lstlisting}[mathescape = true,  basicstyle = \small]
SELECT T.city,  
 CASE WHEN S.W THEN T.week ELSE NULL END, 
 CASE WHEN S.M THEN T.month ELSE NULL END, 
 CASE WHEN S.C THEN T.country ELSE NULL END, 
 CASE WHEN S.P THEN T.profit ELSE NULL END, 
 CASE WHEN S.V THEN T.revenue ELSE NULL END, 
 S.R1,  S.R2,  S.C1,  S.C2,  S.score 
FROM sales T
(SELECT R1,  R2,  P1,  P2,  W,  C,  V,  M,  score
FROM sales R
\optr [((R.Region  =  Asia) AS R1,  (R.city) AS C1)
@\cs@ ((R.Region  =  Europe) AS R2,  (R.city) AS C2)]
[(R.week AS W,  AVG(R.revenue) AS V),  (R.country AS C,  AVG(R.profit)) 
AS P,  (R.month AS M,  V)]
USING SUM OVER DIFF(2) AS score
ORDER BY score ASC
LIMIT 1) AS S
WHERE  (T.city = S.C1 AND T.Region =  S.R1) OR (T.city = S.C2 AND
T.Region =  S.R2)
\end{lstlisting}	
\end{center}
}

\begin{center}
\vspace{-10pt}
\small
\begin{lstlisting}[mathescape = true,  basicstyle = \small]
SELECT T.city, S.R1,  S.R2,  S.C1,  S.C2, 
 CASE WHEN S.W THEN T.week ELSE NULL END, 
 ...
 CASE WHEN S.V THEN T.revenue ELSE NULL END, 
 S.score 
FROM sales T JOIN  
(SELECT * FROM @\bf \color{black}COMPAREXPR2B@ 
ORDER BY score
LIMIT 1) AS S
WHERE  (T.city = S.C1 AND T.Region =  S.R1) OR (T.city = S.C2 AND
T.Region =  S.R2)
\end{lstlisting}	
\end{center}
The SELECT clause only outputs the values of  columns for which corresponding {\groups} has the highest score,  setting NULL for other columns to indicate that those columns were not part of top-1 pair of {\groups}. This idea of setting NULL is borrowed from prior work on CUBE~\cite{gray1997data}.  Nevertheless, an alternative is to output values of all columns,  and add  (S.W,  S.M,  S.C,  S.P,  S.V) (as in the previous example) to the output to indicate which columns were part of the comparison between top-1 pair of {\groups}.

\vspace{-5pt}

\section{Optimizing Comparative Queries}
\label{sec:phyop}
\rev{In this section, we discuss how we optimize a logical query plan consisting of a \optr operation. We extend the Microsoft \db optimizer to replace \optr with a sub-plan of existing physical operators using two steps. First, we transform \optr into a sub-plan of existing logical operators. These logical operators are then transformed into physical operators using existing rules to compute the cost of \optr. The cost of the sub-plan for \optr is combined with costs of other physical operators to estimate the total cost of the query.} We state our problem formally:

\begin{problem}
Given a logical query plan consisting of \optr operation: $\Phi(R,$ [$c_1$\cs$c_2$] [($d_1$, $m_1$), $...$, ($d_n$, $m_n)]$, $\mathcal{F})$ $\rightarrow$ $R'$, replace \optr with a sub-plan of physical operators with the lowest cost.
\end{problem}

\noindent
For ease of exposition, we assume that both {\groupsets} contain the same set of {\groups}, one for each unique value of $c$, i.e., $c_1 $ = $c_2$ = $c$.

\subsection{Basic Execution}  
We start with a simple approach that transforms \optr into a sub-plan of logical operators. The sub-plan is similar to the one generated by database engines when comparative queries are expressed using existing SQL clauses (discussed in Section 1). We perform the transformation using the following steps:

\eat{
(1) $\forall (d_i, m_i)$:  $R_i \leftarrow \text{Group-by}_{c,d_i}\text{Agg}_{m_i}(R)$

(2) $\forall$ $R_i$:  $R_{ij} \leftarrow \Join_{R_i.c !=  R_i.c, R_i.d_i = R_i.d_i} (R_i)$

(3) $\forall$ $R_{ij}$  $R_{ijk} \leftarrow \text{Partition}(R_{ij}) \text{ ON } $(c^i, c^j)$

(4) $\forall$ $R_{ijk}$  $score_{ijk} \leftarrow  \text{UDA}_{\mathcal{F}}(R_{ijk})$
}

\noindent
(1) $\forall (d_i, m_i)$:  $R_i \leftarrow \text{Group-by}_{c,d_i}\text{Agg}_{m_i}(R)$

\noindent
(2) $\forall$ $R_i$:  $R_{ij} \leftarrow \Join_{R_i.c !=  R_i.c, R_i.d_i = R_i.d_i} (R_i)$

\noindent
(3) $\forall$ $R_{ij}$:  $R_{ijk} \leftarrow  \text{Group-by}_{c^i,c^j}\text{Agg}_{\text{UDA}_{\mathcal{F}}}(R_{ij})$ // $c^i$, $c^j$ are aliases of column $c$

\noindent
(4) $R' \leftarrow  \underset{i,j,k}{\text{Union All}} (R_{ijk}$)

\rev{
First, we create trendsets for each ({\indexi}, {\measure}) combination (e.g., GROUP BY product, week, AGG on AVG(reve- nue)). Next, we join tuples between each pair of {\groups} that are compared, i.e., tuples with different constraints but same value of {\indexi} (e.g., $\Join_\text{R'.product !=\\  R'.product, R'.week = R'.week})$). 
The score between each pair of trends is computed by applying $\mathcal{F}$ specified as an user-defined aggregate (UDA). This is done by first partitioning the join output to create a partition for each pair of {\groups}. Each partition is then aggregated using $\mathcal{F}$. Finally, the scores from comparing each pairs of {\groups} are aggregated via Union All.
}

\eat{
First, for each  ({\indexi}, {\measure}), we create a group-by aggregate (e.g., GROUP BY product, week: AVG(revenue)) expression consisting of  {\groupsets}.
Then, a join (e.g., $\Join_{R'.product !=  R'.product, R'.week = R'.week]} )$ is performed on the output of group-by aggregate  to join tuples between two {\groups} (e.g., different products) that have the same value of {\indexi} (e.g., $R'.week$). Finally, the join  output is  partitioned to group together each pair of {\groups}  that are compared. Each partition consisting of pair of {\groups} is then passed to a \scorer, an UDF that computes a score between the two \groups. Finally, the scores for all pairs of {\groups} are aggregated via UNION ALL.
}

Unfortunately, this approach has two issues that make it less efficient as the size of the input dataset and the number of ({\indexi}, {\measure}) combinations become large.  First, aggregations across ({\indexi}, {\measure}) are performed separately, even when there are overlaps in the subset of tuples being aggregated. Second, the cost of join increases rapidly as the number of {\groups} being compared and the size of each {\group}  increases (see Figure~\ref{fig:joinmotiv}). We next discuss how we address these issues via merging and partitioning optimizations

%while each {\group}  is compared with other {\groups}  independently, the second step above typically involves a join at the granularity of {\groupsets}, i.e., a relation consisting of one {\groupset} is joined with a relation consisting of another {\groupset}. The cost of join increases rapidly as the number of {\groups} being compared and the size of each {\group}  increases (see Figure~\ref{fig:joinmotiv}). We next discuss how we address these issues via merging and partitioning optimizations.

\begin{figure}	
  \vspace{-10pt}
	\hspace{-0.2cm}
	%\techreport{
	\begin{subfigure}{0.42\columnwidth}
		\centerline{
			\hbox{\resizebox{\columnwidth}{!}{\includegraphics{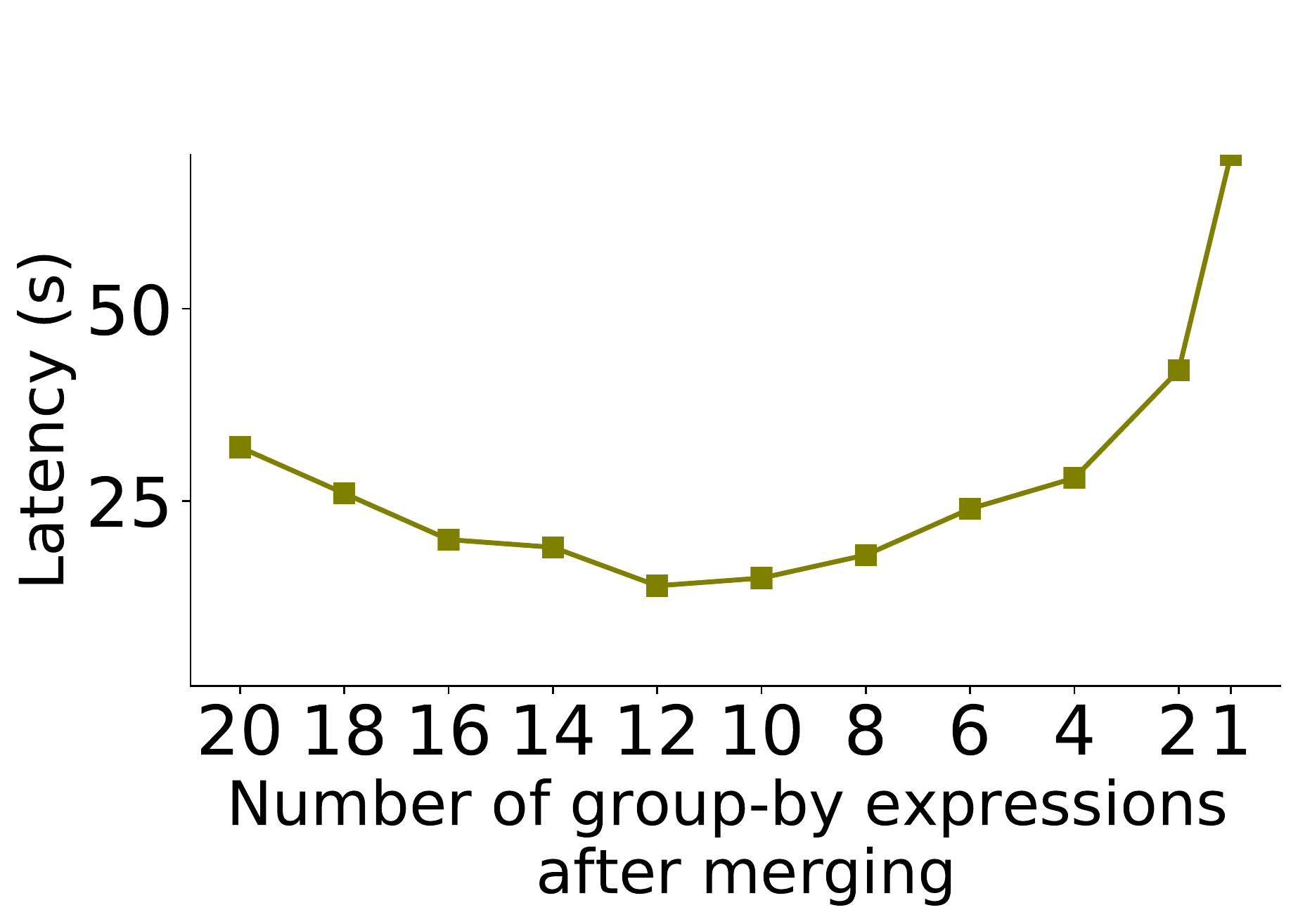}}}}
		\caption{Variation in performance as we merge group-by aggregates to share computations}
		%	\caption{\smallcaption{A complex SQL query in PowerBI to fetch only relevant data for visualization}}
		\label{fig:merging}
	\end{subfigure}
	\vspace{0.5cm}
	\begin{subfigure}{0.57\columnwidth}
		\centerline {
			\hbox{\resizebox{\columnwidth}{!}{\includegraphics{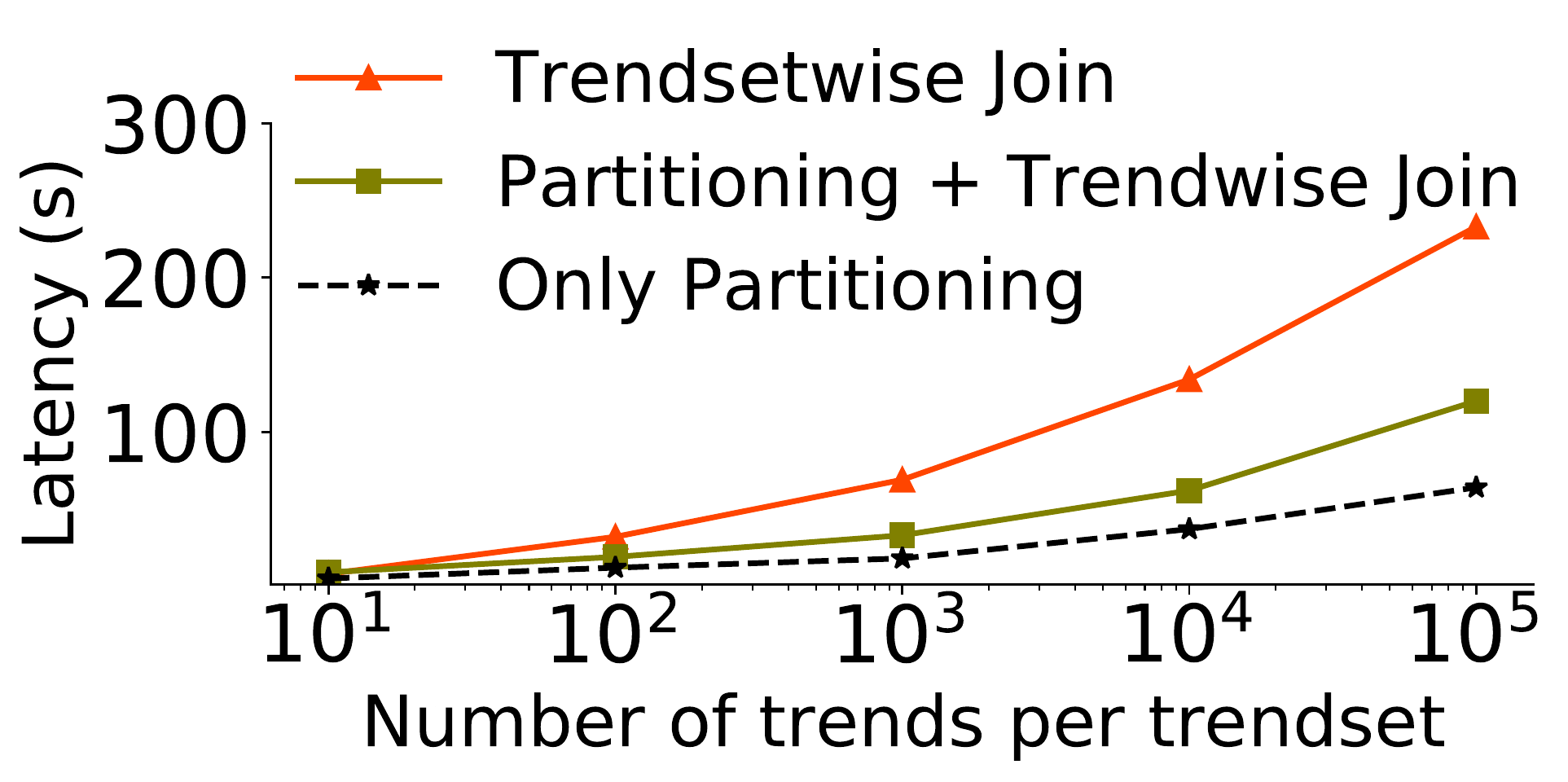}}}}	
		\caption{Improvements due to {\group}wise join after partitioning {\groupset} into {\groups} (the size of each {\group} is fixed to 1000 tuples)}
	%	\caption{\smallcaption{Searching states with comparable housing price {\groups} in Zenvisage~\cite{zenvisagevldb} }}
	\label{fig:joinmotiv}
	\end{subfigure} 
	\hspace{0.2cm}
	\vspace{-25pt}
   \caption{Improvement in performance due to merging group-by aggregates and trendwise comparison (via partitioning)}
	\label{fig:motiv}	
	\vspace{-20pt}
\end{figure} 

\subsection{Merging and Partitioning Optimization}
To generate a more efficient plan, we adapt the sub-plan generated above using two optimizations. We first describe each of these optimizations and then present an algorithm that incorporates both of these optimizations to find an overall efficient plan.

%in two ways. First, we \emph{merge} group-by aggregates into a fewer number of group-by aggregates. Next, we \emph{partition} the output of  group-by aggregates to create one relation for each {\group}, which allows  parallel join and comparison for scoring.
%We describe each of these optimizations and then present an algorithm that jointly optimizes them to find an overall efficient plan.

\stitle{Merging group-by aggregates}.
The first optimization shares the computations across a set of group-by aggregates, one for each ({\indexi}, {\measure}), by merging them into fewer group-by aggregates. We observe that ({\indexi}, {\measure}) often share a common {\indexi} column, e.g., [(day, AVG(revenue), (day, AVG(profit)] or have correlated {\indexi} columns (e.g., [(day, AVG(revenue), (month, AVG(revenue)]) or have high degree of overlapping tuples across {\groups}. For example,  we considered a set of $20$ group-by aggregates in the flights~\cite{airlinedata} dataset, computing AVG(ArrivalDelay), AVG(DepDelay), ..., AVG(Duration)  grouped by day, week, ..., airport. As depicted in Figure~\ref{fig:merging}, by merging them (using an approach discussed shortly) into $12$ aggregates, the latency improves by 2$\times$. However, merging is helpful only up to a certain point,  after which the performance degrades due to less sharing and much larger increase in the output size of group-by aggregates. 

Finding the optimal merging of group-by aggregates is NP- Complete~\cite{agarwal1996computation}. Prior work on optimizing GROUPING SETs computation~\cite{groupingsets} have proposed best-first greedy approaches that merge  those group-by aggregates first that lead to maximum decrease in the cost. Unfortunately, in our setting, we also need to consider the impact of merging on the cost of subsequent comparison between {\groups}; ignoring which can lead to sub-optimal plans as we describe shortly. We first introduce the second optimization for comparison.

\stitle{Trendwise Comparison via Partitioning.} 
The second optimization is based on the observation that \emph{pairwise joins of multiple smaller relations is much faster than the a single join between two large relations}.
%rev{Formally, consider a comparison between two {\groupsets} each of size $n$ and consisting of $p$ {\groups} (say of size $n/p$), then performing  $p$ joins for each {\group} is much faster than performing a single join of size $n$.} 
This is because the cost of join increases super-linearly with the increase in the size of the {\groupsets}. 
In addition to improvement in complexity,  {\group}wise joins are  more amenable to parallelization than a single join between two {\groupsets}.
Figure~\ref{fig:joinmotiv} depicts the difference in latency for these two approaches as we increase the number of {\groups} from $10$ to $10^5$ (each of size $1000$). The black dotted line shows the partitioning overhead incurred while creating partitions for each {\group}, showing that the overhead is small (linear in $n$) compared to the gains due to {\group}wise join. Moreover, this is much smaller than the  overhead incurred when partitioning is performed on the join output ($\propto$ $n^2$) in the basic plan (see step 3 in Section 4.1). 

Figure 7 depicts the query plan after applying the above two optimizations on the basic query plan. First,we merge multiple group-by aggregates to share computations (using the approach discussed below). Then, we partition the output of merged group-by aggregates into smaller relations, one for each {\group}. This is followed by joining and scoring between each pair of  {\groups} independently and in parallel. Observe that the merging of group-by aggregates results in multiple {\groups} with overlapping ({\indexi}, {\measure}) in the output relation. Hence, we apply the partitioning in two phases. In the first phase, we partition it vertically, creating one relation for each ({\indexi}, {\measure}). In the second phase, we partition horizontally, creating one relation for each {\group}. 
\tar{Can we cite something on vertical partitioning}
%We determine the ordering between the two phases can be interchanged based on the costing.

\eat{
\begin{figure}
	\centerline {
		\hbox{\resizebox{0.7\columnwidth}{!}{\includegraphics{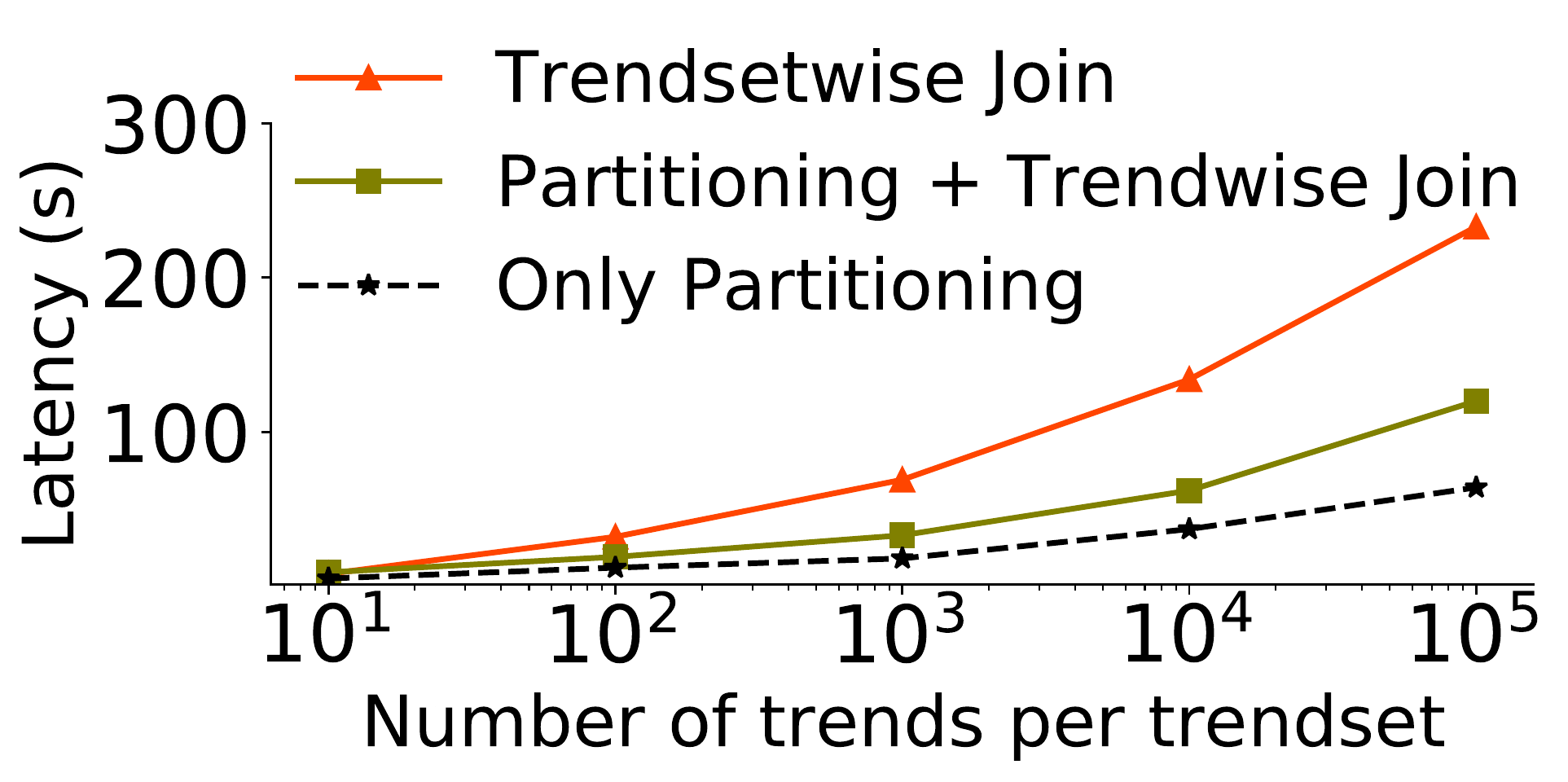}}}}
	\vspace{-10pt}
	\caption{Runtime improvements due to {\group}wise join (the size of each {\group} is fixed to 1000 tuples)}
	\label{fig:joinmotiv}
	 \vspace{-15pt}
\end{figure}
}

\stitle{Joint Optimization of Merging and Partitioning.}
As depicted in Figure 6b, the cost of partitioning increases with the increase in the size of its input. The input size is proportional to the number of unique group-by values, which increases with the increase in the number of merging of group-by aggregates. Thus, when the input becomes large, the cost of partitioning dominates the gains due to merging. It is therefore important to merge group-by aggregates such that the overall cost of computing group-by aggregates, partitioning and trendwise comparison together is minimal.

\begin{figure}
	\papertext{\vspace{-15pt}}
	\centerline {
		\hbox{\resizebox{\columnwidth}{!}{\includegraphics{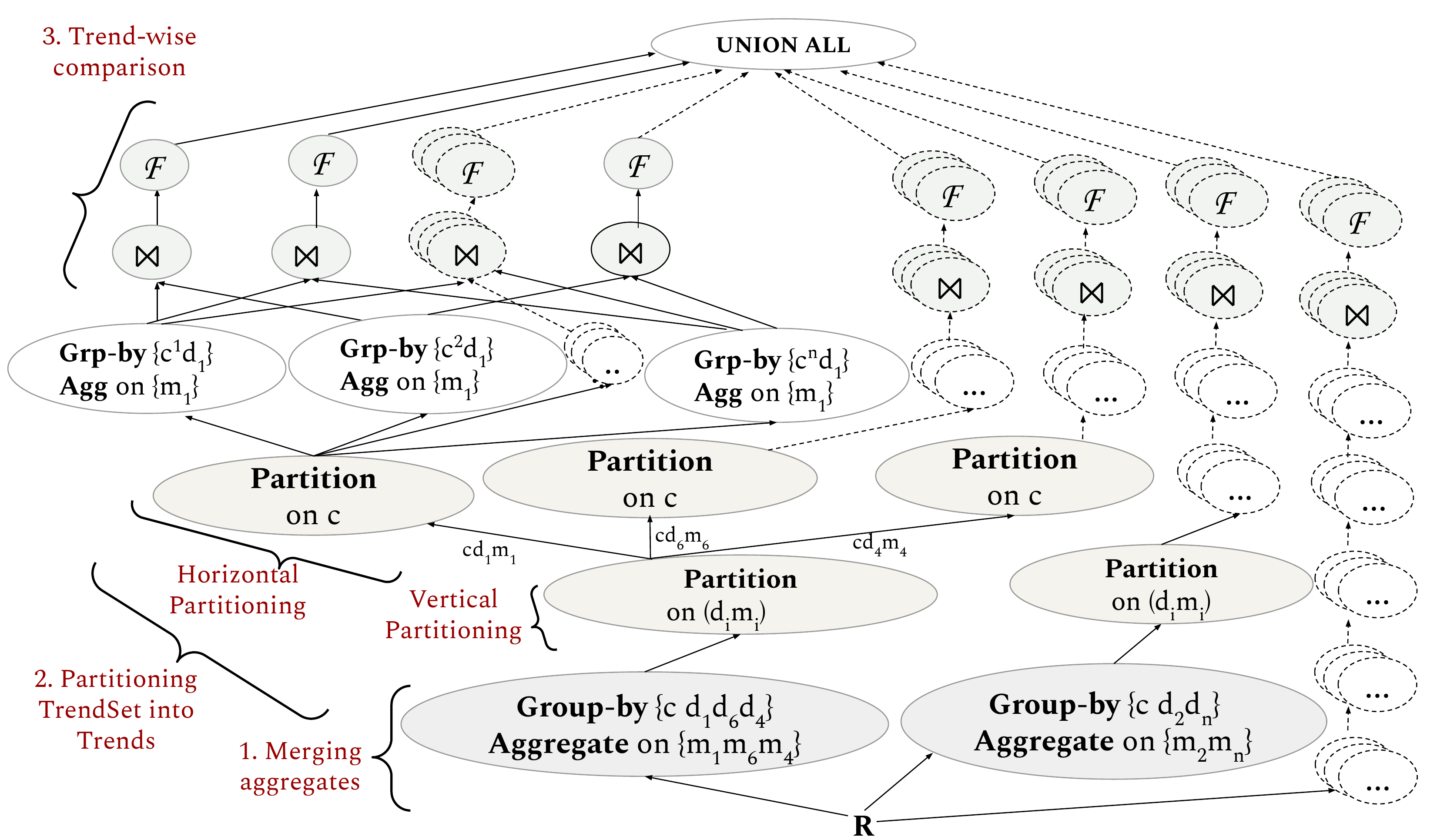}}}}
	\vspace{-10pt}
	\caption{\smallcaption{Optimized query plan generated after applying merging and partitioning on basic query plan in Figure 7.}}
	\label{fig:examples}
	\vspace{-15pt}
\end{figure}

In order to find the optimal merging and partitioning, we follow a greedy approach as outlined in Algorithm 1. \emph{Our key idea is to merge at the granularity of sub-plans instead of the group-by aggregates}. We start with a set of sub-plans, one for each (\indexi, \measure) as generated by the basic execution strategy discussed earlier and merge two sub-plan at a time that lead to the maximum decrease in cost.

Formally, if the two sub-plans operate over ($d_1$, $m_1$) and ($d_2$, $m_2$) respectively, we merge them using the following steps (illustrated in  Figure 6):

\noindent
(1) $R1 \leftarrow \text{Group-by}_{c,d_1, d_2}\text{Agg}_{m_2, m_2}[R]$ // merge group-by aggregates

\noindent
(2) $\forall$ $(d_i, m_i)$: $R_{i} \leftarrow  \Pi_{(d_i, m_i)}(R1)$ // vertical partitioning

\noindent
(3) $\forall$ $i$: $R_{ij} \leftarrow \text{Partition } R_{i} \text{ ON } c$ // horizontal partitioning, one partition for each value of c

\noindent
(4)  $\forall$ $i,j$: $R_{i'j'} \leftarrow \text{Group-by}_{c_j,d_i}\text{Agg}_{m_i}[R_{ij}]$ // aggregate again

\noindent
(5)  $\forall$ $i', j', k$: $R_{i'j'k} \leftarrow R_{i'j'} \Join_{d_i} R_{i'k}$ //partitition-wise join
			
\noindent
(6) $\forall$ $i', j', k$:  $R^{'}_{i'j'k} \leftarrow  \text{Agg}_{\text{UDA}_{\mathcal{F}}}(R_{i'j'k})$  // compute scores

\noindent
(7) $R' \leftarrow \underset{i',j',k}{\text{Union All}}(R^{'}_{i'j'k}$)

 We first merge group-by aggregates to share the computation, followed by creating one partitions for each {\group} using both vertical and horizontal partitioning. Then, we join pairs of {\groups} and compute the score as discussed in Section 4.1. 
 For computing the cost of the merged  sub-plan, we use the optimizer cost model. The cost is computed as  a function of available database statistics (e.g., histograms, distinct value estimates), which also captures the effects of the  physical design, e.g., indexes as well as degree of parallelism (DOP).  We merge two sub plans at a time until there is no improvement in cost. 

\begin{center}
	\begin{algorithm}
		\small
		\caption{Merge-Partition Algorithm}
		\label{algo:scoringalgo}
		\small
		\begin{algorithmic}[1]
			\State Let $B$  be a basic sub-plan computed from $\Phi$ as described in Section 4.1
			\While {true}
			\State $C$ $\leftarrow$ OptimizerCost(B)
			\State Let $s_i$ $\in$ $S$ be a sub-plan in B  consisting of a sequence of group-by aggregate, join and partition operations over ($d_i$, $m_i$)
			\State Let $MP$ = Set of all sub-plans obtained by merging a  pair of sub-plans in $S$ as described in Section 4.2
			\State Let $B_{new}$ be the sub-plan  in $MP$ with lowest cost ($C_{new}$) after merging two sub-plans $s_i, s_j$
			\If{$C_{new} > C$}
			\State break;
			\EndIf
			\State $C$ $\leftarrow$ $C_{new}$
			\State $B$  $\leftarrow$ $B_{new}$
			\EndWhile
			\State Return B
		\end{algorithmic}
	\end{algorithm}
	\vspace{-10pt}
\end{center}

\section{Optimizing DIFF-based Comparison}
While the approach discussed in the previous section works for any arbitrary {\scorer} (implemented as UDA), we note that for top-$k$ comparative queries involving aggregated distance functions (defined in Section 2.2) such as Euclidean distance, we can substantially reduce the cost of comparison between pairs of \groups. We first outline the three properties of DIFF(.) function that we leverage for optimizations.

\stitle{1. Non-negativity:} DIFF( $m_1, m_2, p$) $\geq 0$

\stitle{2. Monotonicity:}  DIFF($m_1, m_2, p$) varies monotonically with the increase or decrease in $|m_1 - m_2|$. 

\stitle{3. Convexity:}  DIFF( $m_1, m_2, p$) are convex for all $p$.

\subsection{Summarize $\rightarrow$ Bound  $\rightarrow$  Prune}
\label{sec:segaggregates}

\stitle{Overview.} \rev{We introduce a new physical operator that minimizes the number of {\groups} that are compared using the following three steps (illustrated in Figure 8)}. 1. We summarize each {\group} \emph{independently} using a set of three aggregates: \asum, {\amin} and {\amax} and a bitmap corresponding to the  {\indexi} column. 2. Next, we intersect the bitmaps between {\groups} to compute the {\acount} of matching tuples between trends, which together with three aggregates help compute the upper and lower bounds on the score between the two {\groups}. 
Given bounds on scores for each pair of trends, we  find a pruning threshold {\cdt} on the lowest possible top $k$ score, as the $k$th largest lower bound score. Any pair with its upper bound score smaller than  {\cdt} can thus be pruned. 3. Finally, we perform join only between those {\groups} that are not pruned.

\begin{figure}[H]
    \vspace{-5pt}
	\centerline {
		\hbox{\resizebox{\columnwidth}{!}{\includegraphics{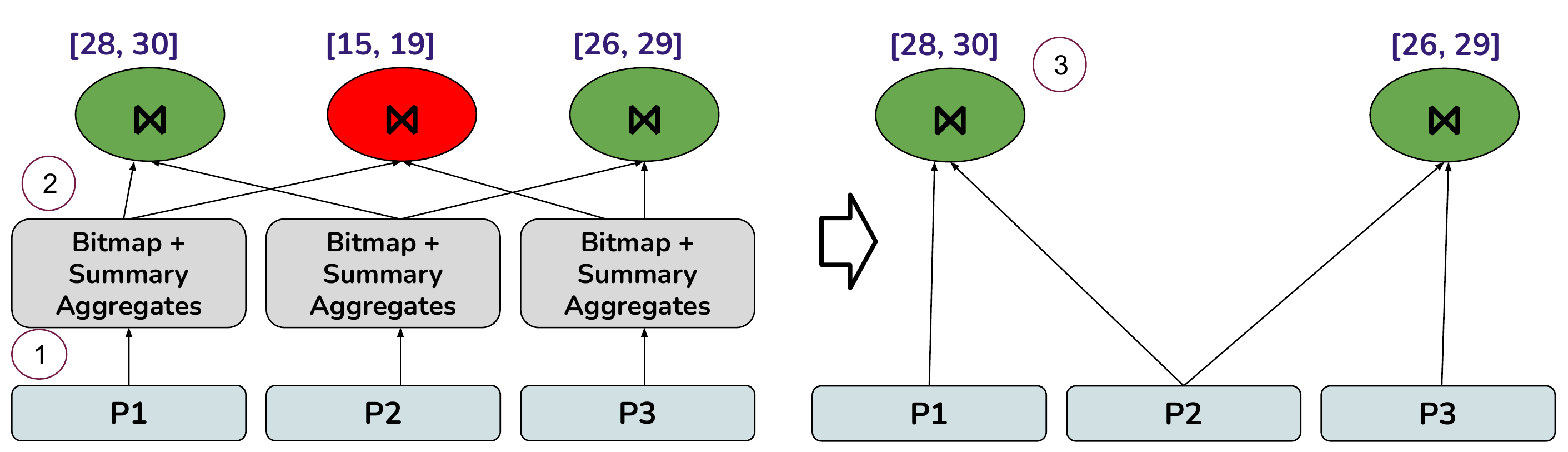}}}}
	\techreport{\vspace{-10pt}}
	\caption{\smallcaption{Illustrating pruning for DIFF-based comparisons}}
	\vspace{-10pt}
	\label{fig:examples}
\end{figure}

{
	\center
	\begin{figure*}
		\centering
		\resizebox{0.8\textwidth}{!}{%
			\begin{tikzpicture}[node distance = 3pt and 0.5cm]
				
				\matrix[table, ampersand replacement=\& ,style={
					nodes={rectangle,draw=black,text width=4ex,align=center}}] (mat12) 
				{  	
					|[fill=lightgray]| 18 \& |[fill=lightgray]| 18 \& |[fill=lightgray]| 14 \&  |[fill=lightgray]| 18\& |[fill=lightgray]| 18 \&  |[fill=lightgray]| 16 \& |[fill=lightgray]|14\& |[fill=lightgray]| 14\&  |[fill=lightgray]| 10 \& |[fill=lightgray]| 14 \&	|[fill=lightgray]|12 \& |[fill=lightgray]| 10 \& |[fill=lightgray]| 13 \&  |[fill=lightgray]| 13 \& |[fill=lightgray]| 14 \&  |[fill=lightgray]| 14  \\
				};

				\matrix[table, ampersand replacement=\& ,below=1cm of mat12,
				,style={
					nodes={rectangle,draw=black,text width=4ex,align=center}}
				] (mat15) 
				{    
					|[fill=blueiii]| 26 \& |[fill=blueiii]| 23 \& |[fill=blueiii]| 23 \&  |[fill=blueiii]| 29\& |[fill=blueiii]| 30 \&  |[fill=blueiii]| 28 \& |[fill=blueiii]| 24 \& |[fill=blueiii]| 25\&  |[fill=blueiii]| 27 \& |[fill=blueiii]| 24 \&	|[fill=blueiii]| 24 \& |[fill=blueiii]| 20 \& |[fill=blueiii]| 21 \&  |[fill=blueiii]| 25 \& |[fill=blueiii]| 20 \&  |[fill=blueiii]| 22  \\
				};

				\node[text width =4cm, font=\Large,  above right = .3cm and -7.1cm of mat15] (f3) {Score $=1717$};

				\matrix[table, ampersand replacement=\& ,right=2cm of mat12,style={
					nodes={rectangle,draw=black,text width=22ex,align=center}}] (mat21) 
				{  	
					|[fill=lightgray]| 16, 229, 10, 18 \\
				};

				\matrix[table, ampersand replacement=\& ,below= 1cm of mat21, style={
					nodes={rectangle,draw=black,text width=22ex,align=center}}] (mat25) 
				{    |[fill=blueiii]| 16, 394, 20, 30 \\   
				};

				\node[text width =5cm, font=\Large, above right = .20cm and -4.2cm of mat25] (f32) {Bounds $=[1700,6400]$};

				%	\draw [->] (mat21-1-1.east) -- (f2.west);
				%	\draw [->] (mat25-1-1.east) -- (f2.west);

				\matrix[table, ampersand replacement=\& ,right=3cm of mat21, style={
					nodes={rectangle,draw=black,text width=18ex,align=center}}] (mat31) 
				{  	
					|[fill=lightgray]| 8, 129, 13, 18  \&[2mm]	|[fill=lightgray]|  8, 100, 10, 14  \\
				};

				\matrix[table, ampersand replacement=\& ,below=1cm of mat31, style={
					nodes={rectangle,draw=black,text width=18ex,align=center}}] (mat35) 
				{    	|[fill=blueiii]|  8, 211, 23, 30 \&[2mm]	|[fill=blueiii]|  8, 183, 20, 27 \\  
				};

				\node[text width =5cm, font=\Large, above right = .20cm and -5.4cm of mat35] (f3) { Bounds $=[1702, 4624]$};

				%	\draw [->] (mat31-1-2.east) -- (f3.west);
				%	\draw [->] (mat35-1-2.east) -- (f3.west);

				%\node[text width =1.5cm, right = 0.01cm  of f3] (f31) {$=[18,25]$}; 
				
				\node[align = center,font=\Large, text width = 10.5 cm, below = .3 cm  of mat15] (t1) {(a) Exact score on comparing two {\groups}}; 
				\node[align = center,font=\Large, text width =6.4 cm, right =  1.5 cm  of t1] (t2) {(b)  Bounds on score using a single summary};
				\node[align = center, font=\Large, text width =7.4 cm, right =0.9 cm  of t2] (t3) {(c)  Bounds on score using two-segment summaries};

			\end{tikzpicture}%
		}
		\vspace{-10pt}
		\caption{\small Using summaries to bound scores. $\mathcal{F}$ = {\asum} OVER DIFF($2$). Each value in (a) corresponds to a single tuple in a {\group}.}
		\label{fig:segmengttree}
	\end{figure*}
}

\techreport{While the pruning incurs an overhead of first computing the summary aggregates and bitmap for each candidate {\group}, the gains from skipping  tuple comparisons for pruned {\groups}  offsets the overhead. Moreover, the summary aggregates of each {\group} can be computed independently in parallel.}

%In order to reduce the number of tuples joined between two {\groups}, we summarize each {\group} \emph{independently} using a set of four aggregates: \acount(), \asum(), \amin(), and \amax() together. As we describe below, these summary aggregates are then used to compute the upper and lower bounds on each \group's score.

%\stitle{Pruning Threshold.} Given the upper and lower bound scores of each candidate {\group}, we  find a pruning threshold {\cdt} on the lowest possible top $k$ score, as the $k$th largest lower bound score across all candidate {\groups} .Any {\group} with its upper bound score $<$  {\cdt} can thus be pruned. We empirically show in Section~\ref{sec:exp} that while the pruning incurs an overhead of first computing the summary aggregates for each candidate {\group}, the gains from skipping  tuple comparisons for pruned {\groups}  offsets the overhead. Moreover, the summary aggregates of each {\group} can be computed independently in parallel.

\stitle{Computing Bounds.} The simplest approach is to create a single set of summary aggregates for each {\group} as depicted in Figure~\ref{fig:segmengttree}b. The gray and yellow blocks depict the summary aggregates for two {\groups} respectively, consisting of \acount (computed using bitmaps), \asum, \amin, and {\amax} in order.

First, for deriving the lower bound, we prove the following useful property based on the convexity property of DIFF functions (see ~\cite{techreport} for the proof).

\vspace{4pt}
\stitle{Theorem 1.}  $\forall$ DIFF$(m_1, m_2, p)$, \\
\hspace{1cm} \aavg(DIFF$(m_1, m_2, p))  \geq$ DIFF(\aavg$(m_1)$,\aavg$ (m_2), p)$ \\ 

\vspace{-8pt}
This essentially allows us to apply DIFF on the average values of each {\group} to get a sufficiently tight lower bounds on scores. For example, in Figure~\ref{fig:segmengttree}b, we get a lower bound of $1700$ for a score of $1717$ for the two {\groups}  shown in Figure~\ref{fig:segmengttree}b.

For the upper bound, it is easy to see that the maximum value of DIFF($m_1$, $m_2$, 2) between any pairs of tuples in R and S is given by:
\amax(
|\amax({\cdcc}) $-$ \amin({\cdca})|, |\amax({\cdca}) $-$ \amin({\cdcc})|). Given that \cdiff is Non-negative and Monotonic, we can compute the upper bound on \asum \xspace by multiplying the the \amax(\cdiff) by {\acount}. For example, in Figure~\ref{fig:segmengttree}b, we get an upper bound of $6400$.

\stitle{Multiple Piecewise Summaries.} Given that the value of {\measure} can vary over a wide range in each {\group}, using a single summary aggregate often does not result in tight upper bound.
Thus, to tighten the upper bound, we create multiple summary aggregates for each {\group}, by logically dividing each {\group} into a sequence of $l$ \emph{segments}, where segment 
$i$ represents tuples from index: $(i-1)\times\frac{n}{l} + 1$ to $i\times\frac{n}{l}$ where $n$ is the number of tuples in the \group.
%Recall that we assume that there are no missing values or duplicates on week, so segments across {\groups}  are automatically alinged on week.
Instead of creating a single summary, we compute a set of same summary aggregates over \emph{each} segment, called \emph{segment aggregates.}  For example, Figure~\ref{fig:segmengttree}c depicts two  segment aggregates for each {\group}, with each segment representing a range of $8$ tuples. The bounds between a pair of matching segments is computed in the same way as we described above for a single summary aggregates. Then, we sum over the bounds across all pairs of matching segments to get the overall bound (see ~\cite{techreport} for formal description). %As we can see in Figure~\ref{fig:segmengttree}c, increasing the number of segments from 1 to 2 improves the bounds  from [$1700$, $6400$] to [$1702$, $4624$], substantially reducing the upper bound.
To estimate the number of summary aggregates for each {\group}, we use Sturges formula, i.e., ($\left\lfloor1 + log_2(n) \right\rfloor$)~\cite{scott2009sturges}, which assumes the normal distribution of {\measure} values for each {\group}. Because of its low computation overhead and effectiveness in capturing the distribution or trends of values, Sturges formula is widely used in the statistical packages for automatically segmenting or binning data points into fewer groups. We empirically evaluate the effectiveness of Sturges formula in Section~\ref{sec:exp}.

\vspace{4pt}

\subsection{Early Termination}
\label{sec:termination}

When selecting top-$k$ trends, we can further reduce the computation by ordering the comparison of trends that are not pruned in the previous step. To do so, we assign an utility to each of the {\groups}  that tells how likely they are going to be in the top-$k$. For estimating the utility of {\groups}, we use the bounds computed using segment aggregates. Specifically, for selecting top-$k$ {\groups}  in descending order of their scores, \emph{a {\group} with higher upper bound score has a higher utility} and for ascending order of scores, a {\group} with the smallest lower bound has a higher utility. The processing of higher utility {\groups}  leads to the faster improvement in the pruning threshold, thereby minimizing wastage of tuple comparisons over low utility {\groups}.

Furthermore, the utility of a {\group} can vary after comparing a few tuples in a candidate \group. Hence, instead of processing the entire {\group} in one go, we process one segment of a {\group} at a time, and then update the bounds to check (i) if the {\group} can be pruned, or (ii) if there is another {\group} with better utility that we can switch to. Incrementally comparing high utility {\groups}  leads to pruning of many {\groups}  without processing all of their tuples.

\subsection{Putting It All Together}
\rev{We implemented a new physical operator, $\Phi_p$, that takes as input the trends, and replaces the join and $\mathcal{F}$  in query plan discussed in Section 4. It outputs a relation consisting of tuples that identify the top-$k$ pairs of {\groups} along with their scores.
The algorithm used by the operator 
makes use of four data structures: 
(1) \segagg: An array where index $i$ stores summary aggregates for segment $i$. There is one \segagg per \group.
(2) \ptstate: It consists of the current upper and lowers bounds on the score between two trends, as well as the next segment within the trends to be compared next. There is one \ptstate for each pairs of trends, and is updated after comparing each pairs of segment. 
(3) \pqp: a max priority queue  that keeps track of the {\group} pairs with the highest upper bound. It is updated after comparing each segment.
(4) \pqs: a min priority queue that keeps track of the {\group} pairs with the smallest lowest bound. It is updated after comparing each segment.}
\vspace{-5pt}

\eat{
\begin{packed_enum}
	\item \segagg: An array where index $i$ stores summary aggregates for segment $i$. There is one \segagg per \group.
	\item \ptstate: The  state of a {\group}, consisting of the current upper and lowers bounds on the score, as well as the next segment within the {\group} to be compared next. There is one \ptstate for each {\group}, and is updated after comparing each segment.
	
	% The state of a {\group}, consisting of 10  elements: a) {\group} key, b) the current upper bound, c) the current lower bound,  d) the left segment index ($e$) and the right segment index ($f$) in the {\group} to process next,	g) the left segment index ($h$) and right segment index (i) in reference {\group} to process next,  ($j$) the next index ($e$ or $f$) from which to access the {\group}, 	($k$) next index ($h$ or $i$) from which to access the segment in $a$.	Here, $i$, $j$, $k$, and $l$ have the same meaning as in the pseudocode in Section~\ref{sec:bidirectional}. There is one \ptstate for each {\group}, updated after scoring of each segment.
	\item\pqp: a \amax \xspace priority queue  that keeps track of the {\group} pairs with the highest upper bound. It is updated after comparing each segment.
	\item \pqs: a \amin \xspace priority queue that keeps track of the {\group} pairs with the smallest lowest bound. It is updated after comparing each segment.
\end{packed_enum}
}

\begin{center}
	\begin{algorithm}
		\small
		\caption{Pruning Algorithm for DIFF-based Comparison}
		\label{algo:scoringalgo}
		\small
		\begin{algorithmic}[1]
			\State Compute \segagg and bitmaps for  each {\group}  $c_i$
			\For{each pair of  {\groups} $c_i$, $c_j$}
			\State Compute bounds on scores  (Section~\ref{sec:segaggregates})
			\State Update \pqs
			\EndFor	
			\For{each pair of trends $(c_i, c_j)$}
			\State \textbf{If} ($(c_i, c_j)$ upper bound $<$ \pqs.Top()) Continue;
			\State Initialize $(c_i, c_j)$ {\ptstate} and push to \pqp
			\EndFor
			\While{size of \pqp  $>$ $k$  }
			\State $(c_i, c_j) =$ \pqp.Top()
			\State Compare a segment of $c_i$ with that of $c_j$ 
			\State Update bounds and \pqs
			\State \textbf{If} ($(c_i, c_j)$ upper bound $<$ \pqs.Top())  Continue;
			\State Push $(c_i, c_j)$ to \pqp 
			\EndWhile
			\State Return Top $k$ {\group} pairs  of trends and their scores from \pqp	
		\end{algorithmic}
	\end{algorithm}
\end{center}

Algorithm~\ref{algo:scoringalgo} depicts the pseudo-code for a single threaded implementation. We first compute the segment aggregates for {\groups}  (line $1$). For each pair of {\groups}, we compute the bounds on scores as discussed in Section~\ref{sec:segaggregates}, and  update \pqs to keep track of top $k$ lower bounds (lines $2$---$5$). The upper bound for each pair of {\group} is compared with  \pqs.Top() to check if it can be pruned (line $7$). If not pruned, the \ptstate is initialized and pushed to \pqp (line $8$). Once the \ptstate of all unpruned {\groups}  are pushed to 
\pqp, we fetch the pair of {\groups} with the highest upper bound score ((line $11$)), and following the process outlined in Section 5.2, compare a pair of segments (line $12$). After the comparison, we check if the current pair of {\groups} is pruned or if there is another pair of {\groups} with higher upper bound (line $14$--$15$). This process is continued  until we are left with $k$ pairs of {\groups} . Finally, we output values of $k$ pairs of {\groups}  with highest scores (line $17$).

\stitle{Memory Overhead.}  Given a relation of $n$ tuples consisting of $p$ {\groups}, $\Phi_p$ creates $p \times log(n/p)$ segment aggregates (assuming tuples are uniformly distributed across trends), with each segment aggregate consisting of fixed set of aggregates. In addition, the operator maintains a \ptstate consisting of bounds on scores between each pair of trends as well as the priority queues to maintain top-k pairs of trends.
Thus, the overall space overhead is $O(p \times log(n/p) +  p^2)$. 
\section{Additional Algebraic Rules}
\label{sec:logopt}
The query optimizer in Microsoft \dbb relies on algebraic equivalence rules for enumerating  query plans to find the plan with the least cost. \rev{When \optr occurs with other logical operators, we  present five transformation rules (see Table~\ref{tab:equivrules})  that reorder $\Phi$ with other operators to generate more efficient plans.}

\stitle{R1. Pushing $\Phi$ below join.} 
Data warehouses often have a snowflake or star schema, where the input to \optr operation may involve a PK-FK join between fact and dimension tables.  If one or more columns in $\Phi$ are the PK columns or have  functional dependencies on the PK columns in the dimension tables , $\Phi$ can be pushed down below the join on fact table by replacing the dimension tables columns with the corresponding FK columns in the fact table (see Rule $R_1$ in 	Table~\ref{tab:equivrules}.)  For instance, consider example 1a  in Section 2.1 that finds a product with a similar average revenue over week trend to `Asia'. Here, revenue column would typically be in a fact table along with foreign key columns for region, product and year. In such cases, we can push $\Phi$  below the join by replacing dimension table columns (e.g., product, week) values with corresponding PK column values.

\eat{Data warehouses typically have a snowflake or star schema
% where the fact table contains all quantitative  data such as sales, profits, quantity, along with the foreign key columns for dimension attributes such as product, region,Week. The details of dimension attributes are kept in dimension tables.
A \optr operation typically involves columns from both fact and dimension tables, and thus input to a \optr operation is  primary key--foreign key (PK-FK) join between these tables. For instance,  for the example query 1 that finds top $10$ products that are similar to `Inspiron' in sales over year, sales column would typically be in the fact table with foreign key columns for product (product-fk) and year (year-fk), while the product name and year are stored in their respective fact tables.
 
In such cases, we push $\Phi$ operation below the fact table by replacing column names (e.g., product, year) with the primary key column names (product-fk, year-fk) in the \optr, and the reference partition value ('Inspiron') with the key value by performing a lookup in the corresponding fact table. More formally, if the input to a $\Phi$ operation is a Primary key (PK)--Foreign key (FK) join $R$ $\Join_T$ $S$ where $R$ is the FK table and $S$ is the PK table for column $T$. 
Let $T_k$ and $T_a$ denote the foreign key and actual column names for $T$. Then, we push $\Phi$ below $R$ if by renaming $T_a$ in $\Phi$ to $T_k$ if (i) $T_d$ $\subseteq$ $\{$\cp,{\cw}$\}$, (ii) $\{$\cp,\cw, {\cdc}$\}$ - $T_d$ are in $R$.
}

\stitle{R2. Pushing Group-by Aggregate ($\Upsilon$) below $\Phi$ to remove duplicates}.
When an aggregate operation occurs above a \optr operation, in some cases we can push the aggregate operation below the \optr to reduce the size of each partition. In particular, consider an aggregate operation  $\Upsilon_{G,A}$ with group by attributes $G$ 
 and aggregate function $A$ such that all columns used in $\Phi$ are in $G$. Then,  if all aggregation functions in $\Phi$  $\in$ \{\amax, \amin\}, we can push $\Upsilon$ below $\Phi$ as  per the Rule $R_2$ in 	Table~\ref{tab:equivrules}. Pushing aggregation operation below $\Phi$ reduces the size of each partition by removing the duplicate values.

\stitle{R3. Predicate pushdown.} A  filter operation ($\sigma$) on partition column (e.g., \cp) can be pushed down below $\Phi$, to reduce the number of partitions to be compared. While predicate pushdown in a standard optimization, we notice that optimizers are unable to apply such optimizations when the \optr are expressed via complex combination of operations as described in Section~\ref{sec:intro}. Adding an explicit logical \optr, with a predicate pushdown rule makes it easier for the optimizer to apply this optimization. Note that if $\sigma$ involves any attribute other than the partitioning column, then we cannot push it below $\Phi$. This is because the number of tuples for partitions compared in $\Phi$ can vary depending on its location.

\stitle{R4. Commutativity.} Finally, a single query can consist of a chain of multiple \optr operations  for performing comparison based on different metrics (e.g., comparing products first on revenue, and then on profit). When multiple $\Phi$ operations on the same partitioning attribute, we can swap the order such that more selective \optr operation is executed first.

\stitle{R5. Reducing comparative sub-plans to $\Phi$}. Finally, we extend the optimizer to check for an occurrence of the comparative sub-expression specified using existing relational operators to create an alternative candidate plan by replacing the sub-expression with $\Phi$. In order to do so, we add the equivalence rule R5 where the expression on the left side represents the sub-expression using existing relational operators. This rule allows us to leverage physical optimizations for comparative queries expressed without using SQL extensions.

\vspace{-15pt}
\rev{
\section{Discussion}
We discuss the  generalizability and robustness of our proposed optimizations as well as potential applications of \optr.

\stitle{Generalizability of optimizations.} Our proposed optimizations in Section 4 deal with replacing \optr to a sub-plan of logical and physical operators within existing database engines. These optimizations can be incorporated in other database engines supporting cost-based optimizations and addition of new transformation rules. Concretely, given a \optr expression, one can generate a sub-plan using Algorithm 1 and transformation rules implementing steps outlined in Section 4.1 and Section 4.2. Furthermore, we discuss additional transformation rules (see Table 3) in Section 6 that optimize the query when \optr occurs along with other logical operators such as join, group-by, and filter. We show that DIFF-based comparisons can be further optimized by adding a new physical operator that first computes the upper and lower bounds on the scores of each trend, which can then be used for pruning partitions without performing costly join. 

%, we added a new physical operator to the engine. \rev{In section 5.3, we provide implementation details such as data structures and algorithm used by the operator, which can be similarly implemented in other database engines}.

\stitle{Robustness to physical design changes.} 
A large part of \optr execution involves operators such as group-by, joins and partition (See Figure 6). Hence, the effect of physical design changes on \optr is  similar to their effect on these operators. For instance, since column-stores tend to improve the performance of group-by operations, they will likely improve the performance of \optr. Similarly, if indexes are ordered on the columns used in {\constraints} or {\indexi}, the optimizer will pick merge join over hash-join for joining tuples from two trends.  Finally, if there is  a materialized view for a part of the \optr expression, modern day optimizers can match and replace the part of the sub-plan with a scan over the materialized view. We empirically evaluate the impact of indexes on \optr implementation in Section 8.

\stitle{Applications of \optr.}
\optr is meant to be used by data analysts as well as applications to issue comparative queries over large datasets stored in relational databases. It has two advantages over regular SQL and middleware approaches (e.g., Zenvisage, Seedb). First, it allows succinct specification of comparative queries which can be invoked from  data analytic tools supporting SQL clients. Second, it helps avoid data movement and serialization and deserialization overheads, and is thus more efficient and scalable. We classify the applications into three categories:

\vspace{2pt}
\noindent
\emph{BI Tools}. BI  applications such as Tableau and Power BI do not provide an easier mechanism for analysts to compare visualizations. However, for supporting complex analytics involving multiple joins and sub-queries, these tools support SQL querying interfaces. For comparative queries, users currently have to either write complex SQL queries as discussed in Introduction, or generate all possible visualizations and compare them manually. With \optr, users can now succinctly express such queries (as illustrated in Section 3) for in-database comparison.

%Furthermore, applications may have to parse the output of COMPARE differently, very similar to how we need to parse the output of Grouping Sets differently. We already show how it can be done via top-k examples

\vspace{2pt}
\noindent
\emph{Notebooks.} For large datasets stored in relational databases, it is inefficient to pull the data into notebook and use dataframe APIs for processing. Hence, analysts often use a SQL interface to access and manipulate data within databases. While  one  can also expose Python APIs for comparative queries and automatically  translate them  to SQL, such features are limited to the users of the Python library. SQL extensions, on the other hand, can be invoked from multiple applications and languages that support SQL clients. Furthermore, in the same query, one can use \optr along with other relational operators such as join and group-by that are frequently used in data analytics (see Section 3.2).

%SQL can be invoked from different tools 

%is relative more ubiquitous, and thus our extensions can be leveraged across multiple applications and languages that support SQL client to \db.

%While  one  can also expose Python APIs to easily express comparative queries and automatically  translate them  to SQL, such features are limited to the users of the Python library. SQL is relative more ubiquitous, and thus our extensions can be leveraged across multiple applications and languages that support SQL client to \db.

\vspace{2pt}
\noindent
\emph{Visual analytic tools.} Finally, there are visual analytic tools such as as Zenvisage and Seedb that perform comparison between subsets of data in a middle-ware. With \optr, such tools can scale to large datasets and decrease the latency of queries as we show in Section 8.
}

 \begin{table*}[t]
 	\centering
 	\vspace{-15pt}
 	\caption{Queries over Flight and TPC-DS datasets}
 	\vspace{-8pt}
 	\resizebox{\textwidth}{!}{%
 		\begin{tabular}{  c | p{2cm}| p{2cm} | p{3cm} | p{0.8cm} | p{2cm}| p{3cm}| p{0.8cm} | p{2cm} | p{3cm} |p{0.8cm}| p{2cm}| p{3cm} |p{0.8cm}  } \hline
 			 ID & \pbox{1cm}{Type} & \multicolumn{6}{c|}{\bf Flight} & \multicolumn{6}{c}{\bf TPC-DS} \\ \hline 
 			   & & \multicolumn{3}{c|}{{\groups}et 1} & \multicolumn{3}{c|}{{\groupset} 2} &
 			   \multicolumn{3}{c|}{{\groupset} 1} &
 			   \multicolumn{3}{c}{{\groupset} 2} \\ \hline
 			   & &
 			   {\constraint}, \# & ({\indexi},{\measure}), \#
 			   &  \# {\groups}
  			   & {\constraint}, \# & ({\indexi},{\measure}), \#
 			   &  \# {\groups}
 			   & {\constraint}, \# & ({\indexi},{\measure}), \#
 		       &  \# {\groups}
   			   & {\constraint}, \# & ({\indexi},{\measure}), \#
  			   &  \# {\groups}  \\ \hline   
 			 Q1 & One to many with fixed attributes & airport=`SFO', 1
 			   & (Days, ArrDelays), 1  & 1
 			 &  all airports, 384 & (Days, ArrDelays)  &  384   & webpage = 1; 1  & (Items, NetProfits), 1 & 1  & all webpages; 2040 & 1 & 2040 \\ \hline
 			 Q2 & Many to many with fixed attributes  & all airports, 384  & (Days, ArrDelays), 1  & 384	 &  all airports, 384  & (Days, ArrDelays)   & 384 & all webpages; 2040 & (Items, NetProfits), 1 & 2040 & all webpages; 2040  &  (Items,NetProfits), 1 & 2040 \\ \hline
 			 Q3 & One to one with varying attributes  & airport=`SFO', 1 &
 			  (Days, ArrDelays),  (Days, DepDelays), (Weeks, ArrDelays),
 			 ...,  (Weeks, WeatherDelays,); 10 & 10  & airport = `SFO', 1 &  (Days, ArrDelays),  (Days, DepDelays)), (Weeks, ArrDelays),
 			 ...,  (Weeks, DepDelays); 10 & 10  & webpage = 1; 1  & (Items, NetProfits), (Days, NetProfits), ..., (Days, Quantity),5 & 5 &  webpage = 1; 1 & (Items, NetProfits), (Days, NetProfits), ..., (Days, Quantity),5 & 5 \\ \hline
 			 Q4 & Many to many with varying attribues & all airports, 384  &    (Days, ArrDelays),  (Days, DepDelays), (Weeks, ArrDelays),
 			 			 ...,  (Weeks, WeatherDelays,); 10 & 3840 & all airports  &   (Days, ArrDelays),  (Days, DepDelays), (Weeks, ArrDelays),
 			 			 			 ...,  (Weeks, WeatherDelays,); 10 & 3840 & all webpages; 2040 & 
 			 (Items, NetProfits), (Days, NetProfits), ..., (Days, Quantity),5
 			  & 10200 & all webpages; 2040 & (Items, NetProfits), (Days, NetProfits), ..., (Days, Quantity),5 & 10200 \\ \hline
 		\end{tabular} %
 	}    
 	\vspace{-10pt}
 	\label{tab:queries}
 \end{table*}

\section{Performance Evaluation}
\label{sec:exp}

\techreport{
\begin{figure*}
\centering
%	\vspace{-15pt}
	\begin{subfigure}{0.53\textwidth}
		\centerline {
			\hbox{\resizebox{\columnwidth}{!}{\includegraphics{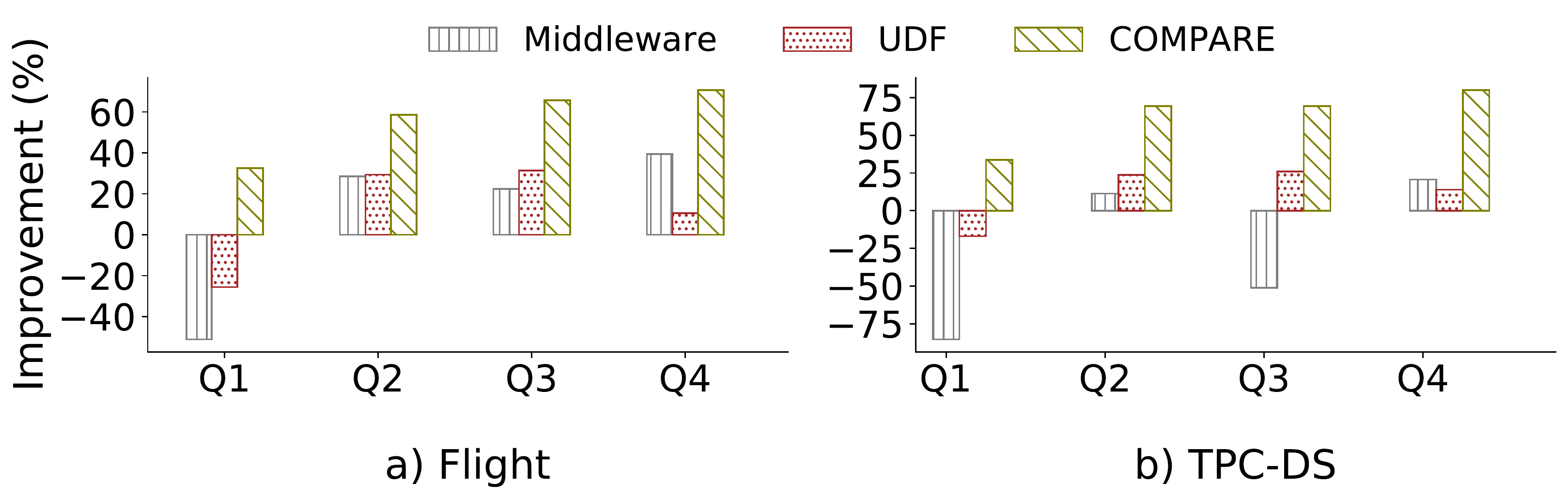}}}}
		\vspace{-5.5pt}
		\caption{\smallcaption{Comparison with Baselines}}
		\label{fig:latency}
		\vspace{-2.5pt}
	\end{subfigure}
	\hspace{-0.1cm}
	\begin{subfigure}{0.43\textwidth}
		\centering
		\hbox{\resizebox{\columnwidth}{!}{\includegraphics{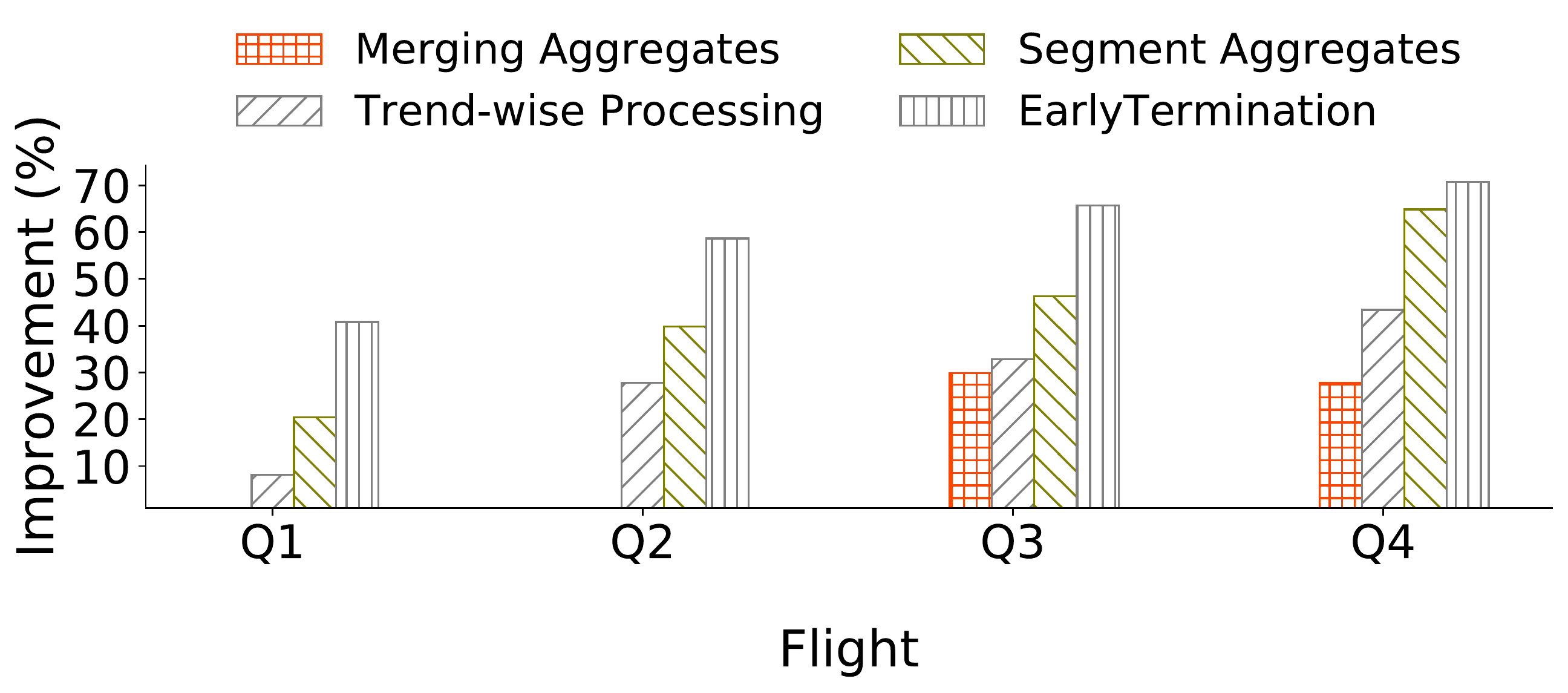}}}
		\vspace{-5.5pt}
		\caption{\smallcaption{Ablative analysis quantifying the impact of each optimization. Each optimization is successively turned on from left to right.}}
		\label{fig:ablativestudy}
		\vspace{-2.5pt}
	\end{subfigure}
	\vspace{-5pt}
	\caption{Improvement in end-to-end latency w.r.t. unmodified \db}
		\vspace{-10pt}
	\label{fig:productresults}	
\end{figure*} 
}

\papertext{
\begin{figure*}
\centering
	\hspace{-0.1cm}
	 \begin{subfigure}{0.25\textwidth}
	    	\centerline {
	    	\hbox{\resizebox{\columnwidth}{!}{\includegraphics{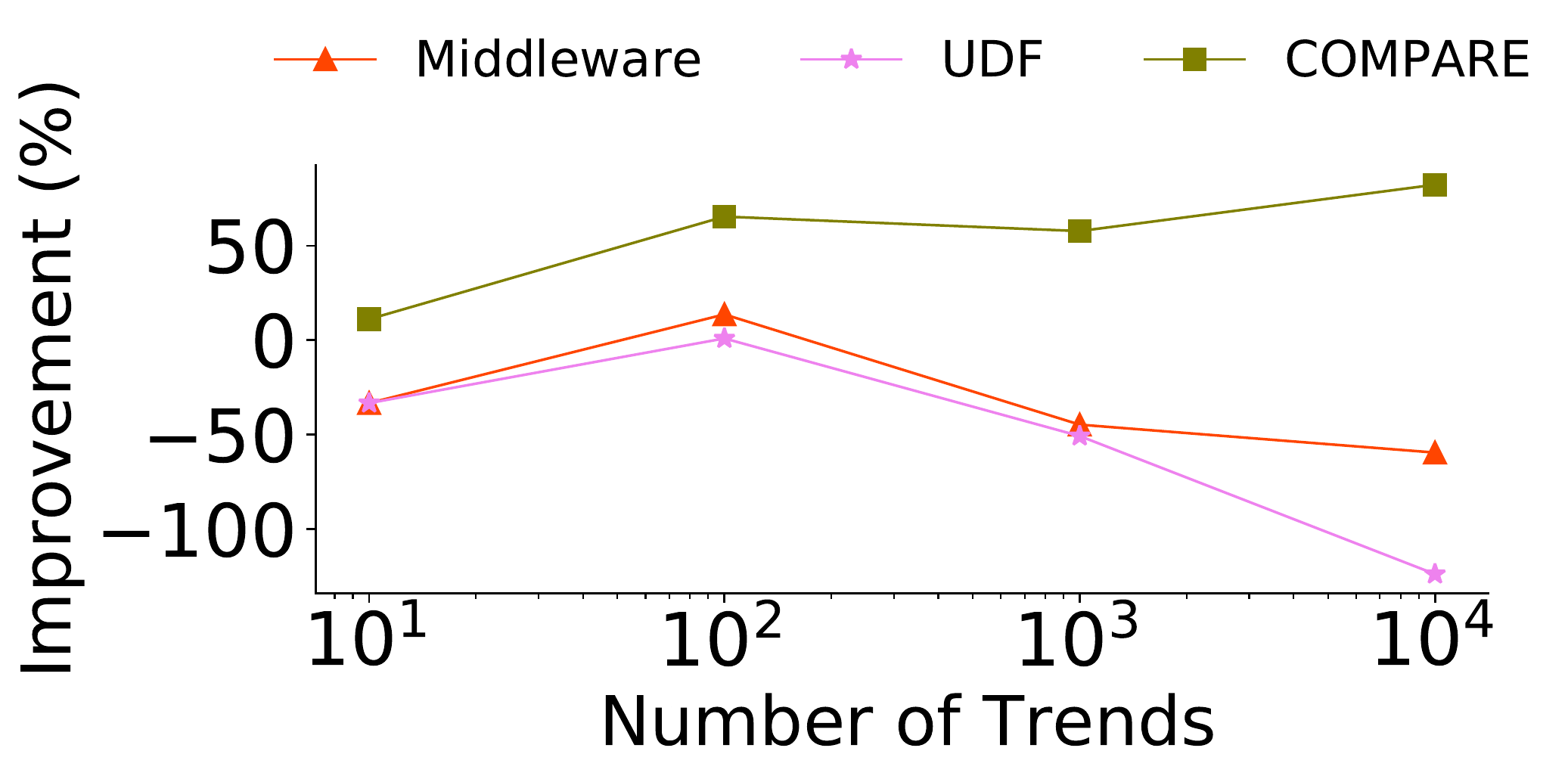}}}}
	    	\vspace{-5.5pt}
	    	\caption{\smallcaption{Varying number of {\groups} with fixed ({\indexi}, {\measure})}}
	    	\label{fig:varyreferencesq2}
	 \end{subfigure}
	  	\hspace{-0.1cm}  
	 \begin{subfigure}{0.25\textwidth}
		    	\centerline {
		    		\hbox{\resizebox{\columnwidth}{!}{\includegraphics{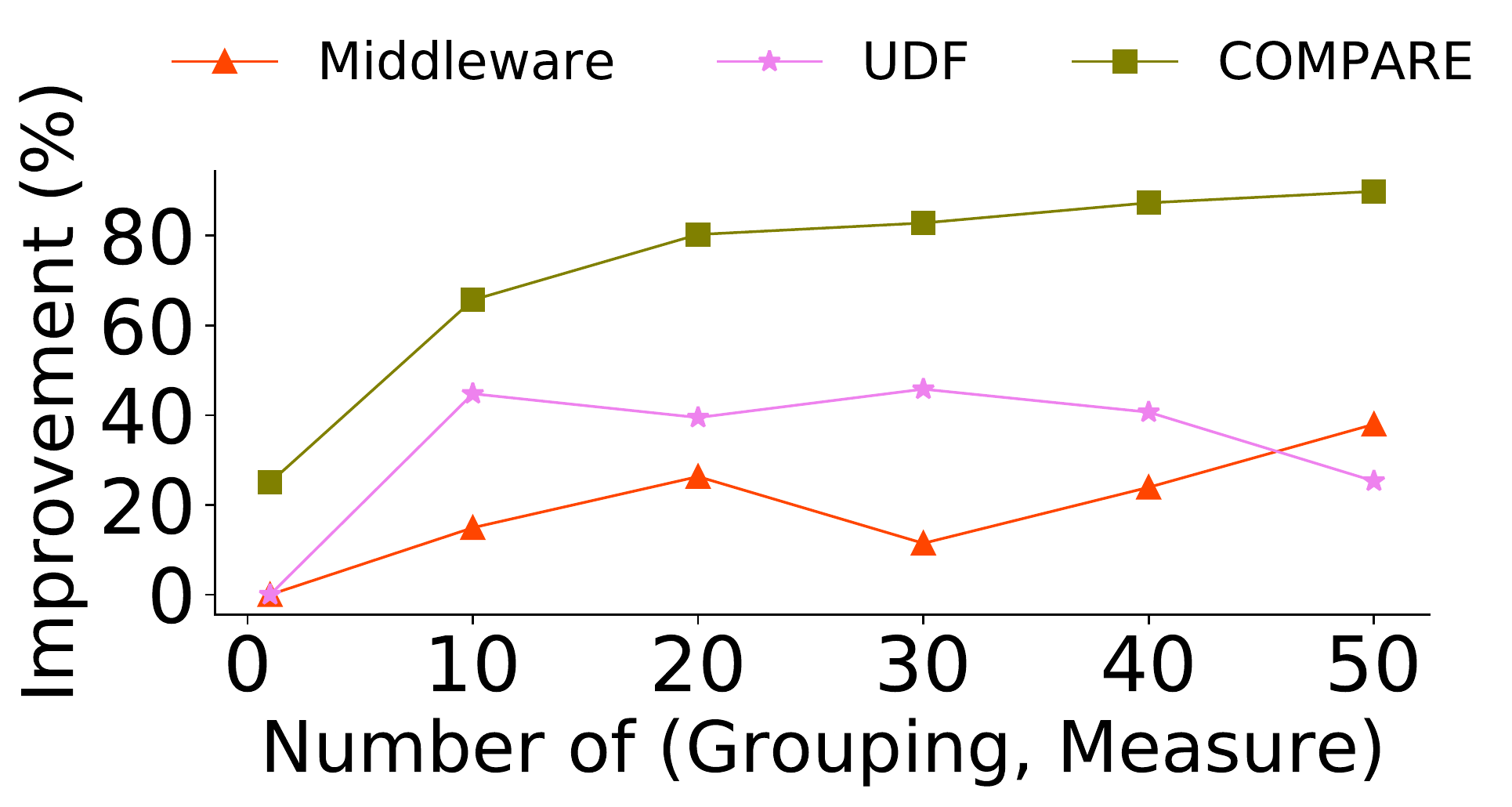}}}}
		    	\vspace{-5.5pt}
		    	\caption{\smallcaption{Varying number of ({\indexi}, {\measure})} }
		    	\label{fig:varyreferencesq2}
     \end{subfigure}
   \hspace{-0.1cm}
	\begin{subfigure}{0.25\textwidth}
		\centerline {
			\hbox{\resizebox{\columnwidth}{!}{\includegraphics{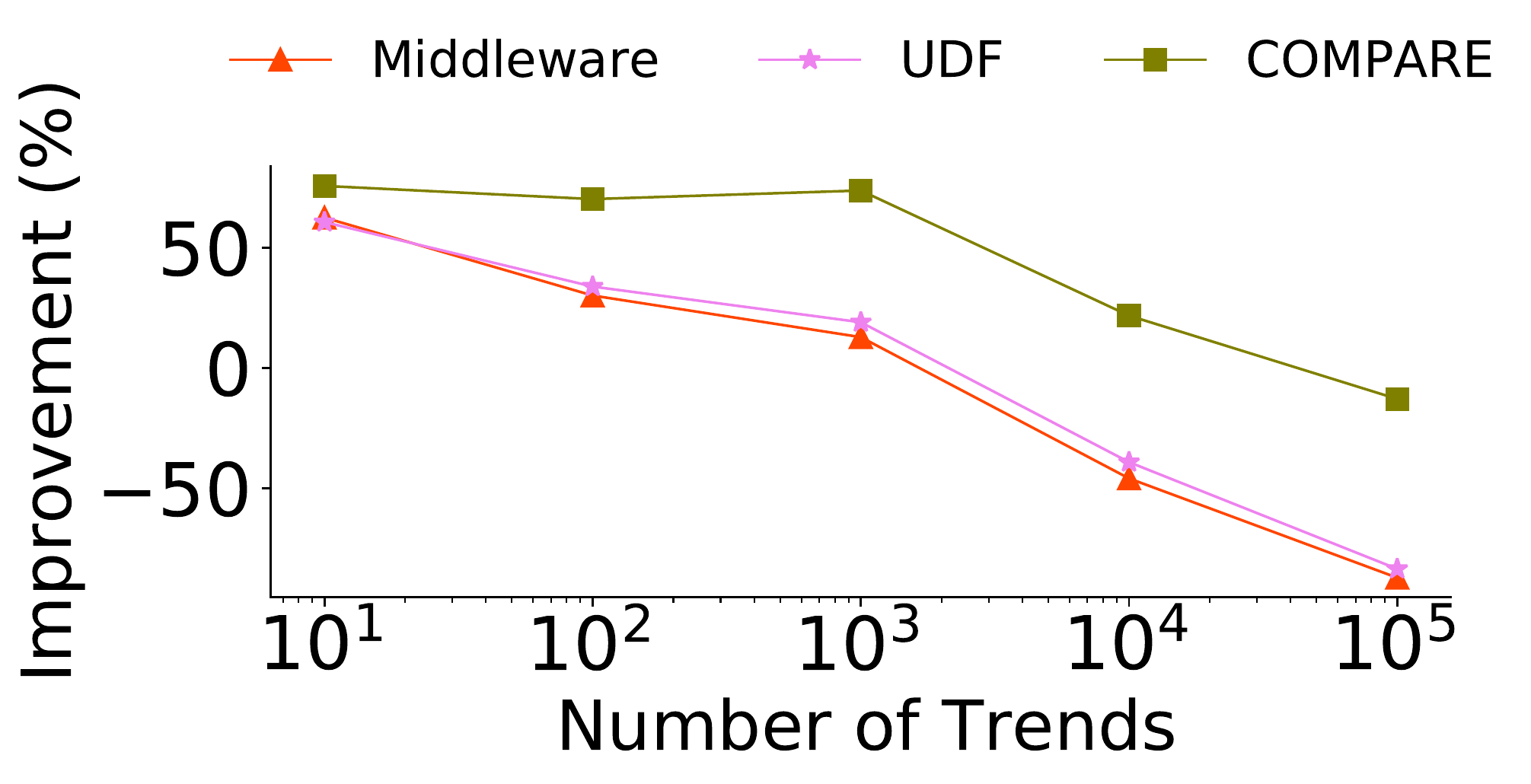}}}}
		\vspace{-5.5pt}
		\caption{\smallcaption{Increasing number of {\groups} with proportional decrease in {\group} size over a fixed data of size $10^5$}}
		\label{fig:varypcountdatafixed}
		\vspace{-2.5pt}
	\end{subfigure}
	\vspace{-5.5pt}
	\caption{Impact on latency on varying the number and size of {\groups} on the flight dataset. \tar{fix}}
	\label{fig:productresults}	
\end{figure*}
}

\techreport{
\begin{figure*}
\centering
	\hspace{-0.1cm}
	 \begin{subfigure}{0.32\textwidth}
	    	\centerline {
	    	\hbox{\resizebox{\columnwidth}{!}{\includegraphics{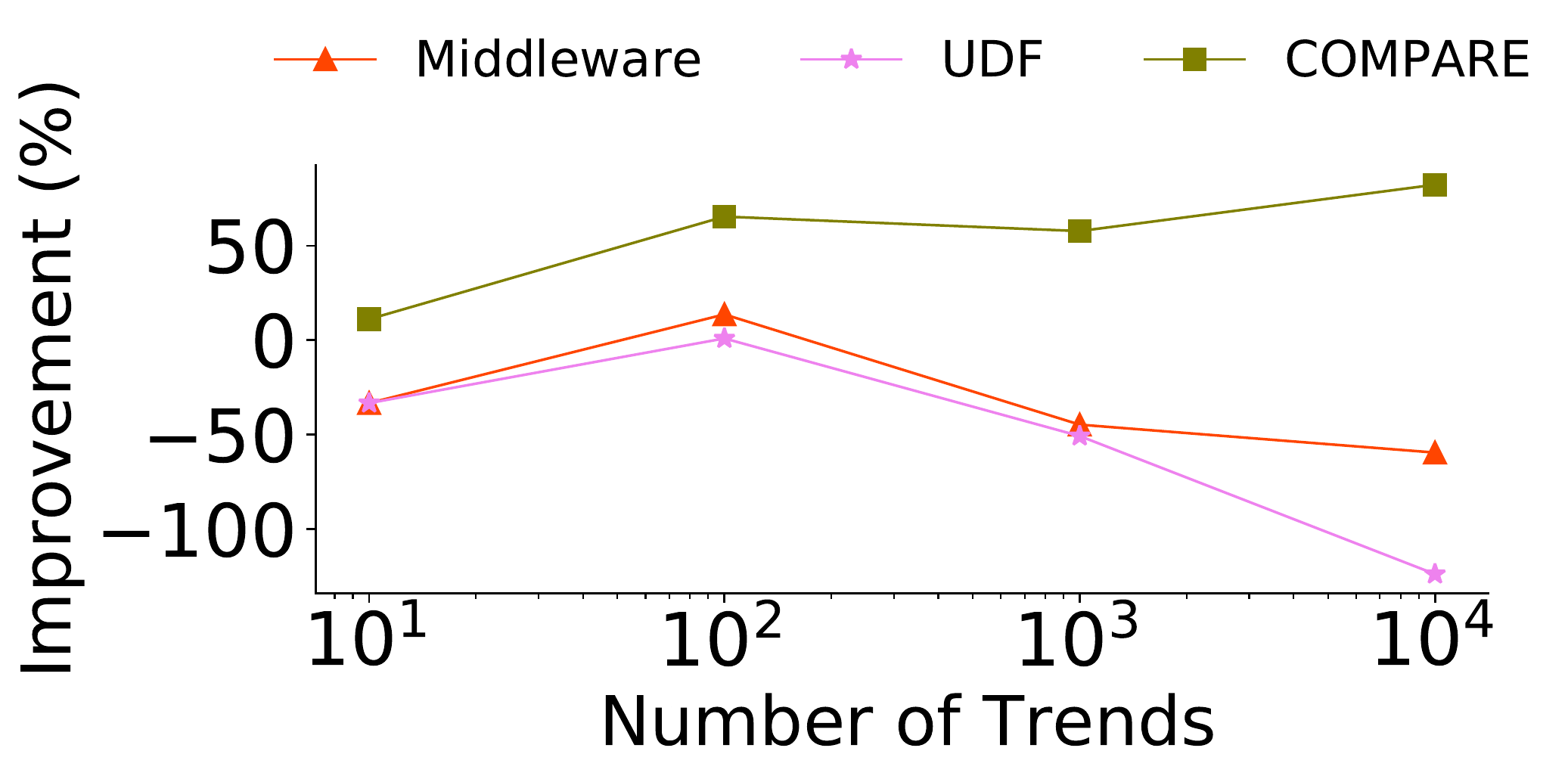}}}}
	    	\vspace{-5.5pt}
	    	\caption{\smallcaption{Varying number of {\groups} with fixed ({\indexi}, {\measure})}}
	    	\label{fig:varyreferencesq2}
	 \end{subfigure}
	  	\hspace{-0.1cm}  
	 \begin{subfigure}{0.32\textwidth}
		    	\centerline {
		    		\hbox{\resizebox{\columnwidth}{!}{\includegraphics{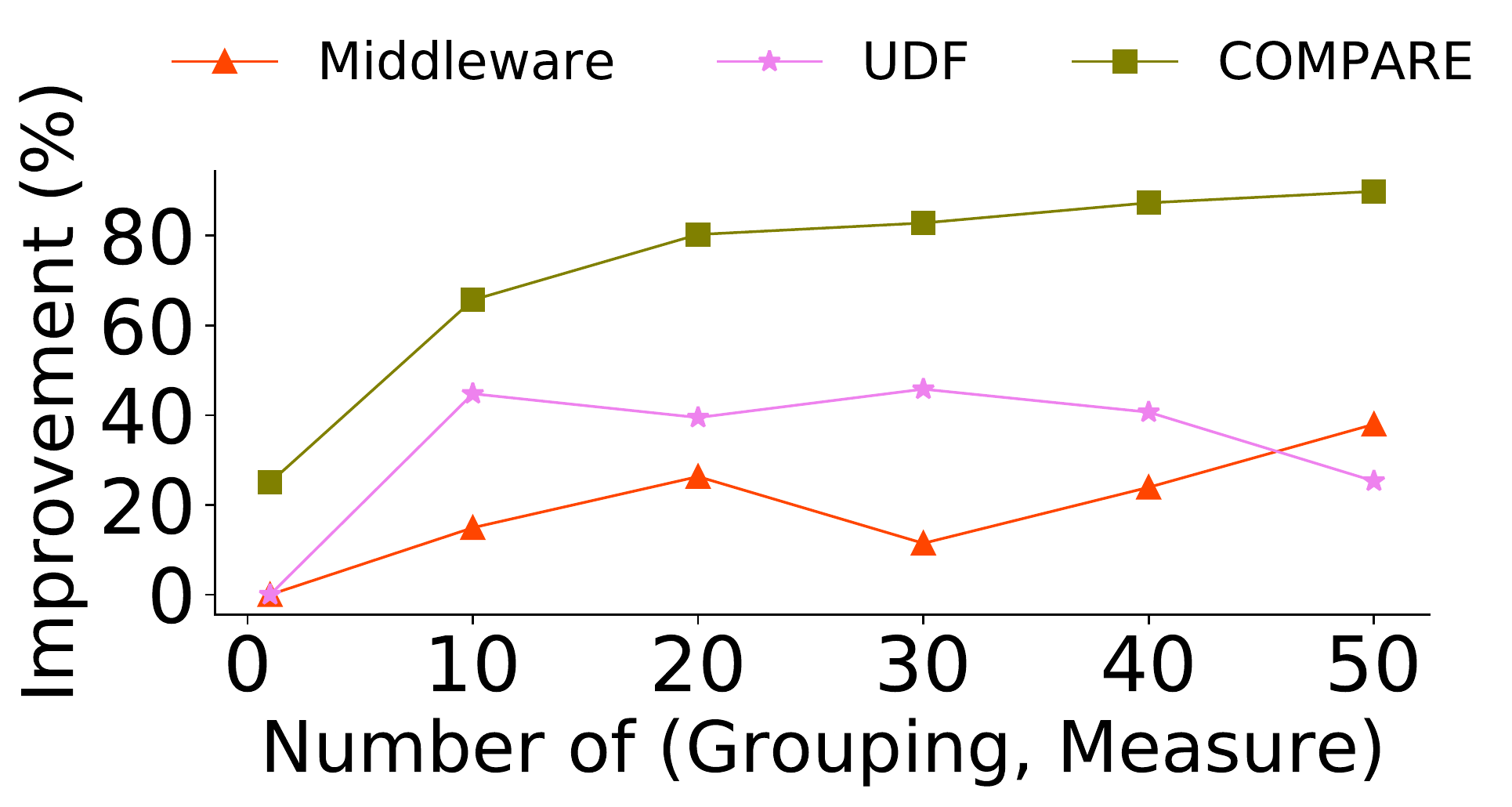}}}}
		    	\vspace{-5.5pt}
		    	\caption{\smallcaption{Varying number of ({\indexi}, {\measure})} }
		    	\label{fig:varyreferencesq2}
     \end{subfigure}
   \hspace{-0.1cm}
	\begin{subfigure}{0.32\textwidth}
		\centerline {
			\hbox{\resizebox{\columnwidth}{!}{\includegraphics{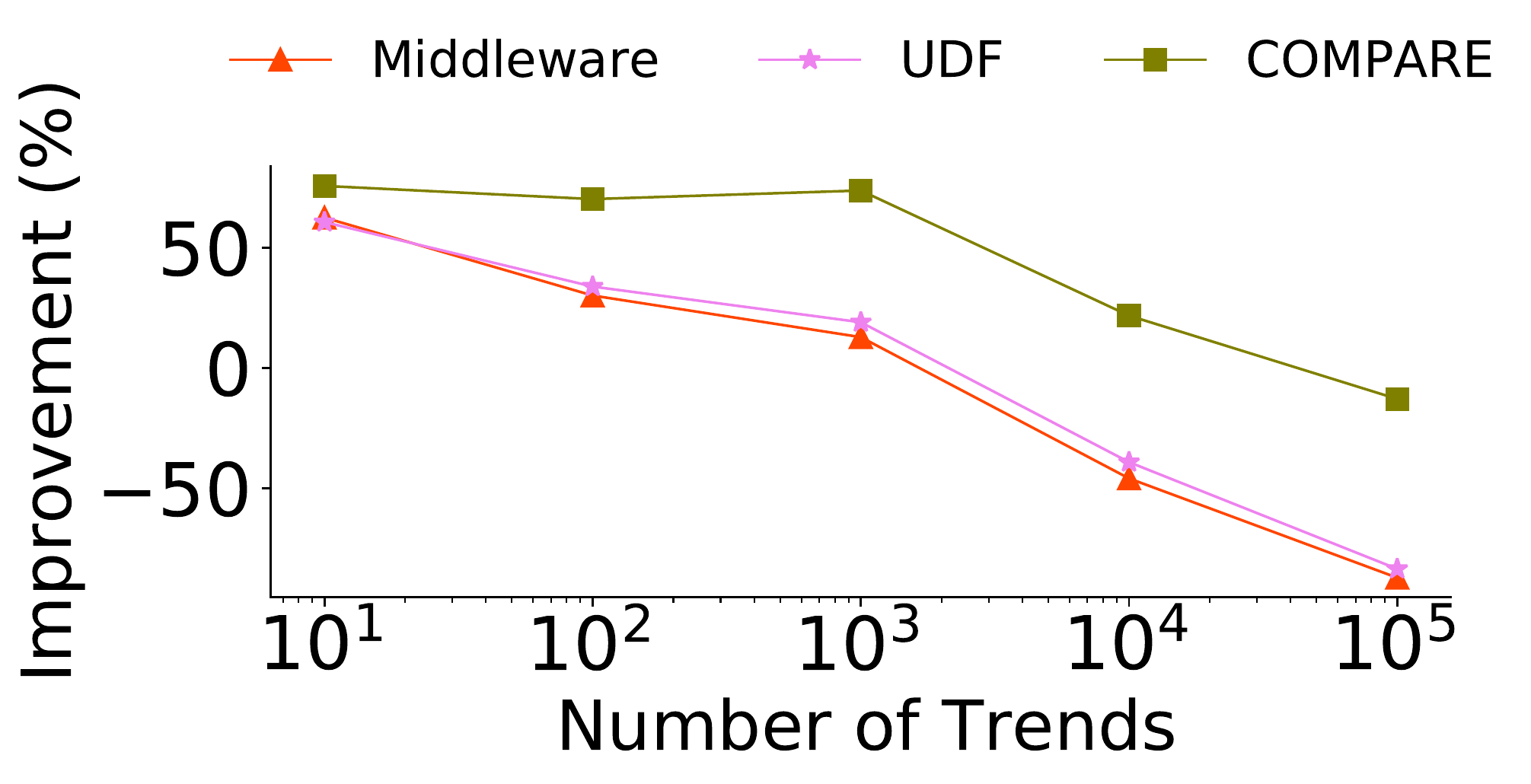}}}}
		\vspace{-5.5pt}
		\caption{\smallcaption{Increasing number of {\groups} with proportional decrease in {\group} size over a fixed data of size $10^5$}}
		\label{fig:varypcountdatafixed}
		\vspace{-2.5pt}
	\end{subfigure}
	\vspace{-5.5pt}
	\caption{Impact on latency on varying the number and size of {\groups} on the flight dataset. \tar{fix}}
	\label{fig:productresults}	
    \papertext{\vspace{-15pt}}
\end{figure*}
}

 Using our prototype implementation on \db (referred as \optr below), we evaluate the improvement in latency with respect to current execution strategy in \dbb as described in Section 4.1. We consider two alternative strategies as baselines: (b) \mdw: Issuing select-aggregate queries to retrieve the data from \dbb over a network (average speed of 10 MB/s) and performing comparison and filtering in a C\# implementation;  
this approach mimics the data retrieval approach followed by visualization tools such as Zenvisage~\cite{zenvisagevldb} while also incorporating {\group}wise comparison and segment-aggregates based pruning optimizations (discussed in Section 5), and (c) an UDF implementation that executes within \dbb. It takes as input the UNION of all group-by aggregates (computed via GROUPING SETs clause) and incorporates {\group}wise comparison and segment-aggregates based pruning optimizations.

\stitle{Datasets and Queries.} We use two datasets: Flight~\cite{airlinedata} and TPC-DS with a scale factor of $100$~\cite{nambiar2006making}(summarized in  Table \ref{tab:datasets}). We use websales table in TPC-DS which has  PK-FK joins with tables webpages and warehouses. As depicted in Table~\ref{tab:queries}, we issue four types of comparative queries (with characteristics similar to  examples discussed in Section 2.1), with the default number of output pair of {\groups} set to $5$. All {\measure} attributes are aggregated using AVG() and we use SUM() OVER DIFF(2) as \scorer.

\begin{table}[H]
	\centering
	\vspace{-5pt}
	\caption{Datasets}
		\vspace{-10pt}
	\resizebox{0.75\columnwidth}{!}{%
		\begin{tabular}{ | c | c | c |c|} \hline
			Dataset & Disk Size  & Buffer Size & Number of rows \\  \hline
			Flight & 8GB & 11GB & 74M \\  \hline
			TPC-DS  & 20GB & 24 GB & 720M \\  \hline
		\end{tabular}%
	}
	\label{tab:datasets}
	\vspace{-10pt}
\end{table}

% We scale the size of TPC-DS by increasing the number of partitions corresponding to warehouses by a factor of 10, and for avoiding duplicate partitions, we update the measure values by adding a random number in the range of $\pm$ stddev.

%\system currently extends the row-store engine of \db, therefore in order to minimize differences between row-store and column-store style query processing, we turn on non-clustered indexes on all the columns in the query. While the query processing time for both \optr and \dbb decreased because of these changes, the relative differences were similar before and after the changes.

%(i) we removed all columns from the tables that are not part of queries, and (ii) created non-clustered indices on the queried columns. While the query processing time for both \optr and \dbb decreased because of these changes, the relative differences were similar before and after the changes. \viv{How dependent on the physical design are the experimental results? Tarique: While my implementation in on rowstore, I think our approach should be equally effective for columnstores}

\stitle{Setup.} All experiments were conducted on a 64-bit Windows 2012 Server with 2.6GHz Intel $\times$eon E3-1240 10-core, 20 logical processors and 192GB of 2597 MHz DDR3 main memory. Unless specified, we use the default settings for the  degree of parallelism (DOP) and buffer memory, where the \db tries to utilize the maximum possible resources available in the system. 
We report the results of warm runs by loading the tables referenced in
the query into memory.
%Moreover, we load  the query tables into buffer via a simple scan before issuing queries to minimize differences between main memory vs disk processing. We show the buffer size of each dataset in Table~\ref{tab:datasets}.

\subsection{End-to-End Latency}
Figure~\ref{fig:latency} depicts the end-to-end improvement in latency of \system, \mdw, and UDF with respect to the unmodified \db runtimes. 
We see that \optr provides a substantial improvement with respect to all approaches, with improvement being proportional number and size of {\groups}.

For Q1 that involves one to many comparisons over a fixed attribute combination, we see a speed-up of about 26\% on Flight and about 36\% on the TPC-DS. The improvement increases substantially as we increase the complexity of the query; for example we see upto 4$\times$ improvement in latency for Q2 and Q4 which involve a large number of {\group} comparisons. 
For \mdw, the main bottleneck is the data transfer and deserialization overhead, which takes up to $70\%$ of the overall execution time. While UDF also incurs an overhead in invocation and reading the input from downstream aggregate operators, a large part of its time ( $>$ 90\%) is spent on processing, indicating that inline execution of \optr via partitioning and join operators is much faster. In summary, we find that {\em \system  gives the best of both worlds: requires minimal data transfer and deserialization overhead, and runs much faster by efficiently comparing tuples within databases.}

\eat{
In particular, we see that \optr  gives up to 2.4$\times$ improvement over \mdw and upto 1.8$\times$ improvement over UDF. As discussed shortly, these gains improve further as we increase the size and number of entities. Thus, we see that {\em \system  gives the best of both worlds: requires minimal data transfer and deserialization overhead, and runs much faster by efficiently comparing tuples within databases.}}

%Figure~\ref{fig:latency} depicts the end-to-end latency of \dbb and \mdw-based approach with respect to variants of \system. Figure~\ref{fig:latency}b depicts the overall number of tuple comparison performed by each of the approaches. Note that we use the log scale for the $y$ axis in both figures.

\stitle{Ablative Analysis.} Next, we conducted an ablative analysis to evaluate the effectiveness of each of the optimizations described in Section 4 and Section 5.
Figure~\ref{fig:ablativestudy} depicts the impact of each optimization as we add them successively from left to right. Each level of \optr optimization provides a substantial speed-up in latency compared to basic execution strategy. For  Q3 and Q4, sharing aggregates improves the runtime by about 30\% (note that there are no sharing opportunities for Q1 and Q2). The trend-wise processing further improves the processing by 25\% on average---more the number of {\group} comparisons, the higher the improvement. Note that both sharing aggregates and trend-wise processing do not depend on the properties of {\scorer} and hence can be applied on arbitrary {\scorer}. The next two optimizations based on segment-aggregates and early termination, although only applicable for {\cdiff}-based comparison, result in the massive improvement ranging between 20-25\% by pruning {\groups} early that are guaranteed to be not in top-k.

%Figure~\ref{fig:latency}b, the key reason behind the lower latency for \optr variants is the reduction in the number of tuple comparisons, which is much more for queries involving all-pairs comparison. While the self-join operation beween reference partition and candidate partitions in  \dbb generates and materializes a large intermediate data, \system-NoOpt performs groupwise processing that minimizes extra tuple comparisons, and skips the materialization of results from individual pairs of tuples. These help \system-NoOpt achieve between 2$\times$ to $40\times$ better latency with respect to \dbb. Next, \system-SA uses segment-aggregates to filter low utility partitions which further reduces the number of tuple comparisons by two orders of magnitude on average, resulting in improvement in latency between 2$\times$ to $10\times$. Finally, the optimizations for prioritizing high utility partitions first (\system-POrd) and bidirectional comparison (\system) lead to faster improvement in bounds of, which in turn lead to early termination of top \cdk \xspace partitions search and help  reducing the latency by a factor of $1.2\times$ to $10\times$.

\subsection{Sensitivity to Data Characteristics}
\label{sec:partionsexp}
We now evaluate the impact of dataset characteristics on the performance of \optr. For these experiments, we use the flight dataset (consisting of real-world trends/distributions) and scale its size as described below.

\stitle{Impact of number of {\groups}.} To evaluate this, we scale the number of {\groups} for query Q2 between $10$ and $10^4$ by randomly removing or replicating the {\groups} corresponding to original $384$ airports. While replicating, we update the original value $m_o$ of each {\measure} column $m$ by a new value $m_n$ where $m_n$ = $m_o$ {\textpm} $stdev(m)$. This ensures that the replicated {\groups} are not duplicates but still represent the original distribution. We find that the increase in the number of {\groups} 
leads to the increase in latency for all approaches; however the increase is much higher for UDF and \mdw due to data movement and deserialization overhead. \system is further able to reduce comparisons due to early pruning of partitions using segment-aggregates.

%Because of the increase in dataset size, all approaches incur increase in latency. However, the increase in latency for \system is relatively less due to higher pruning of partitions due to segment-aggregates as well as fewer number of comparisons across partitions.

\stitle{Impact of number of ({\indexi}, {\measure})}. In this case, we scale the number of ({\indexi}, {\measure}) for query Q3 between $1$ and $50$ by randomly removing or replicating the columns for each {\group} while updating the values of replicated {\measure} column as described above. All approaches incur increase in latency; however, the increase in latency is much higher for \dbb compared to \optr, \mdw and UDF due to higher sharing of aggregate computations.

\stitle{Varying number and size of {\groups} while keeping the overall data size fixed.} Using a similar process as described above, we scale the number of {\groups} between $10$ and $10^5$ while proportionally decreasing the size of each {\group} such that the  size of the dataset is fixed to $10^5$. Here, we see an interesting observation. The latency of \dbb decreases as we increase the number of {\groups} and reduce their size. This is because with the decrease in the size of {\groups}, the number of tuple comparison decreases. As a result of this, the improvement in latency w.r.t \dbb decreases for all of \optr, \mdw, and UDF. However, for \system, the latency initially decreases as sorting and comparison can done faster in parallel as the number of partitions increase. As the number of partitions become too large, the  improvement due to parallelism decreases.

\subsection{Impact of Number of Segment Aggregates}

Recall from Section 5.1 that we use the Sturges formula~\cite{scott2009sturges}, i.e., ($\left\lfloor1 + log_2(n) \right\rfloor$) (where $n$ is the estimated size of {\group}) to estimate the number of segment-aggregates. To measure the efficacy of this formula, we measure the changes in latency  as we increase the number of segment-aggregates for Q2 (Figure~\ref{fig:varysegmentq1}) and Q4 (Figure~\ref{fig:varysegmentq2}). With the increase in number of segments, the
overall latency decreased initially. However, as the number of segments is increased beyond a certain number, the latency starts increasing. This is because of the increase in the number of segment-aggregates comparisons without further pruning. The dotted line shows the results for the number of segments (i.e., ($\left\lfloor1 + log_2(n) \right\rfloor$) that is automatically selected by \system, showing that the latency for selected segments is close to minimal possible latency.

Next, we measure the impact of number of tuples processed per update for early termination (Section~\ref{sec:termination}). Figure~\ref{fig:varyingTuplesFetched} depicts the impact of overall latency for $Q2$ and $Q_4$ as we vary the number of tuples processed for a given {\group} for updating the upper and lower bounds. The dotted black line depicts the performance for the number of tuples that \system automatically decides, i.e., ($\frac{n}{(\left\lfloor1 + log_2(n) \right\rfloor)}$) (i.e., estimated size of a segment). We see that the latency is very high when we only consider a few tuples ($< 10$) at time. This is because of cache misses and many updates to the priority queues for reprocessing the same set of partitions repeatedly. On the other hand, processing too many tuples leads to extra processing, even for low utility partitions that can be pruned earlier. As depicted by the dotted line, the number chosen by \system, although not perfect, is close to the optimal performance that we can get by processing few tuples at a time.

\papertext{
\begin{figure}
	\hspace{-0.2cm}
	\centering
	\begin{subfigure}{0.4\columnwidth}
		\centerline {
			\hbox{\resizebox{\columnwidth}{!}{\includegraphics{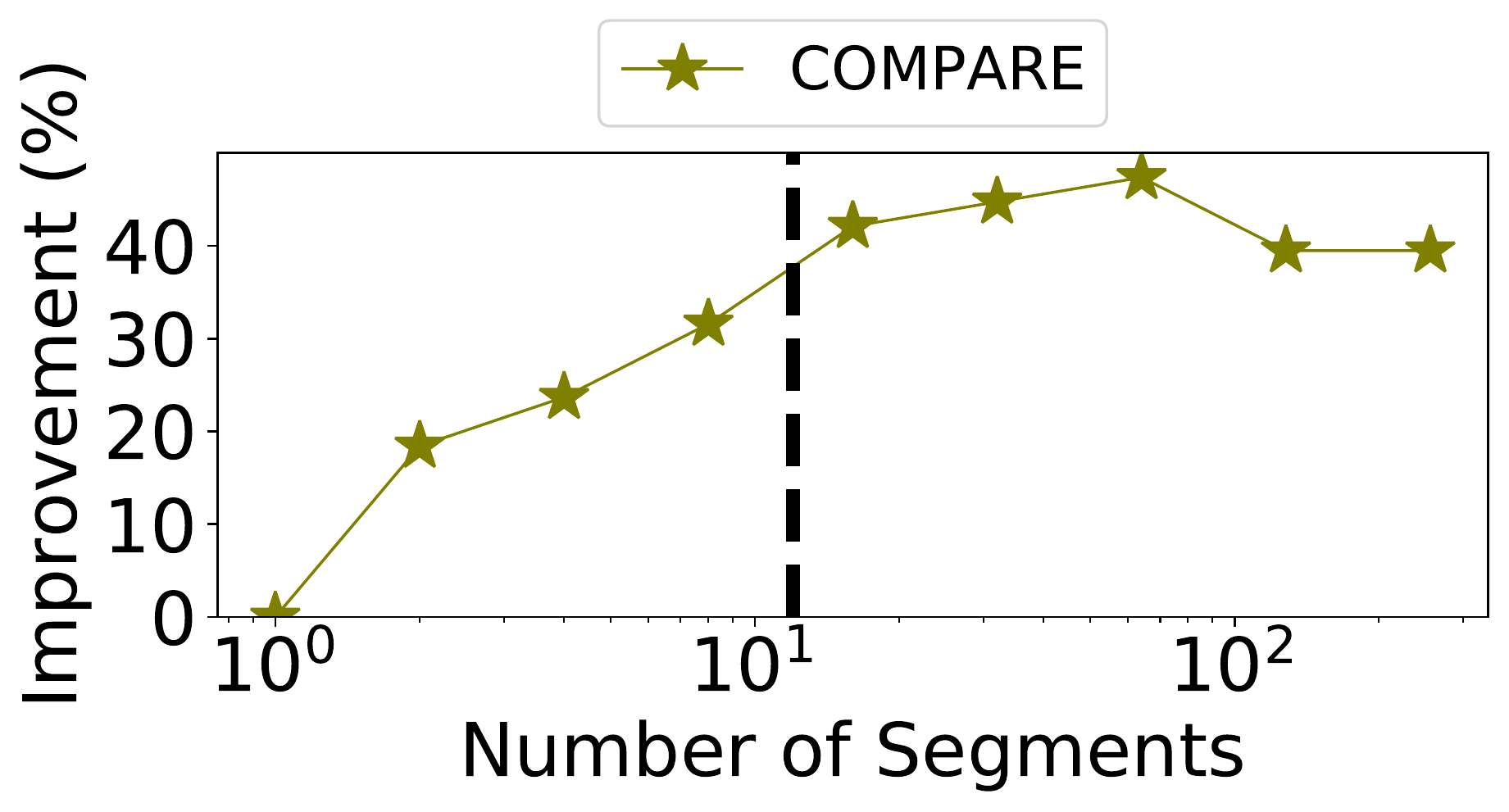}}}}
		\vspace{-1.5pt}
		\caption{\smallcaption{$Q_2$}}
		\label{fig:varysegmentq1}
		\vspace{-2.5pt}
	\end{subfigure}
	\hspace{-0.2cm}
	\begin{subfigure}{0.4\columnwidth}
		\centerline {
			\hbox{\resizebox{\columnwidth}{!}{\includegraphics{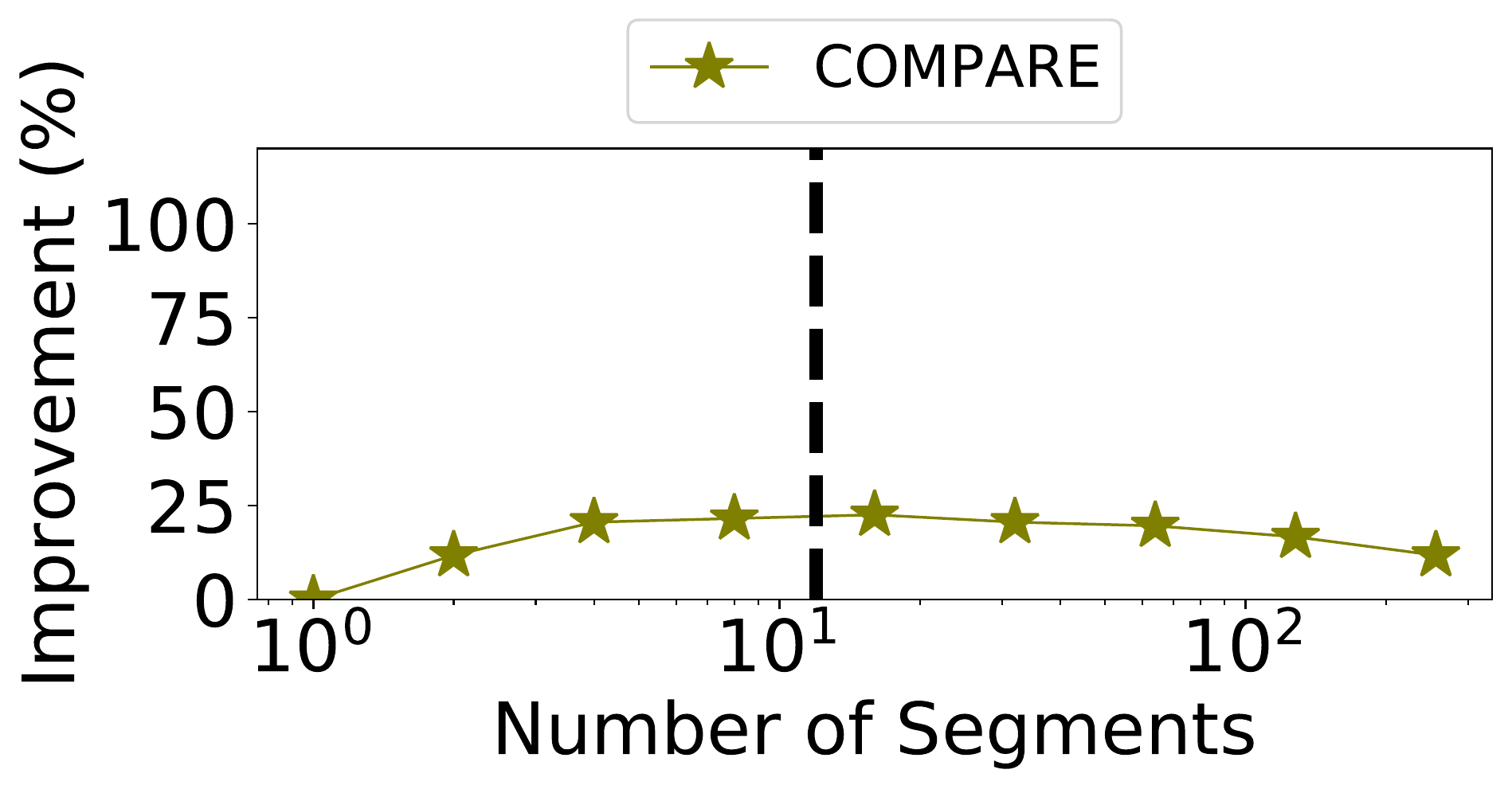}}}}
		\vspace{-1.5pt}
		\caption{\smallcaption{$Q_4$}}
		\label{fig:varysegmentq2}
	\end{subfigure}
	\vspace{-5pt}
	\caption{Varying number of segment-aggregates}
	\label{fig:productresults}	
\end{figure}
}

\techreport{
\begin{figure}
	\vspace{-2.5pt}
	\hspace{-0.2cm}
	\centering
	\begin{subfigure}{0.49\columnwidth}
		\centerline {
			\hbox{\resizebox{\columnwidth}{!}{\includegraphics{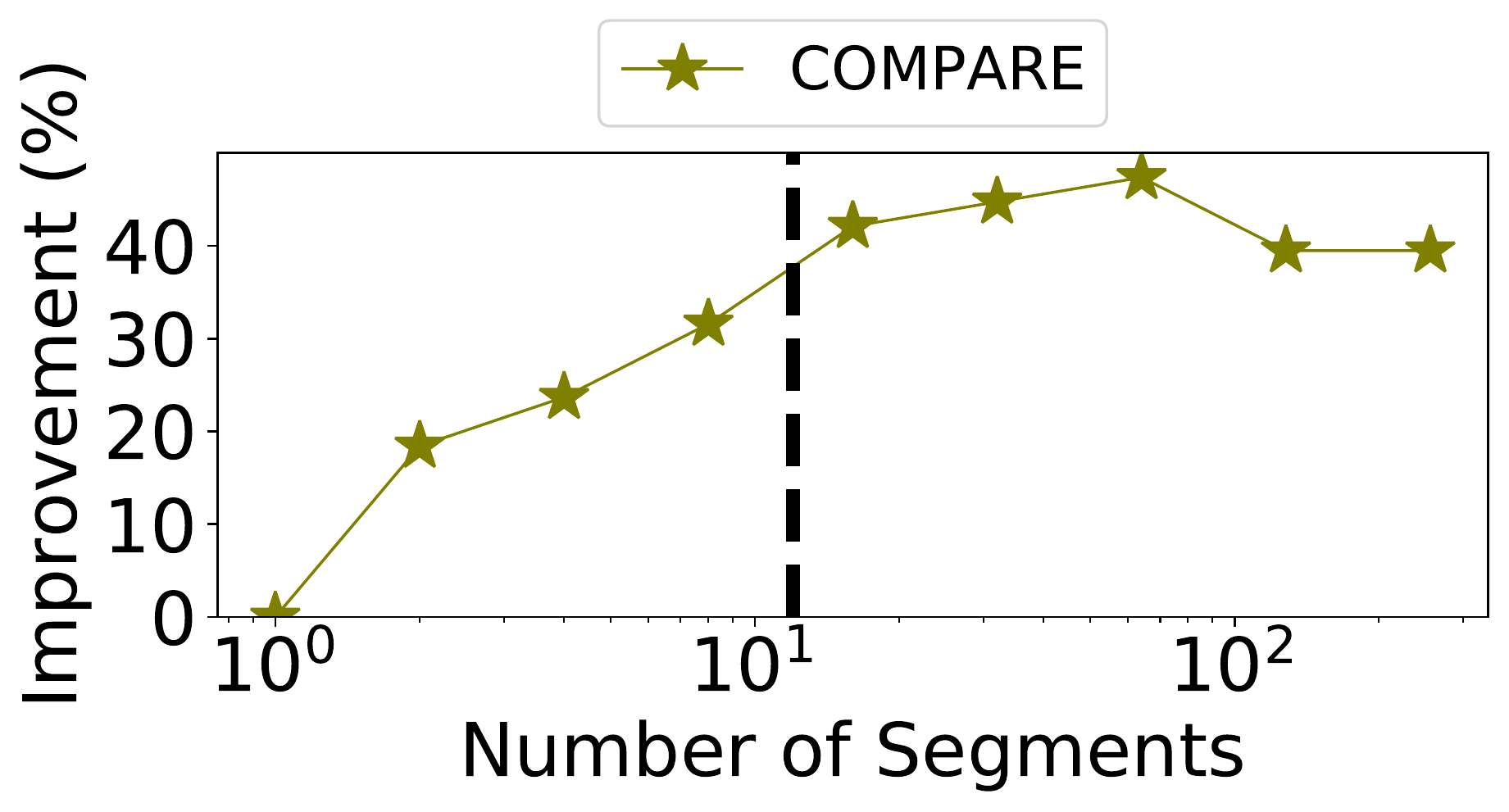}}}}
		\vspace{-1.5pt}
		\caption{\smallcaption{$Q_2$}}
		\label{fig:varysegmentq1}
		\vspace{-2.5pt}
	\end{subfigure}
	\hspace{-0.2cm}
	\begin{subfigure}{0.49\columnwidth}
		\centerline {
			\hbox{\resizebox{\columnwidth}{!}{\includegraphics{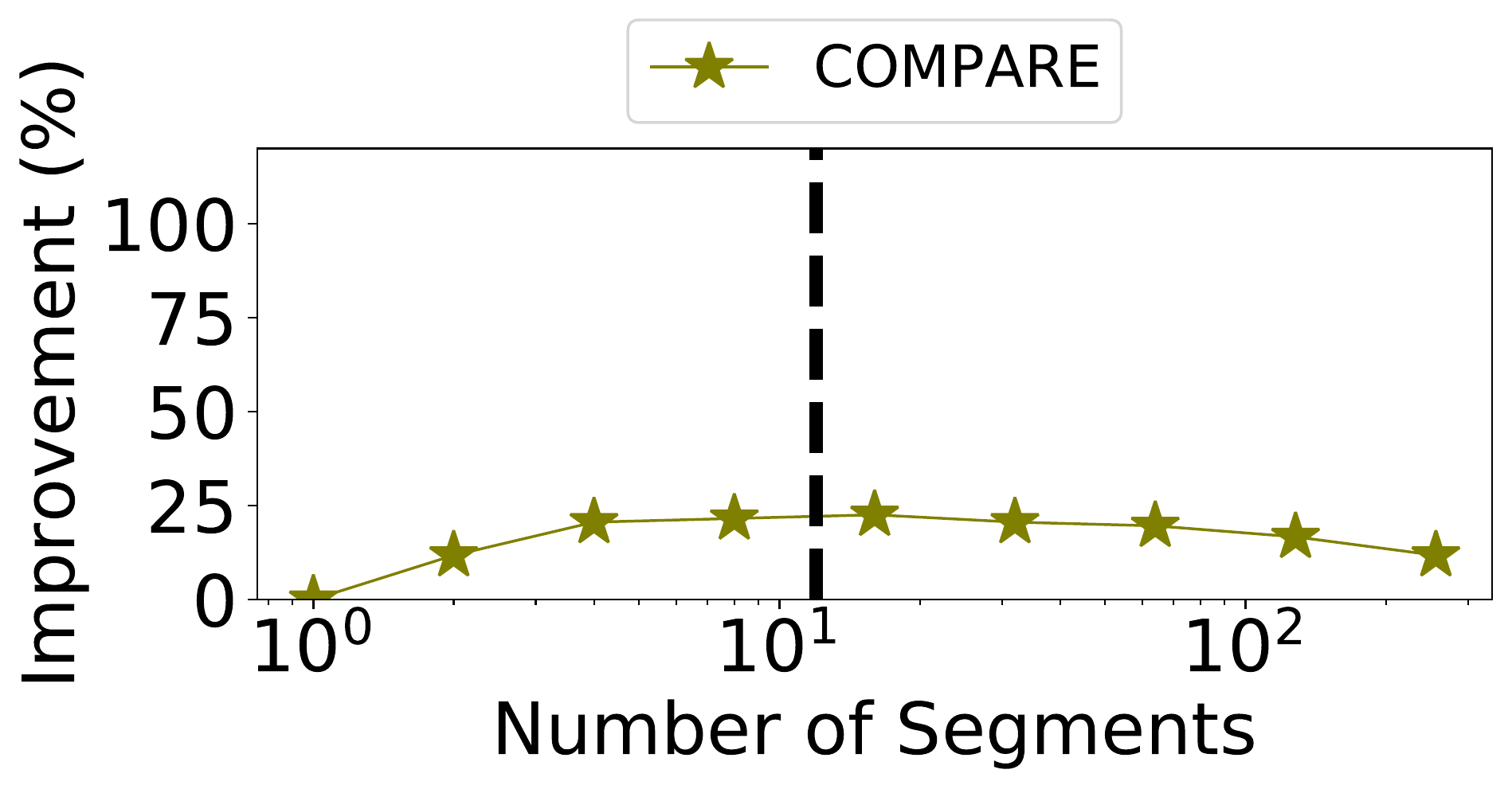}}}}
		\vspace{-1.5pt}
		\caption{\smallcaption{$Q_4$}}
		\label{fig:varysegmentq2}
		\vspace{-2.5pt}
	\end{subfigure}
	\vspace{-5pt}
	\caption{Varying number of segment-aggregates}
	\label{fig:productresults}	
\end{figure}
}

\papertext{
\begin{figure}
	\centering
	\begin{subfigure}{0.40\columnwidth}
		\centerline {
			\hbox{\resizebox{\columnwidth}{!}{\includegraphics{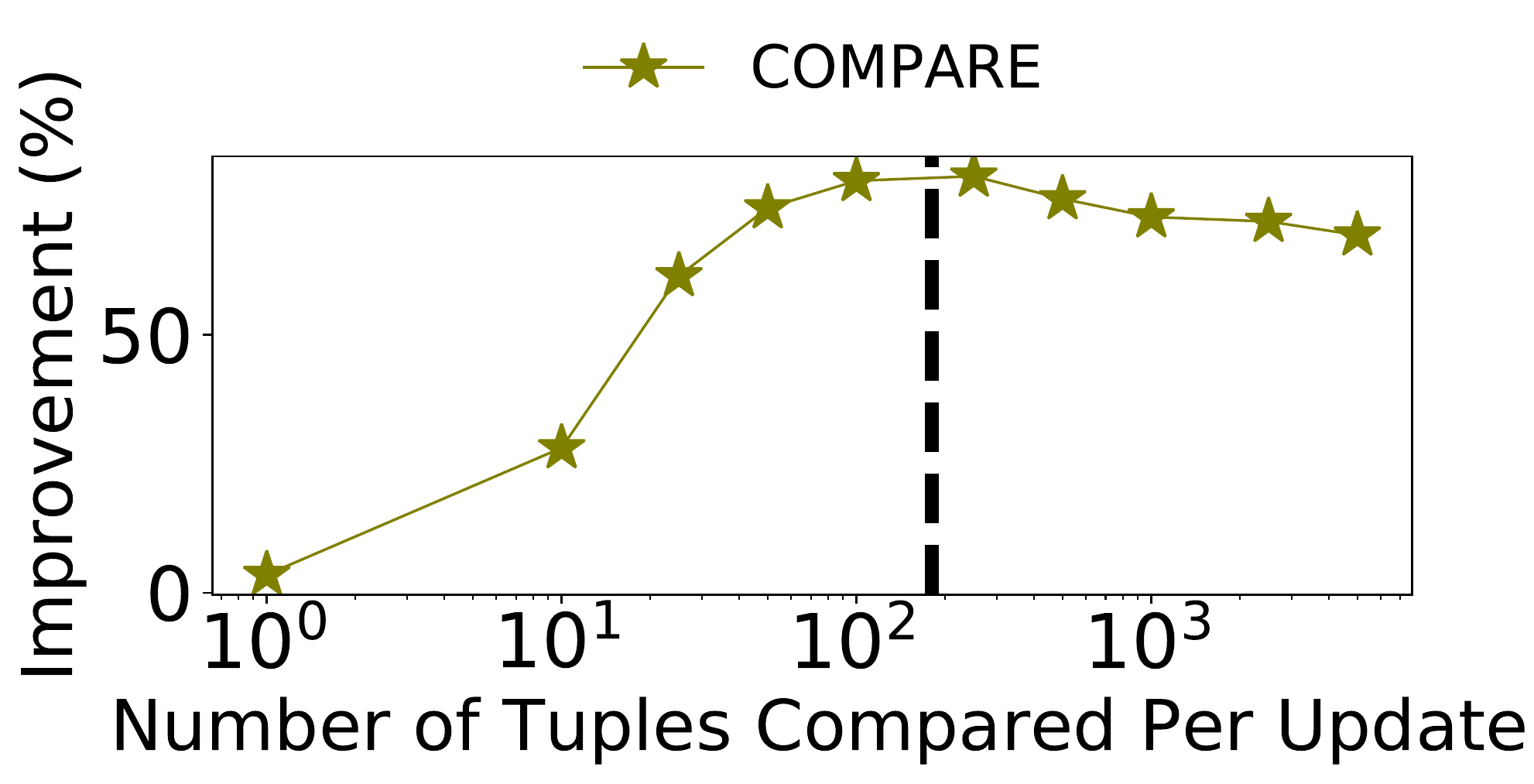}}}}
		\vspace{-1.5pt}
		\caption{\smallcaption{$Q_2$}}
	\end{subfigure}
	\begin{subfigure}{0.40\columnwidth}
		\centerline {
			\hbox{\resizebox{\columnwidth}{!}{\includegraphics{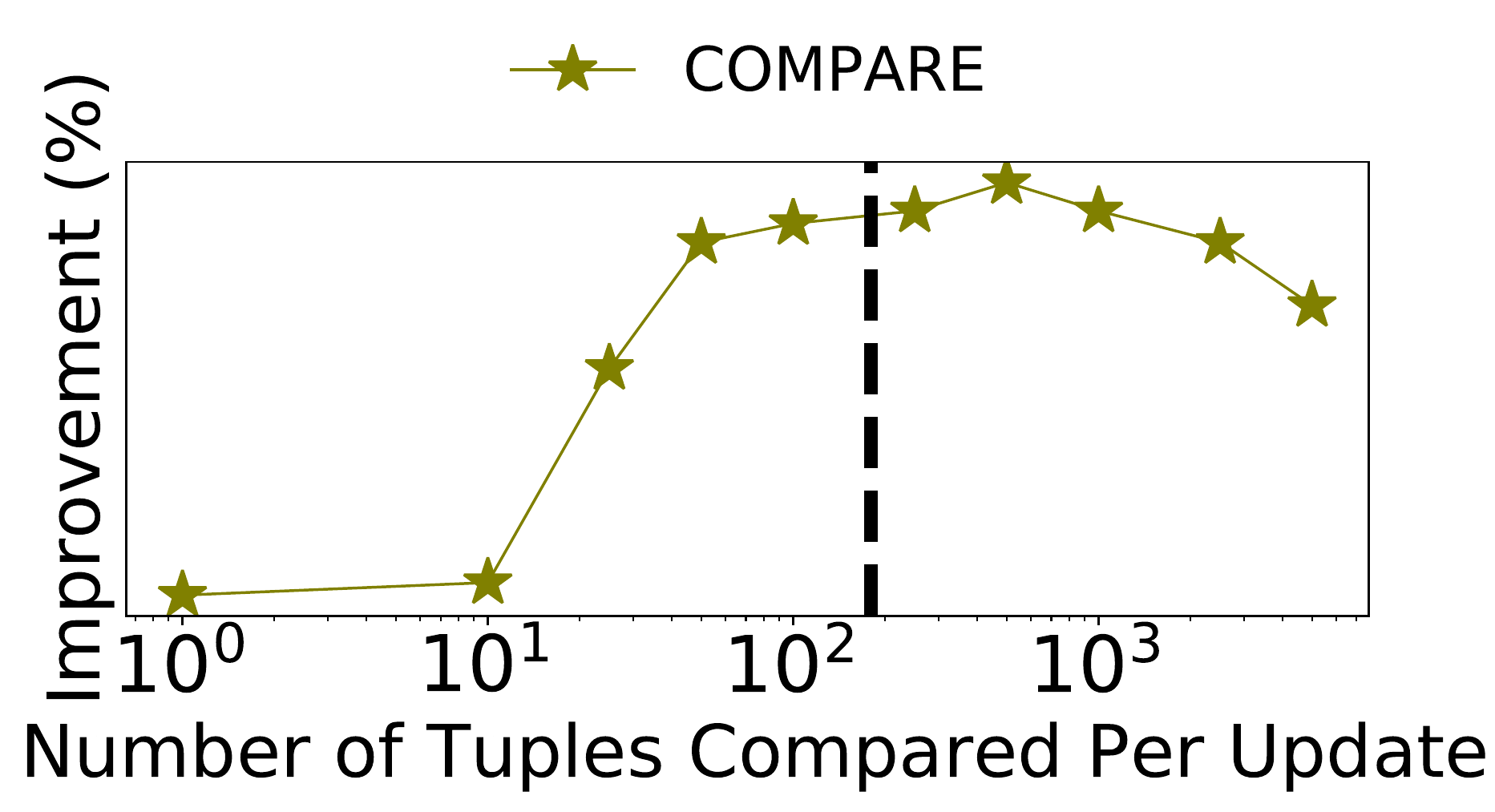}}}}
		\vspace{-1.5pt}
		\caption{\smallcaption{$Q_4$}}
		\vspace{-2.5pt}
	\end{subfigure}
	\vspace{-5pt}
	\caption{Varying number of tuples compared per update during early termination}
	\label{fig:varyingTuplesFetched}
	\vspace{-10pt}	
\end{figure} 
}

\techreport{
\begin{figure}
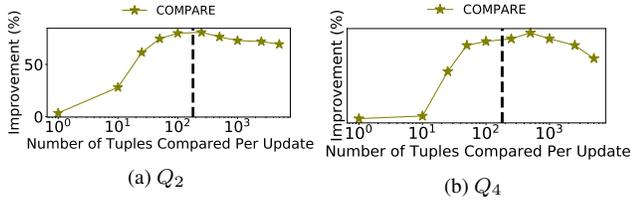

	\vspace{-10pt}
	\centering
	\begin{subfigure}{0.49\columnwidth}
		\centerline {
			\hbox{\resizebox{\columnwidth}{!}{\includegraphics{figs/varytuplefetchedq1.pdf}}}}
		\vspace{-1.5pt}
		\caption{\smallcaption{$Q_2$}}
	\end{subfigure}
	\begin{subfigure}{0.49\columnwidth}
		\centerline {
			\hbox{\resizebox{\columnwidth}{!}{\includegraphics{figs/varytuplefetchedq2.pdf}}}}
		\vspace{-1.5pt}
		\caption{\smallcaption{$Q_4$}}
		\vspace{-2.5pt}
	\end{subfigure}
	\vspace{-5pt}
	\caption{Varying number of tuples compared per update during early termination}
	\label{fig:varyingTuplesFetched}
	\vspace{-20pt}	
\end{figure}
}

\subsection{Impact of Transformation Rules}
Figure~\ref{fig:logopt} depicts the performance results on  pushing $\Phi$ below PK-FK joins ($\Join$) and pushing Aggregate ($\Upsilon$) below $\Phi$.
We omit the results on other logical optimizations such as  predicate pushdown and reordering of multiple $\Phi$ operations as the gains in these cases are always proportional to the selectivity of predicates and $\Phi$ operation pushed down.

\stitle{Pushing $\Phi$ below $\Join$.} We consider $Q_3$ and $Q_4$ over websales table of TPC-DS dataset which has PK-FK joins with two other tables. We observe that by pushing $\Phi$ below join leads to the improvement in the runtime of both queries due to reduction in amount of time taken by join. 
For $Q_3$, $\Phi$ reduces of size of websales to $\frac{1}{30}$th of the original size, which improves the overall latency by about $18\%$. On the other hand, the selectivity of $\Phi$ for $Q_4$ is more ($\frac{1}{200}$th of the original size), which leads to a relatively higher improvement of about (32\%) in latency. Thus, the amount of gain increases with the increase in the  selectivity of $\Phi$.

\stitle{Pushing $\Upsilon$ (aggregation) below $\Phi$.} In order to evaluate this, we use {\amax} as aggregation function for {\measure} and {\scorer} in Q1 and Q2 over the Flight dataset. We added a simple aggregation operation $\Upsilon_{G,A}$ on top of $\Phi$, setting $G$ = \{Days, ArrDelays\} and $A$ = \acount(*). While $\Upsilon$ needs to process more tuples compared to when it is above $\Phi$, the pushdown helps improve the overall latency by reducing duplicate values of $G$, which minimize the number of all pair comparisons for $\Phi$ above. In particular, we observe that pushing $\Upsilon$ down reduces the input to $\Phi$ by about $24$\% leading to an improvement of of about $14$\% for $Q1$ and $19$\% for Q2.

\eat{
\stitle{Pushdown below $\cup$.} For $\cup$, we use $Q_1$ and $Q_2$, splitting the Flight dataset into two smaller datasets, each consisting of equal number of partitions ($\approx$ $191$). Here, the gain is higher, we observe about $30-40$\% improvement in the performance compared to when we do not push $\Phi$ below the $\cup$. \tar{fix}
}

\vspace{5pt}
\subsection{Impact of Indexes}
%Our prototype currently extends the row-store engine of \db.
To evaluate the changes in physical design on \optr, we made the following changes on Flight data set. We removed all columns from the tables that are not part of queries, and created non-clustered indexes on the queried columns. 
%This helps is capture the characteristics of column-store style query processing, as well as understand the effect of indexes. 
Adding indexes results between 20\% to 38\%  improvement in overall runtime across queries; the major changes in physical plan include the use of index scan and the replacement of hash join with merge join. As depicted in Figure 14, due to overall decrease in runtime, the performance improvement for \optr when indexes are used is less than when indexes are not used. However, compared to regular SQL, \optr is still between $2-3\times$ faster. This is primarily because of the reduction in CPU time due to sharing of aggregates, trend-wise processing and pruning of trend comparisons.

\subsection{Parallelism and Memory Overhead}
\noindent
Figure~\ref{fig:varydop} shows the improvement in latency of \system w.r.t. \dbb on $Q_1$ as we vary the Degree of Parallelism (DOP) from $1$ to $64$. Both \db and \optr benefit significantly from increasing DOP up to a point, after which they experience diminishing returns. For any given DOP, COMPARE is usually faster (between $2\times$ to $3$$\times$) similar to what we see in previous experiments. 

\papertext{
\begin{figure}
\vspace{-10pt}
    \centering
	\begin{subfigure}{0.4\columnwidth}
		\centerline {
			\hbox{\resizebox{\columnwidth}{!}{\includegraphics{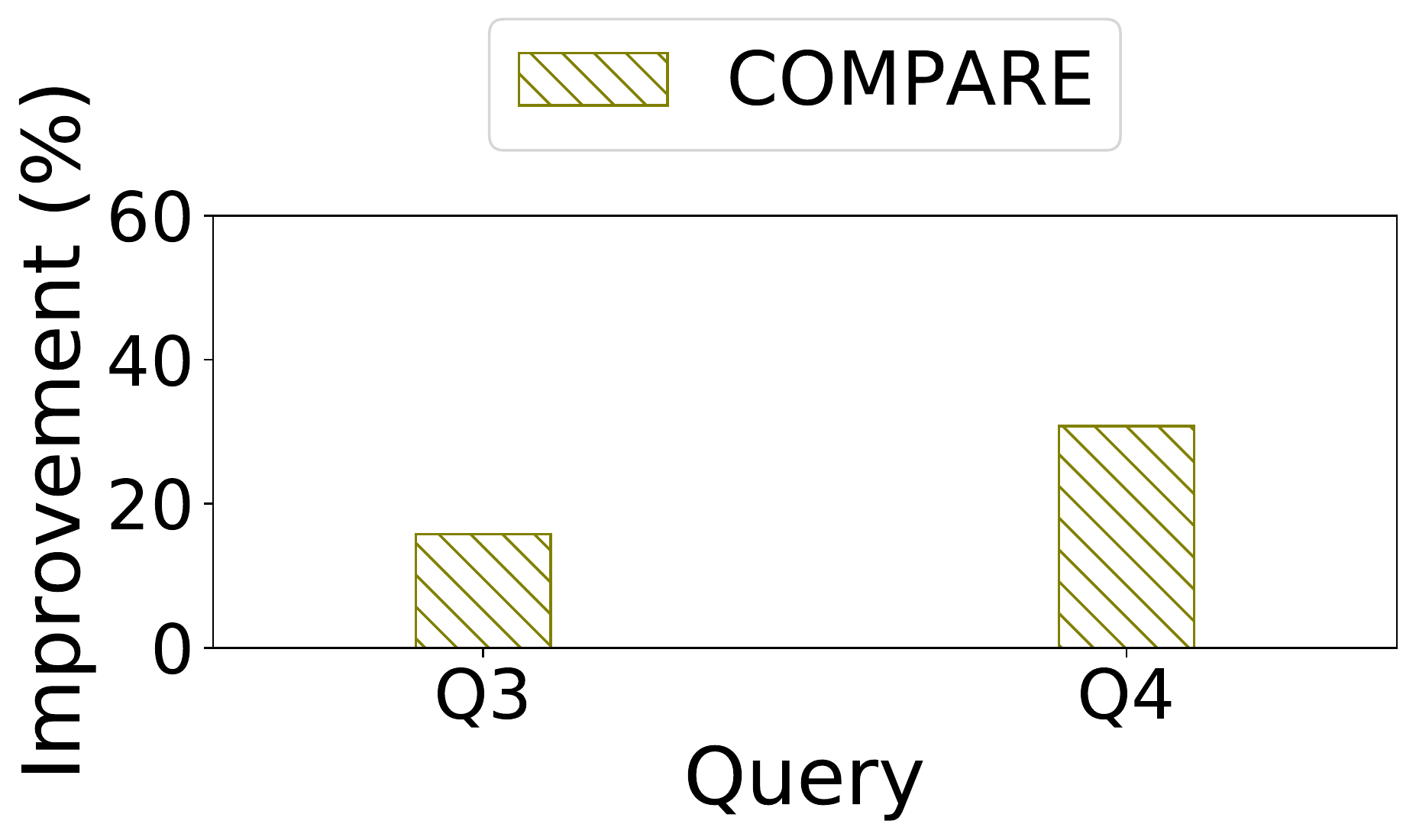}}}}
		\vspace{-1.5pt}
		\caption{\smallcaption{Join pushdown}}
		\label{fig:joinpushdown}
		\vspace{-2.5pt}
	\end{subfigure}
	\begin{subfigure}{0.4\columnwidth}
		\centerline {
		   \hbox{\resizebox{\columnwidth}{!}{\includegraphics{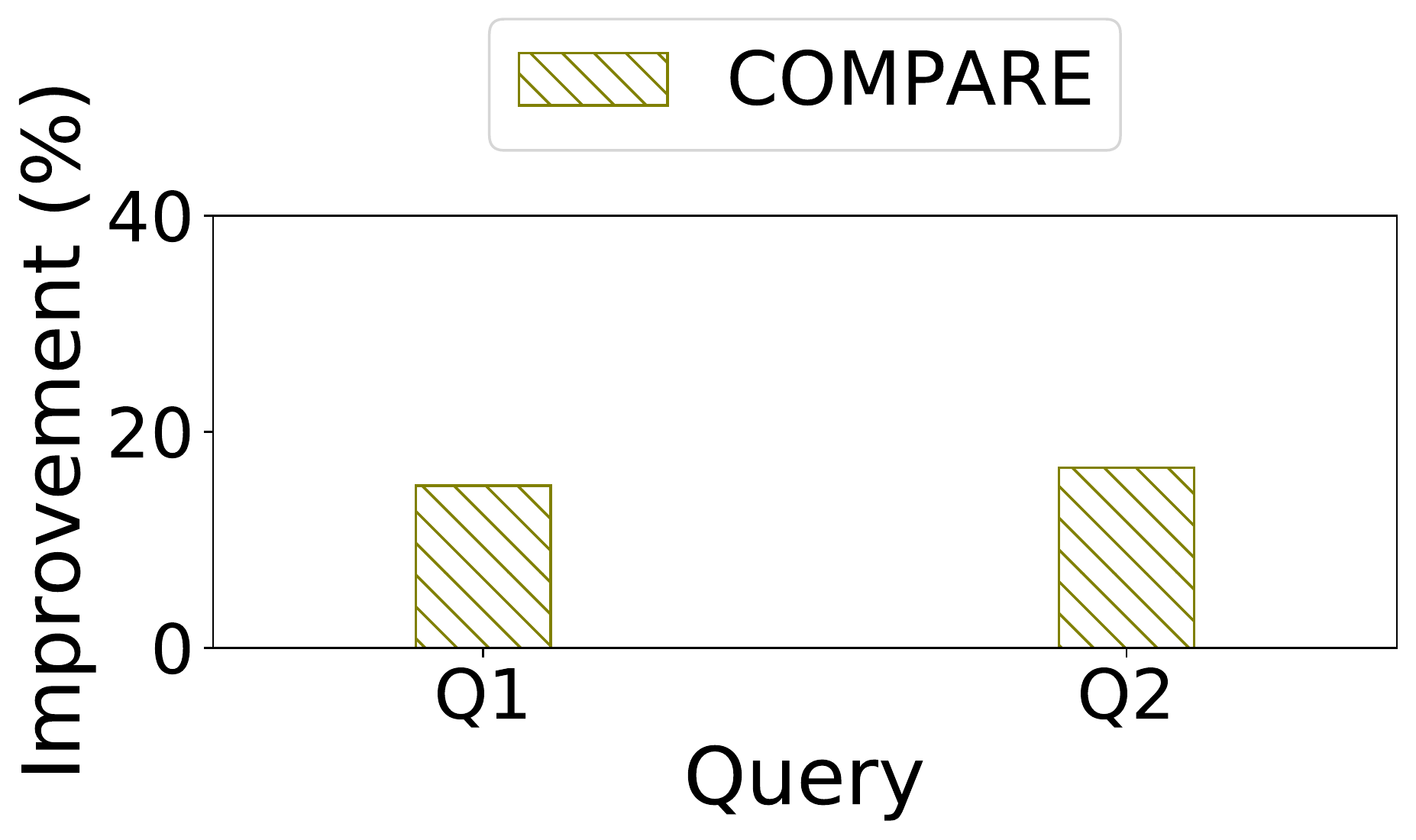}}}}
		\vspace{-1.5pt}
		\caption{\smallcaption{Aggregate pushdown}}
		\label{fig:aggregatepushdown}
		\vspace{-2.5pt}
	\end{subfigure}
	\hspace{-0.2cm}
	\vspace{-5pt}
	\caption{Pushdown logical optimizations}
	\vspace{-5pt}
	\label{fig:logopt}	
\end{figure}
}

\techreport{
\begin{figure}
\vspace{-10pt}
    \centering
	\begin{subfigure}{0.49\columnwidth}
		\centerline {
			\hbox{\resizebox{\columnwidth}{!}{\includegraphics{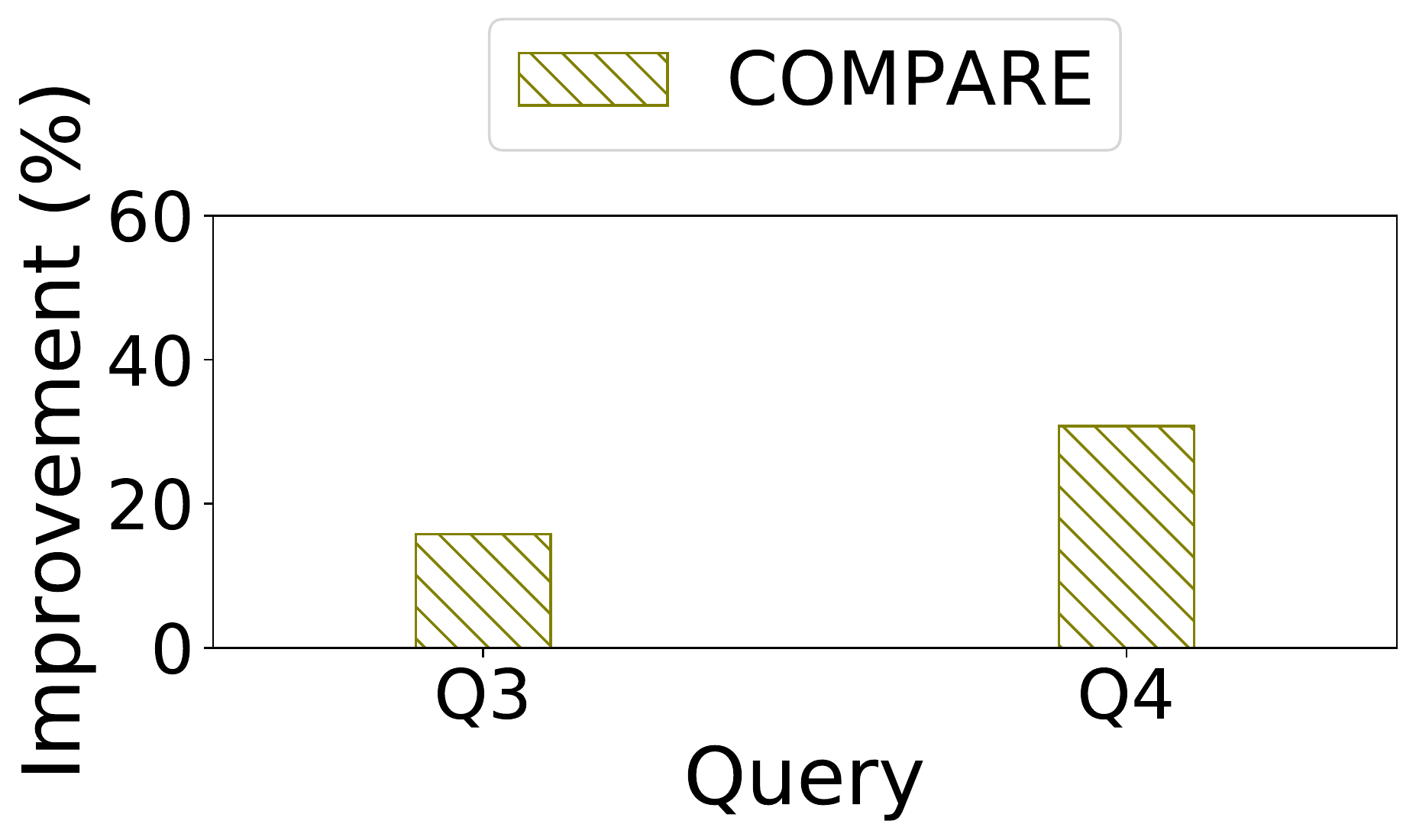}}}}
		\vspace{-1.5pt}
		\caption{\smallcaption{Join pushdown}}
		\label{fig:joinpushdown}
		\vspace{-2.5pt}
	\end{subfigure}
	\begin{subfigure}{0.49\columnwidth}
		\centerline {
		   \hbox{\resizebox{\columnwidth}{!}{\includegraphics{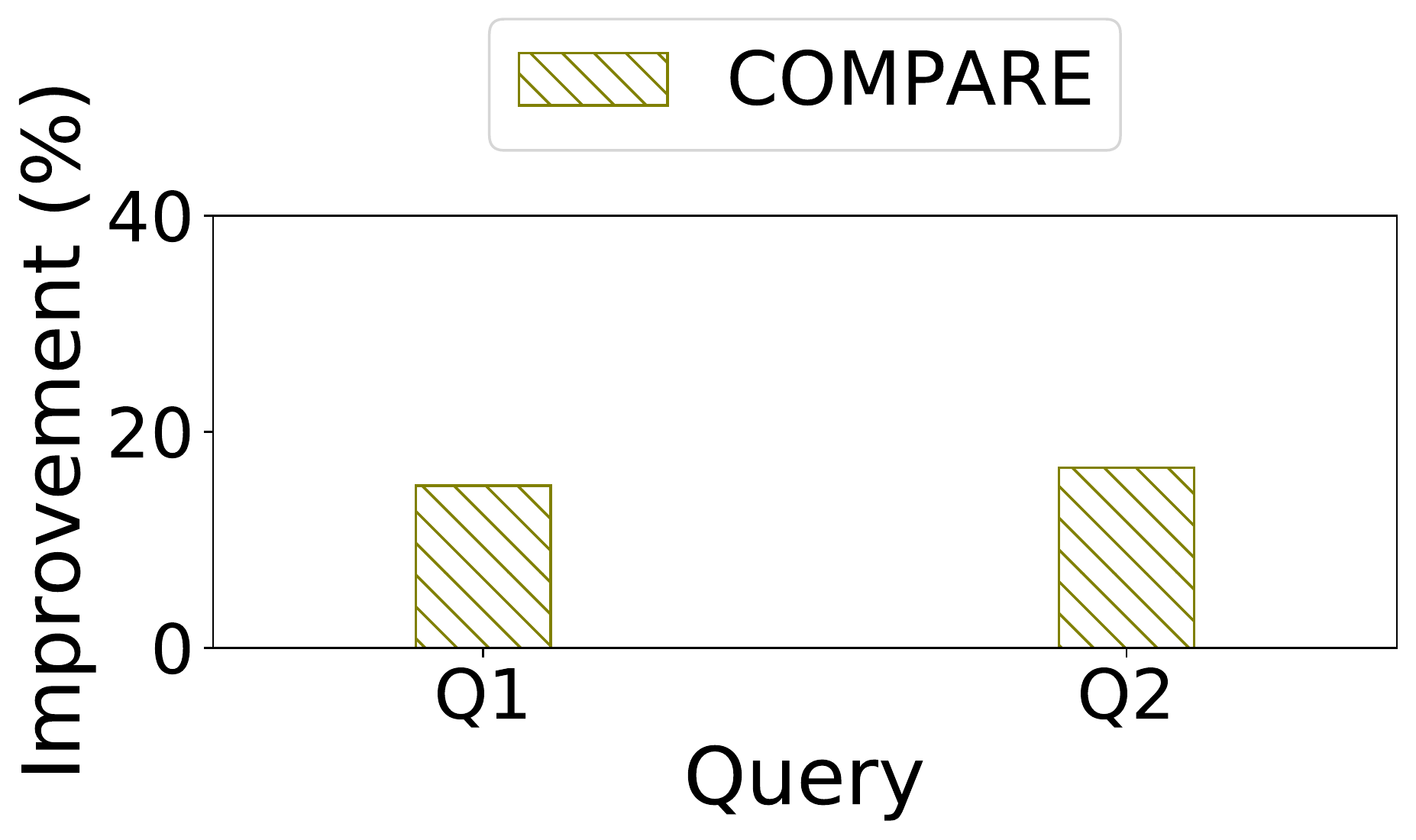}}}}
		\vspace{-1.5pt}
		\caption{\smallcaption{Aggregate pushdown}}
		\label{fig:aggregatepushdown}
		\vspace{-2.5pt}
	\end{subfigure}
	\hspace{-0.2cm}
	\vspace{-5pt}
	\caption{Pushdown logical optimizations}
	\vspace{-5pt}
	\label{fig:logopt}	
\end{figure}
}

\papertext{
\begin{figure}
        \vspace{-5pt}
        \centerline {	    	\hbox{\resizebox{0.50\columnwidth}{!}{\includegraphics{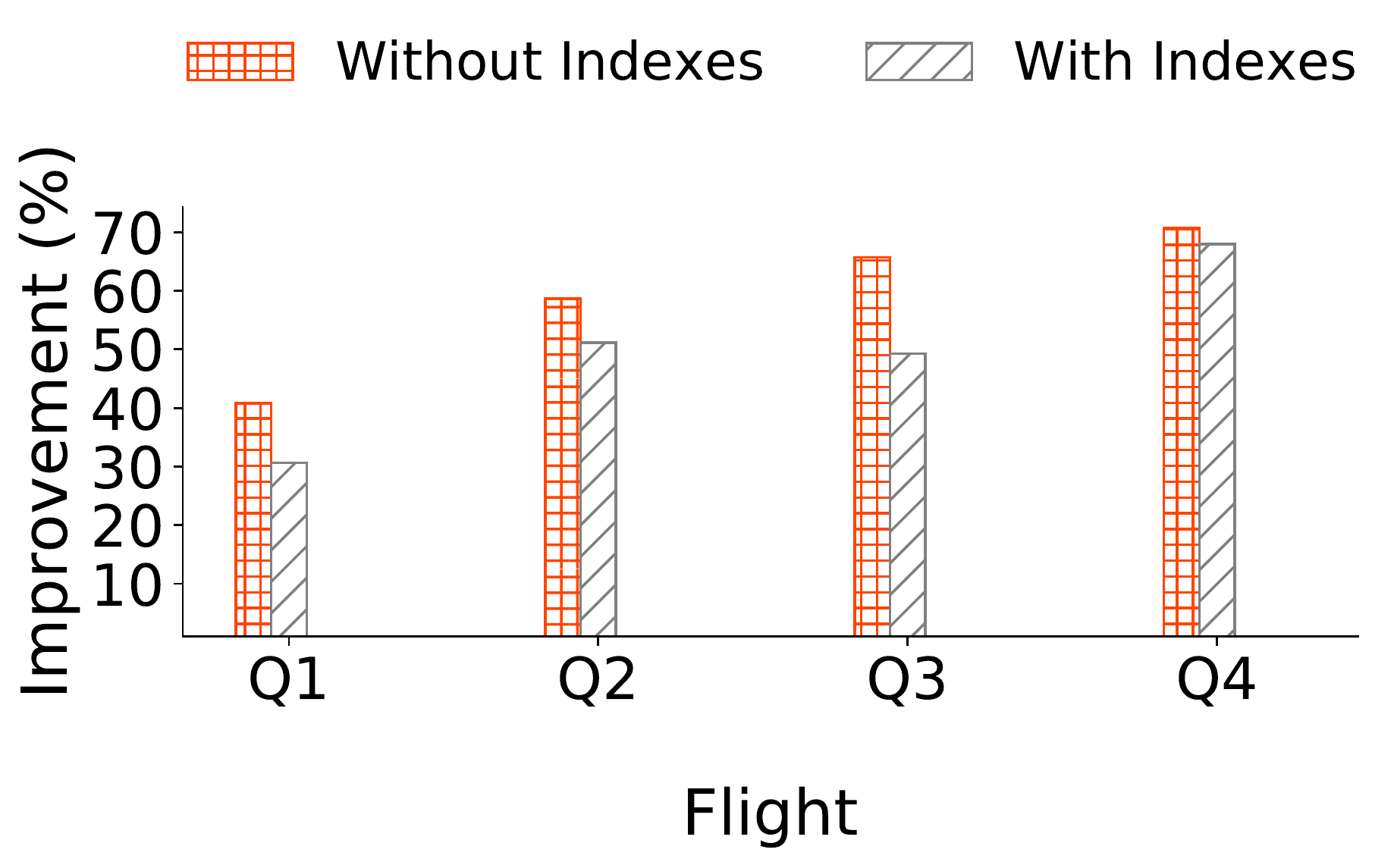}}}}
	    	\vspace{-7.5pt}
	   	\caption{\smallcaption{\rev{Impact of adding non-clustered indexes on referenced columns and removing other columns}}}
	    	\label{fig:phydesign}
	    \vspace{-15pt}
\end{figure}
	 }

\techreport{
\begin{figure}
        \centerline {	    	\hbox{\resizebox{0.75\columnwidth}{!}{\includegraphics{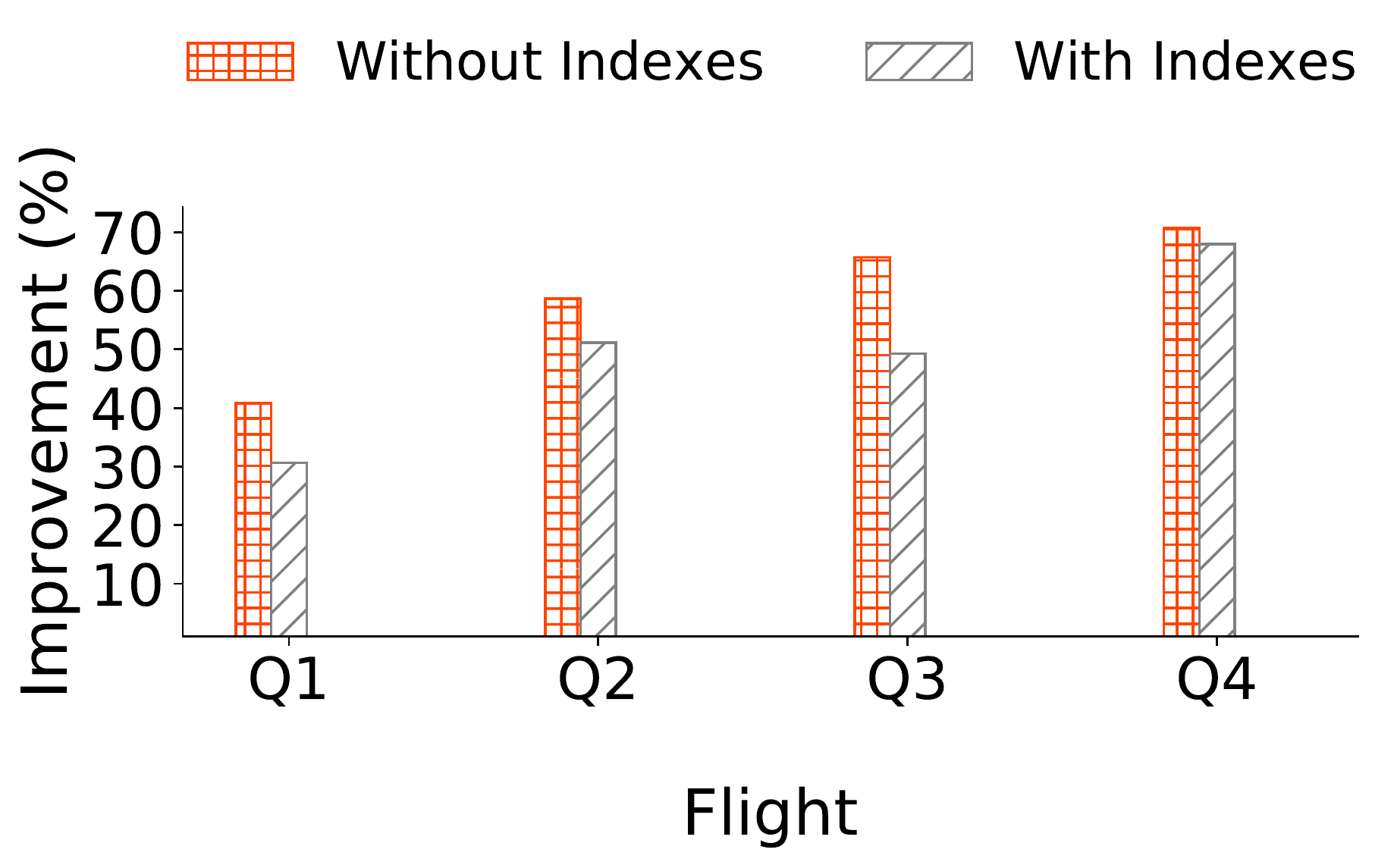}}}}
	    	\vspace{-7.5pt}
	   	\caption{\smallcaption{\rev{Impact of adding non-clustered indexes on referenced columns and removing other columns}}}
	    	\label{fig:phydesign}
	    \papertext{\vspace{-15pt}}
\end{figure}
}

\papertext{
\begin{figure}
    \centering
    \vspace{-4pt}
	\begin{subfigure}{0.4\columnwidth}
		\centerline {
		\hbox{\resizebox{\columnwidth}{!}{\includegraphics{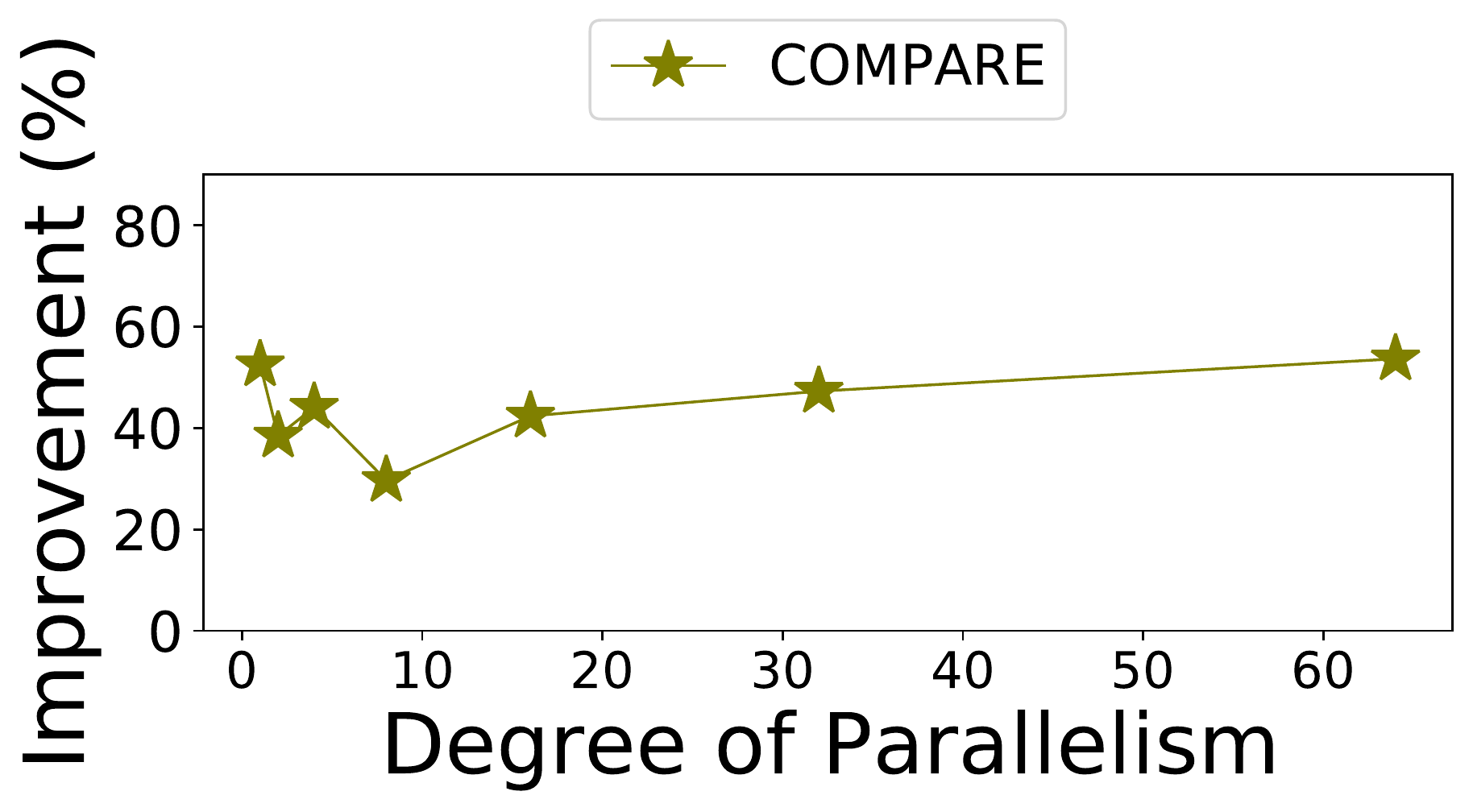}}}}
		\vspace{-5.5pt}
		\caption{\smallcaption{Varying DOP \tar{Add UDF}}}
		\label{fig:varydop}
		\vspace{-2.5pt}
	\end{subfigure}
	\begin{subfigure}{0.4\columnwidth}
		\centerline {
		\hbox{\resizebox{\columnwidth}{!}{\includegraphics{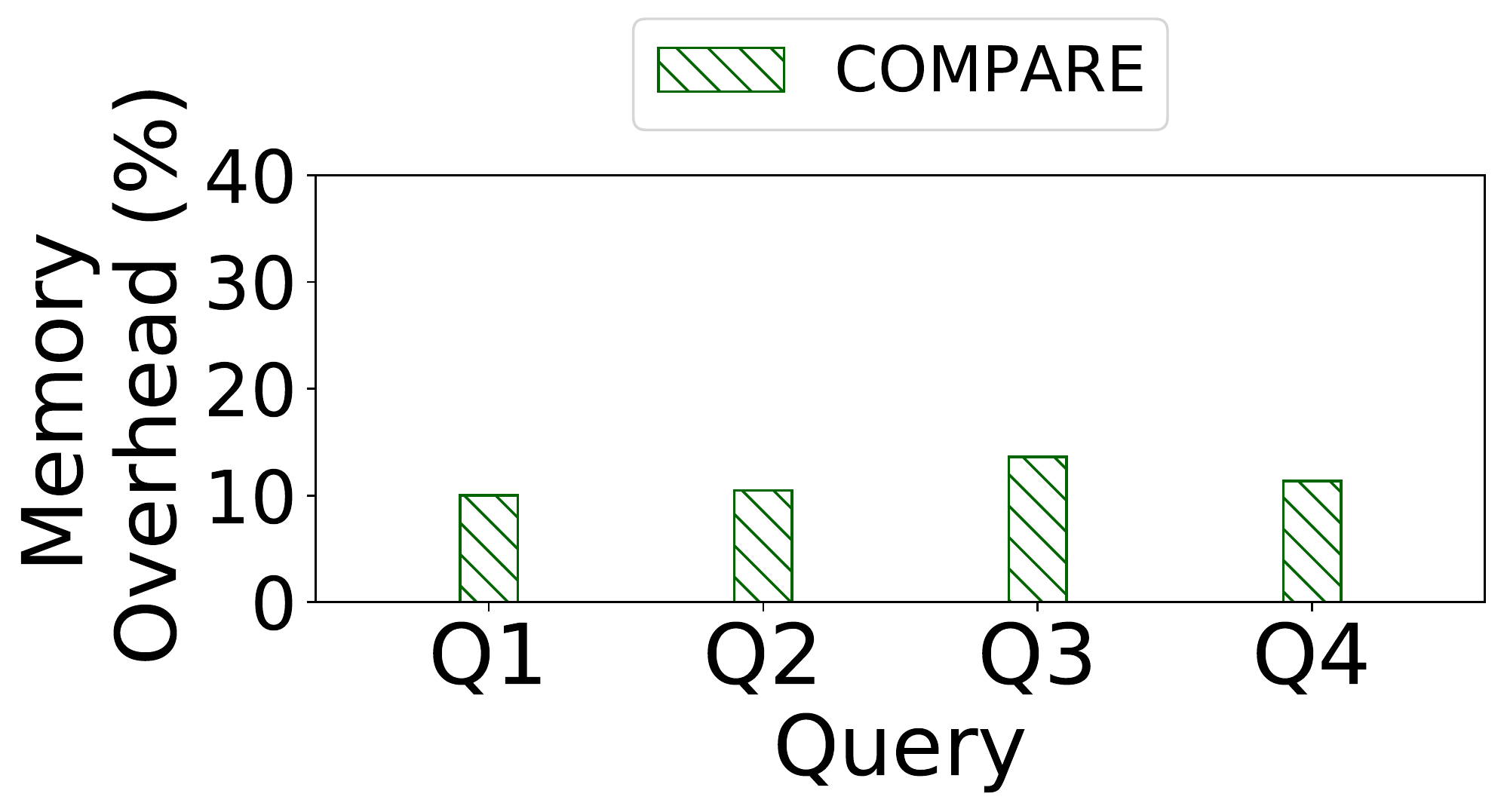}}}}
		\vspace{-5.5pt}
		\caption{\smallcaption{Memory consumption}}
		\label{fig:memoryconsump}
		\vspace{-2.5pt}
	\end{subfigure}
	\vspace{-5pt}
	\caption{Impact of Parallelism and Memory Overhead}
	\vspace{-10pt}
	\label{fig:productresults}	
\end{figure}
}

\techreport{
\begin{figure}
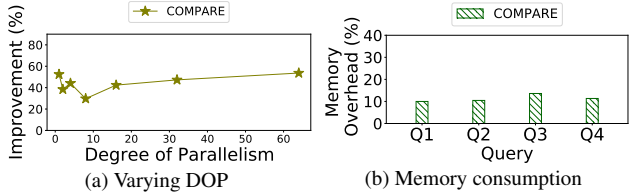

    \centering
    \vspace{-4pt}
	\begin{subfigure}{0.49\columnwidth}
		\centerline {
		\hbox{\resizebox{\columnwidth}{!}{\includegraphics{figs/varydop.pdf}}}}
		\vspace{-5.5pt}
		\caption{\smallcaption{Varying DOP \tar{Add UDF}}}
		\label{fig:varydop}
		\vspace{-2.5pt}
	\end{subfigure}
	\begin{subfigure}{0.49\columnwidth}
		\centerline {
		\hbox{\resizebox{\columnwidth}{!}{\includegraphics{figs/memoryconsump.pdf}}}}
		\vspace{-5.5pt}
		\caption{\smallcaption{Memory consumption}}
		\label{fig:memoryconsump}
		\vspace{-2.5pt}
	\end{subfigure}
	\vspace{-5pt}
	\caption{Impact of Parallelism and Memory Overhead}
	\vspace{-10pt}
	\label{fig:productresults}	
\end{figure}
}

%\vspace{-2pt}
%\subsection{Memory Overhead}
Figure~\ref{fig:memoryconsump} shows the additional overhead in committed memory usage of \system w.r.t. to \dbb for each of the queries. Although \system uses additional data-structures for maintaining segment-aggregates, and bounds in the priority queue, the overhead is minimal ($<$ 13\%) compared to the memory already used by the system for sorting and maintaining aggregates which are common to all approaches. Moreover, the execution engine reuses the memory already committed by the downstream operators in the plan, instead of allocating new memory. Thus, the total memory used during query processing is bounded by the maximum memory used by any operator in the plan.

\eat{
\begin{figure}
	\centerline {
		\hbox{\resizebox{0.75\columnwidth}{!}{\includegraphics{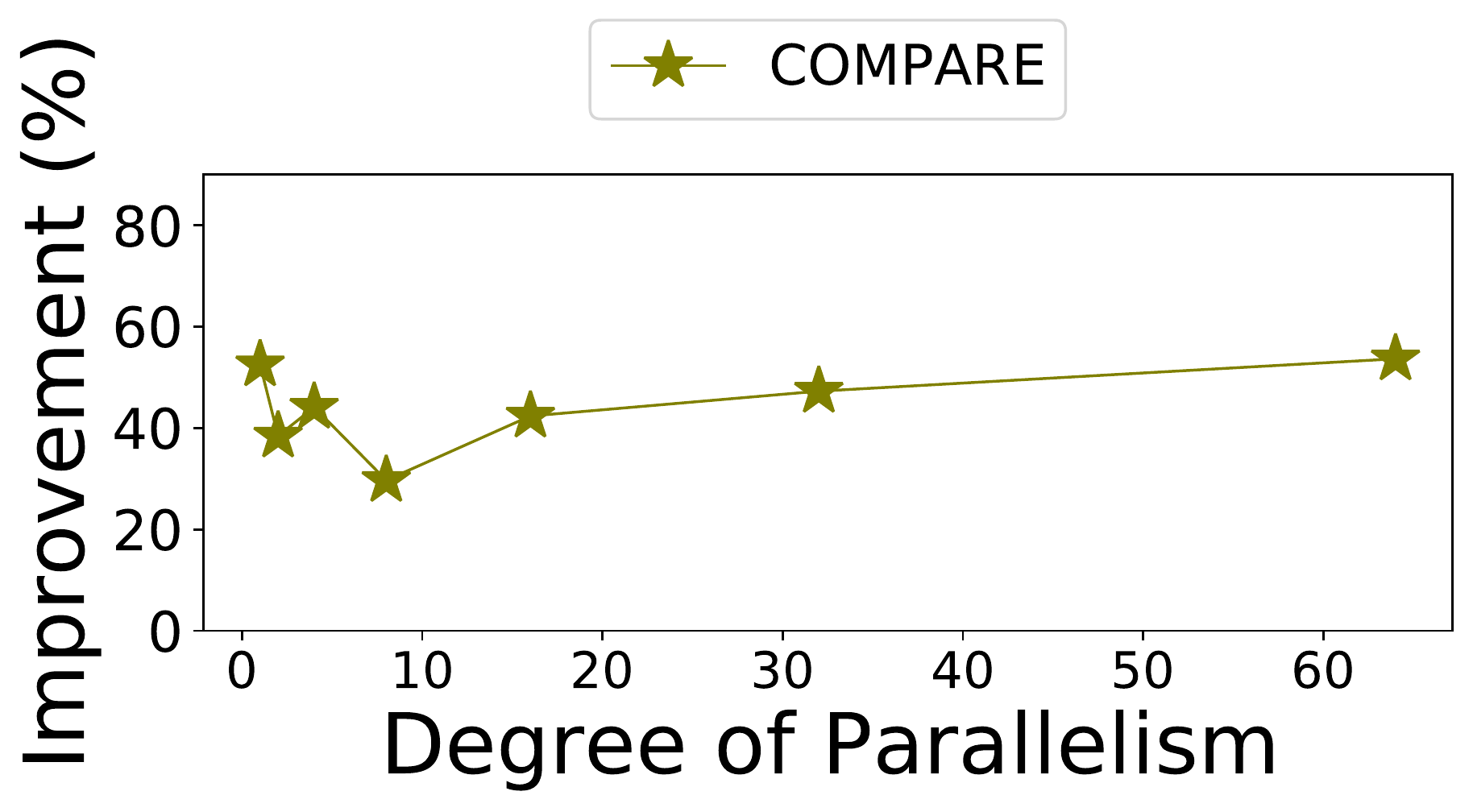}}}}
	\vspace{-10.5pt}
	\caption{\smallcaption{Impact of degree of parallelism(DOP) on latency}}
	\label{fig:varydop}
	\vspace{-2.5pt}
\end{figure}

\begin{figure}
	\centerline {
		\hbox{\resizebox{0.75\columnwidth}{!}{\includegraphics{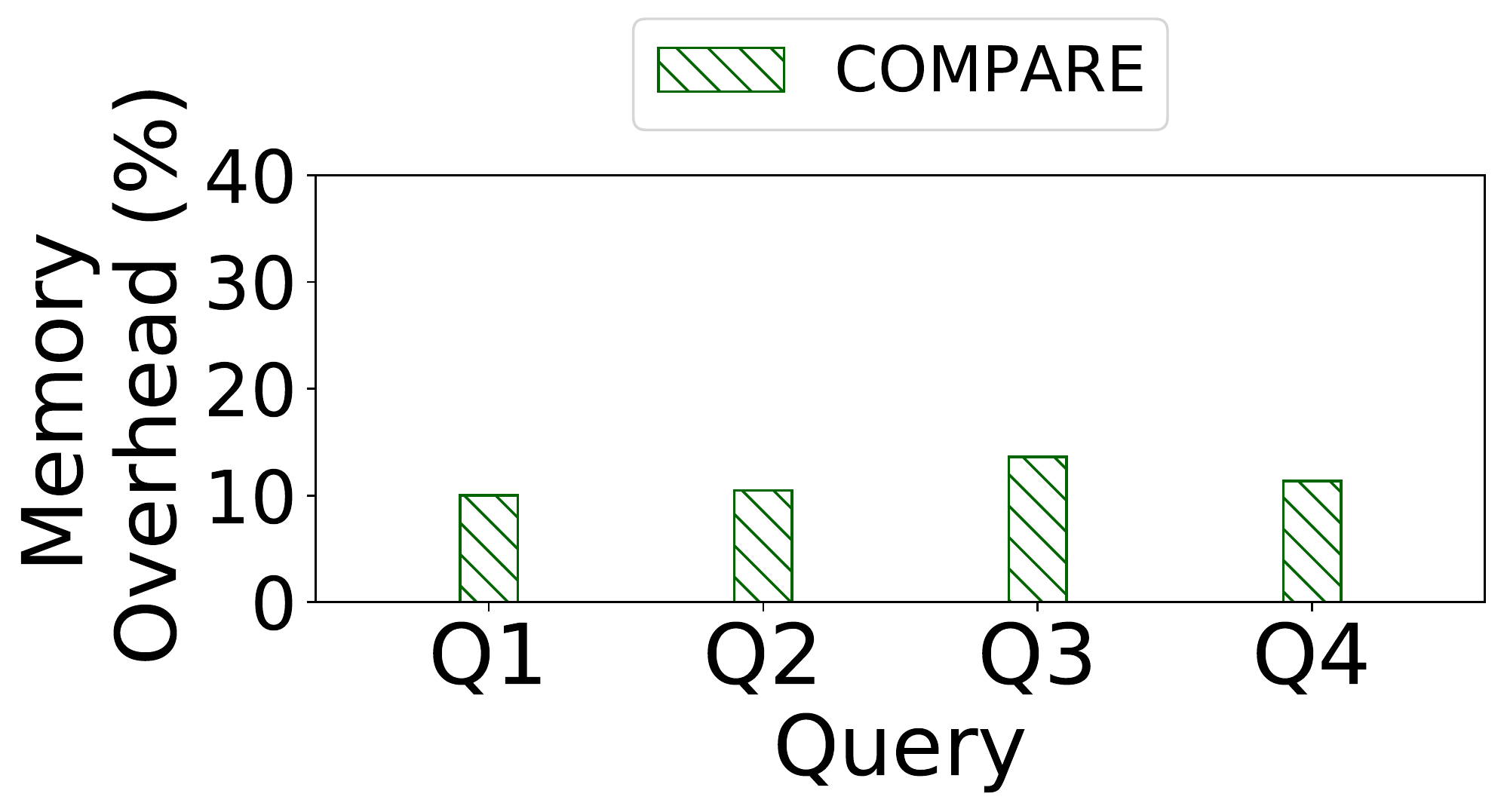}}}}
	\vspace{-10.5pt}
	\caption{\smallcaption{Memory consumption for difference approaches}}
	\label{fig:memoryconsump}
	\vspace{-15pt}
\end{figure}
}

%\vspace{-5pt}

\vspace{10pt}
\section{Related Work}

%multi-level aggregation~\cite{gray1997data} and semantic group by operations~\cite{tang2015similarity},

\stitle{Visual Analytics.} Our work has been motivated by many recent visual analytic tools~\cite{ding2019quickinsights,seedb,zenvisagevldb,wongsuphasawat2017voyager,macke2018adaptive} where comparing subsets or groups of tuples using a deviation-based measures (e.g., $L_p$ norms) is the common theme. Unfortunately, as discussed in Section~\ref{sec:intro} these tools either retrieve the data into a middleware or issue complex SQL queries for comparison, both approaches do not scale to large datasets. As a result, recent work~\cite{tang2019towards,wu2014case,d2018aida} have called for supporting new abstractions and query optimization techniques for addressing the impedance mismatch between relational databases and analytic tasks---our work is a concrete step in this direction.

%A number of SQL extensions have been proposed by prior work that provide more fine-grained control to users for easily expressing complex decision support queries. For instance,

\stitle{OLAP.}  Damianos et al. have proposed grouping variables and operations such as MD-Join~\cite{chatziantoniou2007using, chatziantoniou1996querying} for succinctly expressing complex aggregate queries such as finding products with sales \emph{above average sales}. Similarly, CUBE~\cite{gray1997data}, GROUPING SETs~\cite{zaharioudakis2000answering}, Semantic Group By~\cite{tang2015similarity} allow flexible specification and optimization of group by queries. In our work, we extend grouping of tuples to support (i) easier and more direct specification of \emph{comparison} between groups of tuples using complex aggregate expressions (e.g., $L_p$ norms), and (ii) jointly optimize both aggregation and comparison between groups of tuples. Sarawagi et al. have proposed techniques for interactive browsing of interesting cells in data cube~\cite{sarawagi1999explaining,sarawagi2000i3}. Similarly, %Nandi et al. have proposed MR-Cube~\cite{nandi2011data} for automatically mining interesting cells in a data-cube using Map-Reduce (MR) framework.
These work suggest raw aggregates that are informative given past browsing, or  those that show a generalization or explanation of a specific cell. In contrast, we provide  extensions to traditional query optimization and execution layers of relational databases to support comparative queries like other SQL queries. \rev{Similar to our approach, there have been database extensions \cite{rao1996providing,han1996dmql,imielinski1999msql, netz2001integrating}, the most recent being the DIFF operator~\cite{abuzaid2018diff},  that support association and frequent pattern mining. While our focus is on aggregate distance measures such as $L_p$ norms (our focus), we share their goal that with an extended syntax, complex analytic queries are easier to write and optimize.}

\stitle{Similarity Join.} There  has been work  on similarity join that use set similarity functions such as edit distance, Jaccard similarity, cosine similarity or their variants to join two relations~\cite{sarawagi2004efficient,gravano2001approximate,ramasamy2000set,bohm2001cost,chaudhuri2006primitive,arasu2006efficient}.  While these work are based on \emph{measuring set overlap or edit distance between strings}, \optr optimizes \emph{aggregate distance functions between groups of tuples} such as Euclidean distance, requiring fundamentally different execution techniques. Similarly, there is a vast
body of work on top-k query processing~\cite{ilyas2008survey}, including ones that extend relational databases ~\cite{chaudhuri1999evaluating, tsaparas2003ranked, li2005ranksql,ilyas2004supporting}. While these work rank each tuple independently based on an aggregate expression, our focus is on ranking \emph{groups of tuples} by \emph{comparing} them with other groups of tuples in the same relation.

\stitle{Spatial Databases.} Finally, spatial databases such as PostGIS~\cite{zhang2010management} extend traditional databases to optimize for storage and querying of spatial data. The similarity search queries supported is spatial databases (e.g., ~\cite{papadias2005aggregate}) operate in a different settings from ours. First, the physical design is typically optimized to store all information (e.g., sales) for each entity (e.g., product) required for distance computation as a single object, thus no grouping or sorting of tuples is typically required at runtime. In addition, spatial indexes such as R-Tree are built to optimize for search at runtime. In contrast, our work is meant for supporting ad hoc similarity search queries over traditional databases, which are typically used as back-end for BI tools such as Power BI and Tableau.

\section{Conclusion}
\fixlater{fix these}

In this work, we introduce \optr, a complex operator that concisely captures comparison between groups of tuples using aggregated distance measures.
We  introduce physical optimizations within the execution engine and extend the query optimizer with new algebraic rules that improve the performance by significantly reducing the number of subset comparisons and intermediate data size. Together, these logical and physical optimizations help address the impedance mismatch problem between data exploration systems and relational databases for supporting comparative queries. There are several avenues for future work such as supporting primitives for easily expressing  comparison metrics such as   Jaccard similarity, cosine similarity,  as well as using sampling-based techniques to tighten the bounds on scores for further reducing the number of comparisons. 

\section*{Acknowledgements}
We would like to thank the anonymous reviewers \techreport{at VLDB 2021}, Arnd
Christian König, Wentao Wu, and Bailu Ding for their valuable feedback.

\tar{
\begin{enumerate}
\item Make captions of figures meaningful throughout.
\item Overhead of bitmap
\item Add UDF to DOP
\item clean up the exec diagrams
\item {\group}ing for scoring vs {\group}ing vs entitywise join
\item color \{grouping, measure\}
\item fix x labels in figures
\end{enumerate}
}

\techreport{

\section*{APPENDIX}
\fixlater{fix these}

\subsection*{A. Proof of Theorem 1}
Here, we provide the proof for Theorem 1 stated in Section~\ref{sec:segaggregates}.

The proof directly derives from the property of convex function. For a  convex $f(x)$,

 $k_1$$f(x_1) + k_2$$f(x_2) + ... + k_n$$f(x_n)$ $\geq$ $f(k_1x_1 + x_2,x_2,...,k_nx_n)$

Let each $x_i$ be the value a $|m_1 - m_2 |$ resulting from comparing a pair of tuples between two {\groups}, and $n$ be the total number of tuple comparisons. On setting, $k_i = 1/n$ and $f(x) = |x|^p$:
 
 $\frac{|m_1 - m_2 |^p}{n} \geq  |\frac{m_1 }{n} - \frac{m_2 }{n}|^p$ 
 
 $\Rightarrow$ \aavg(DIFF$(m_1, m_2 , p))  \geq$ DIFF(\aavg$(m_1)$, \aavg$ (m_2 ), p)$   
 (by def. of DIFF)\\
 $\square$

\eat{
Given $x_1,x_2,...,x_n$ $\in \mathbb{R}$, $k_1,k_2,...,k_n$ $\in \mathbb{R}$ and $\sum_k k_i=1$, we know that if $\Delta(.)$ is a convex function, then
$\Delta(k_1x_1 + x_2,x_2,...,k_nx_n) <= k_1\Delta(x_1) + k_2\Delta(x_2) + ... + k_n\Delta(x_n)$. On setting, each $k_i = 1/n$, we see that $n*\Delta(\overline{x}) <= (\Delta(x_1) + \Delta(x_2) + ... + \Delta(x_n))$. Taking $x_i = (a_i-b_i)$, $k_i=2/n^2$ it is easy to see that $\overline{x} = (\overline{a}-\overline{b})$. Note that $(a-b)$ involves all pairs (i.e., $(n^2/2)$) differences between values of $a$ and $b$, hence we set $k_i=2/n^2$. $\square$ \\

Furthermore, all DIFF functions are convex functions with the following property: DIFF$(k_1x_1 + x_2,x_2,...,k_nx_n) <=$ $k_1$DIFF$(x_1) + k_2$DIFF$(x_2) + ... + k_n$DIFF$(x_n)$. If we set $n$ to be the number of tuple comparisons with each 
each $x_i$ representing a $|$R.sales - Sales$|$ for a single tuple comparison, and $k_i = 1/n$, we get  the following useful result:
}

\subsection*{B. Formal Description of Bounds Computation}
Here, we formally describe how we compute the bounds on scores of \optr using segment aggregates (Section~\ref{sec:segaggregates}).

Let $p_1$ and $p_2$ be two {\groups} having same number of tuples $c$ for which we want to to compute the upper and lower bounds on score. Let $max_{1i}$ and $min_{1i}$ be the maximum and minimum values of attribute $m_1$ in segment $i$ of {\group}  $p_1$, and similarly $max_{2i}$ and $min_{2i}$ be the maximum and minimum values of $m_2 $ in segment $j$ in $p_2$. Let $c_i$ be the number of tuples in segment $i$.
For succinctness, we use $\Delta(m_1,m_2 )$ for DIFF($m_1, m_2 , p$).

We know that the bounds on the $\Delta_{i}(m_1,m_2 )$ between segment $i$ in $p_1$ and $p_2$,  is given by: \\

{
	\small 
	
	\noindent
	\amax$(\Delta_{i}(m_1, m_2 )) \leq \Delta_{i}($\amax$ (|max_{1i} - min_{2i}|, |min_{1i} - max_{2i} $|)) \\
	
	\noindent
	\amin$(\Delta_{i}(m_1, m_2 )) \geq \Delta_{i}(($\aavg$(m_1)$,\aavg$ (m_2 ))$ (From Theorem 1) \\
	
}

From above we get, \\

{
	\small 
	
	\noindent
	$\Delta_{i}$((\aavg$(m_1)$,\aavg$(m_2 ))  \leq$ \aavg$(\Delta_{i}(m_1, m_2 )) \leq$ \amax$(\Delta_{i}(m_1, m_2 ))$ \\

}

Using the non-negativity and Monotonicity property of DIFF, we can replace the value for each tuple comparison with minimum and maximum bounds to get the bounds on \asum. \\

{
	\small 
	\noindent
	$c_{i}.$\aavg$(\Delta_{i}(m_1, m_2 )) \leq$ \asum$(\Delta_{i}(m_1, m_2 )) \leq c_{i}$ \amax$(\Delta_{i}(m_1, m_2 ))$ \\
	
}

The above bounds over a single pair of segments can be  extended to segments using the union bound principle. Let $sum_{i}^u$, $max_{i}^u$, $min_{i}^u$ be the upper bounds, and $sum_{i}^l$, $max_{i}^l$, $min_{i}^l$ be the lower bounds on the score of {\asum}($\Delta(.)$), {\amax}($\Delta(.)$), and {\amin}($\Delta(.)$) on scoring segment $i$ in $p_1$ and $p_2$. Then, the bounds across all segments can be computed as follows:

\fixlater{fix the indentation}

\vspace{5pt}
{
	\small
	\noindent
	$\underset{i}{\text{\aavg}}(sum_{i}^l)$ $\leq$ 
	\aavg($\Delta(.)$)  $\leq$ $\underset{i}{\aavg}(sum_{i}^u)$
	
	\vspace{3pt}
	\noindent		
	$c.\underset{i}{\text{\asum}}(\frac{sum_{i}}{c_i}^l$) 	$\leq$ 
	\asum$\Delta(.)$)   	$\leq c.\underset{i}{\text{\asum}}(\frac{sum_{i}}{c_i}^u)$ 
	
	\vspace{3pt}
	\noindent
	\amin($\Delta(.)$)  $=$ $\underset{i}{\text{\amin}}(min_{i}^l)$
	
	\vspace{3pt}
	\noindent		
	\amax($\Delta(.)$) $=$ $\underset{i}{\text{\amax}}(max_{i}^u)$
}

%Similarly, for all-pairs comparisons, we can derive the overall bounds between {\groups} by combining the bounds across all pairs of segments. 

\vspace{4pt}

}

\setcounter{secnumdepth}{0}
{\scriptsize
	\bibliographystyle{abbrv}
	\bibliography{references}
}

\end{document}